\documentclass[%
	aps,
	pre,
	amsmath,
	amssymb,
	reprint,
	superscriptaddress,
	showpacs,
	floatfix,
]{revtex4-1}

\usepackage{graphicx}% Include figure files
\usepackage{dcolumn}% Align table columns on decimal point
\usepackage{bm}% bold math
\usepackage{multirow}

\begin{document}

\mathchardef\mhyphen="2D

\title{U.S. stock market interaction network as learned by the Boltzmann Machine}

\author{Stanislav S. Borysov}
\email[]{stanislav@smart.mit.edu}
%\thanks{}
\altaffiliation{Present address: Singapore-MIT Alliance for Research and Technology, 1 CREATE Way, \#09-02, Create Tower, 138602 Singapore}
\affiliation{Nordita, KTH Royal Institute of Technology and Stockholm University, Roslagstullsbacken 23, SE-106 91 Stockholm, Sweden}
\affiliation{Nanostructure Physics, KTH Royal Institute of Technology, Roslagstullsbacken 21, SE-106 91 Stockholm, Sweden}

\author{Yasser Roudi}
\affiliation{Nordita, KTH Royal Institute of Technology and Stockholm University, Roslagstullsbacken 23, SE-106 91 Stockholm, Sweden}
\affiliation{The Kavli Institue for Systems Neuroscience, NTNU, 7030 Trondheim}

\author{Alexander V. Balatsky}
\affiliation{Nordita, KTH Royal Institute of Technology and Stockholm University, Roslagstullsbacken 23, SE-106 91 Stockholm, Sweden}
\affiliation{Institute for Materials Science, Los Alamos National Laboratory, Los Alamos, NM 87545, USA}

\date{\today}

\begin{abstract}
We study historical dynamics of joint equilibrium distribution of stock returns in the U.S. stock market using the Boltzmann distribution model being parametrized by external fields and pairwise couplings. Within Boltzmann learning framework for statistical inference, we analyze historical behavior of the parameters inferred using exact and approximate learning algorithms. Since the model and inference methods require use of binary variables, effect of this mapping of continuous returns to the discrete domain is studied. The presented analysis shows that binarization preserves market correlation structure. Properties of distributions of external fields and couplings as well as industry sector clustering structure are studied for different historical dates and moving window sizes. We found that a heavy positive tail in the distribution of couplings is responsible for the sparse market clustering structure. We also show that discrepancies between the model parameters might be used as a precursor of financial instabilities.
\end{abstract}

%\keywords{}

\maketitle

%-------------------------------------------------------------------------------
\section{Introduction}
\label{intro}
%-------------------------------------------------------------------------------
Price formation on a financial market is a complex problem: It reflects opinion of investors about true value of the asset in question, policies of the producers,  external regulation and many other factors. Given the big number of factors influencing price, many of which unknown to us, describing price formation essentially requires probabilistic approaches. In the last decades, synergy of methods from various scientific areas has opened new horizons in understanding the mechanisms that underlie related problems. One of the popular approaches is to consider a financial market as a complex system, where not only a great number of constituents plays crucial role but also non-trivial interaction properties between them \cite{RePEc:oxp:obooks:9780198526650,sornette2009stock}. For example, related interdisciplinary studies of complex financial systems have revealed their enhanced sensitivity to fluctuations and external factors near critical events \cite{RevModPhys.74.47,Kacperski199953,PhysRevLett.96.068701,Barabasi15101999} with overall change of internal structure \cite{Dorogovtsev:2003:ENB:1212782,Bornholdt:2003:HGN:640635}. This can be complemented by the research devoted to equilibrium \cite{Gai08082010,doi:10.1137/S003614450342480,Phase_transitions,Nature.410:268} and non-equilibrium \cite{PhysRevE.74.056108,PhysRevE.67.026120} phase transitions.

In general, statistical modeling of the state space of a complex system requires writing down the probability distribution over this space using real data. In a simple version of modeling, the probability of an observable configuration (state of a system) described by a vector of variables $\mathbf{s}$ can be given in the exponential form
\begin{equation}
	p(\mathbf{s}) = \mathcal{Z}^{-1}\exp\left\{-\beta\mathcal{H}(\mathbf{s}) \right\},
	\label{eq:distrib}
\end{equation}
where $\mathcal{H}$ is the Hamiltonian of a system, $\beta$ is inverse temperature (further $\beta\equiv 1$ is assumed) and $\mathcal{Z}$ is a statistical sum. Physical meaning of the model's components depends on the context and, for instance, in the case of financial systems, $\mathbf{s}$ can represent a vector of stock returns and $\mathcal{H}$ can be interpreted as the inverse utility function \cite{Bury20131375}. Generally, $\mathcal{H}$ has parameters defined by its series expansion in $\mathbf{s}$. Basing on the maximum entropy principle \cite{PhysRev.106.620,PhysRev.108.171}, expansion up to the quadratic terms is usually used, leading to the pairwise interaction models. In the equilibrium case, the Hamiltonian has form
\begin{equation}
	\mathcal{H}(\mathbf{s}) = - \mathbf{h}^\intercal\mathbf{s} - \mathbf{s}^\intercal\mathbf{J}\mathbf{s},
	\label{eq:distrib_eq}
\end{equation}
where $\mathbf{h}$ is a vector of size $N$ of external fields and $\mathbf{J}$ is a symmetric $N\times N$ matrix of couplings ($\intercal$ denotes transpose). The model may also involve hidden states (nodes), which are not directly observable, but here we restrict ourselves to considering the visible nodes case only.

The energy-based models represented by Eq.~(\ref{eq:distrib}) play essential role not only in statistical physics but also in neuroscience (models of neural networks \cite{ising_neural,PhysRevE.79.051915}) and machine learning (also known as Boltzmann machines \cite{Lecun06atutorial}). Recently, applications of the pairwise interaction models to financial markets have been also explored \cite{Bury20131375,Maskawa2002563,bury1,1742-5468-2014-7-P07008}. Given topological similarities between neural and financial networks \cite{fnsys}, these systems can be considered as examples of complex adaptive systems \cite{doi:10.1080/713665542}, which are characterized by the adaptation ability to changing environment, trying to stay in equilibrium with it. From this point of view, market structural properties, e.g. clustering and networks \cite{refId0,1402-4896-2003-T106-011,Networks1}, play important role for modeling of stock prices. Adaptation (or learning) in these systems implies change of the parameters of $\mathcal{H}$ as financial and economic systems evolve. Using statistical inference for the model's parameters, the main goal is to have a model capable of reproducing statistical observables based on time series for a particular historical period. In the pairwise case, the objective is
\begin{equation}
\begin{array}{lcl}
	\langle s_{i} \rangle_\mathrm{data} &=& \langle s_{i} \rangle_\mathrm{model},\\
	\langle s_{i}s_{j} \rangle_\mathrm{data} &=& \langle s_{i}s_{j} \rangle_\mathrm{model},
\end{array}
	\label{eq:corr_eq}
\end{equation}
where $i=1,\dots,N$ and angular brackets denote statistical averaging over time. 

Having specified general mathematical model, one can also discuss similarities between financial and infinite-range magnetic systems in terms of phenomena related, e.g. extensivity, order parameters and phase transitions, etc. These features can be captured even in the simplified case, when $s_i$ is a binary variable taking only two discrete values. Effect of the mapping to a binarized system, when the values $s_i=+1$ and $s_i=-1$ correspond to profit and loss respectively, is also studied in the paper. In this case, diagonal elements of the coupling matrix, $J_{ii}$, are zero because $s_i^2=1$ terms do not contribute to the Hamiltonian. It is worth stressing that the current investigation develops ideas outlined in the previous studies \cite{Bury20131375,bury1,1742-5468-2014-7-P07008,Maskawa2002563} in a way that the effect of binarization, comparison of learning algorithms, evolution and scaling properties of the parameters distributions are studied in a more systematic fashion.

The paper is organized as follows. In Section~\ref{sec:Methods}, basic statistical definitions and inference methods for $\mathbf{h}$ and $\mathbf{J}$ are presented. In Section~\ref{sec:Results}, effect of binarization of stock returns and historical evolution of the model parameters are discussed. Finally, the main findings of our investigation are summarized in Section~\ref{sec:Discussion}.

%-------------------------------------------------------------------------------
\section{Data and methods}
\label{sec:Methods}
%-------------------------------------------------------------------------------
We study historical dynamics of the U.S. stock market using $N=71$ stock prices time series \cite{yahoo} from the Standard \& Poor's 500 index (hereafter S\&P~500) listed in Table~\ref{tab:companies}. We analyze discrete daily closing prices, $S_{i}(t)$, starting from 1990 till 2013 (5828 trading days), which are converted to logarithmic returns, $s_{i}^\mathrm{raw}(t) = \ln[S_{i}(t)/S_{i}(t-1)]$.

%-------------------------------------------------------------------------------
\subsection{Basic statistical analysis of financial time series}
\label{sec:preliminaries}
\begin{figure}
\centering
\resizebox{0.4\textwidth}{!}{%
 \includegraphics{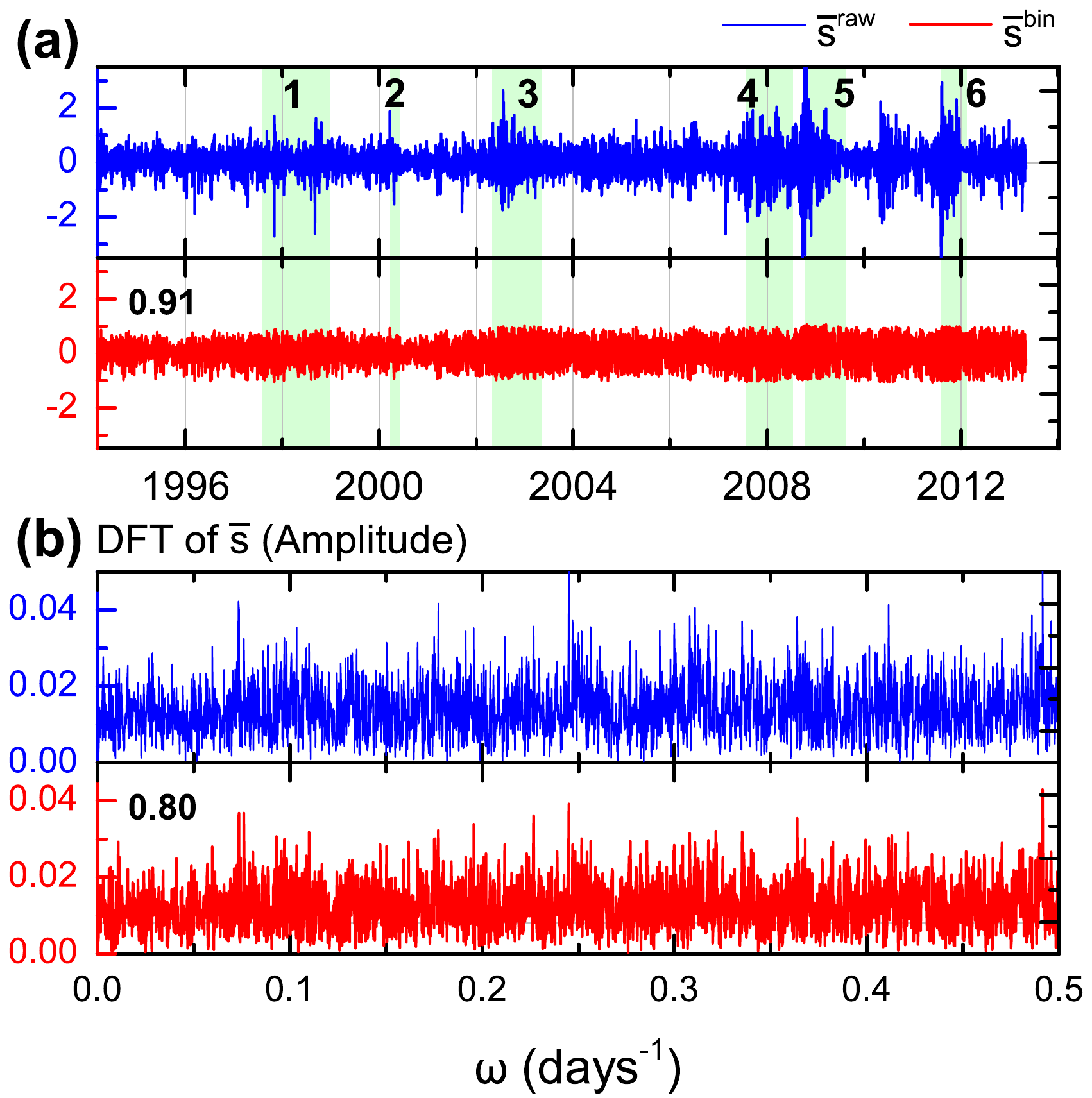}
}
 \caption{Historical dynamics of the mean raw and binary returns (a) and amplitudes of their Discrete Fourier transforms (b). The distance between two labeled dates is 1000 trading days. A few major financial crises are highlighted with the light green background: (1) Asian and Russian crisis of 1997--1998, (2) dot-com bubble, (3) U.S. stock market downturn of 2002, (4) U.S. housing bubble, (5) bankruptcy of Lehman Brothers followed by the global financial crisis, (6) European sovereign debt crisis. The number in each panel shows overall historical correlation between the corresponding series. Historical values of average return, economic cycles and frequency of crashes are preserved for the binary returns despite the maximum magnitude being bounded.}
\label{fig:bin_vs_raw_returns_FT}
\end{figure}

\begin{figure*}
\centering
\resizebox{0.4\textwidth}{!}{%
 \includegraphics{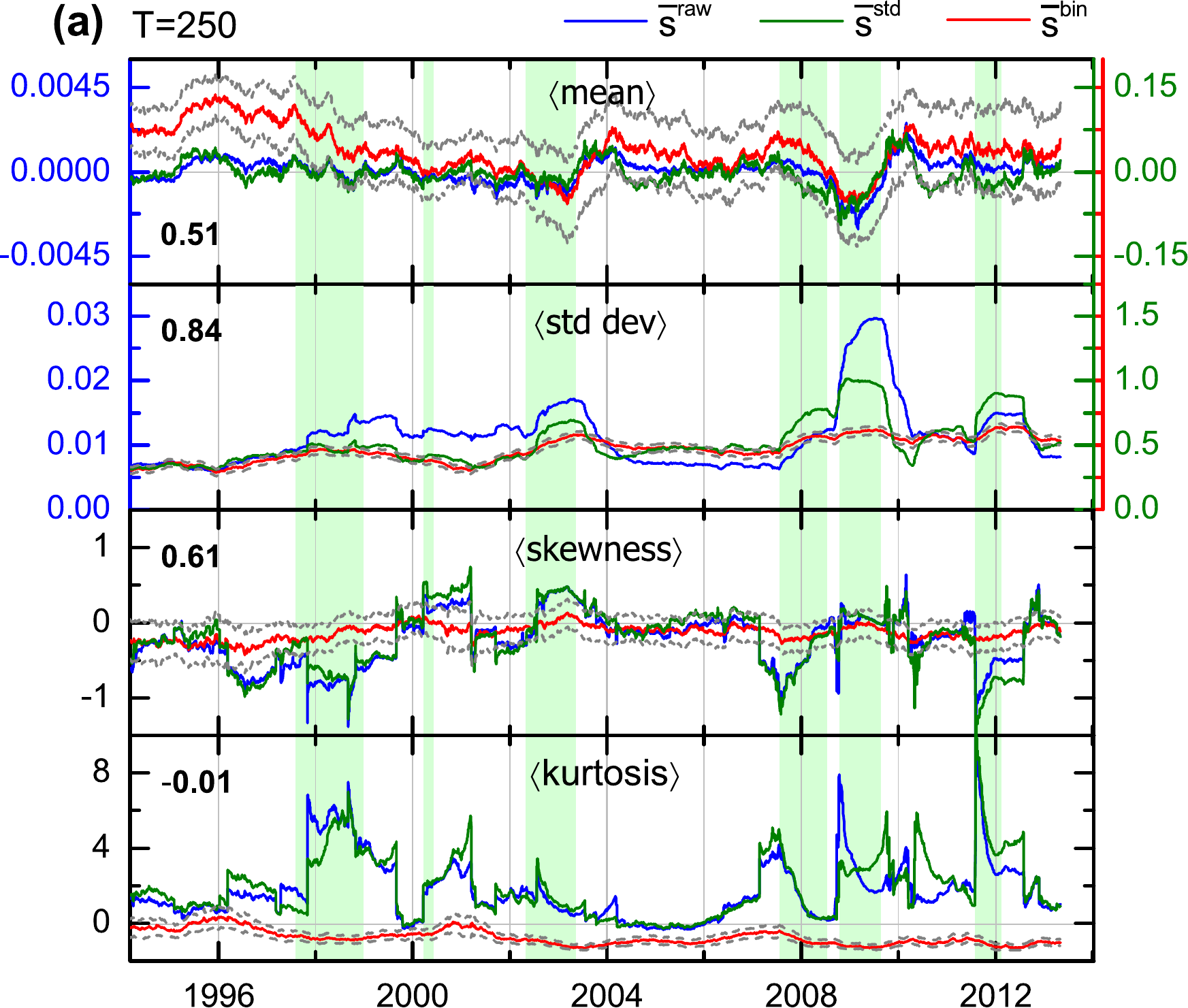}
}
\resizebox{0.4\textwidth}{!}{%
 \vspace{3mm}
 \includegraphics{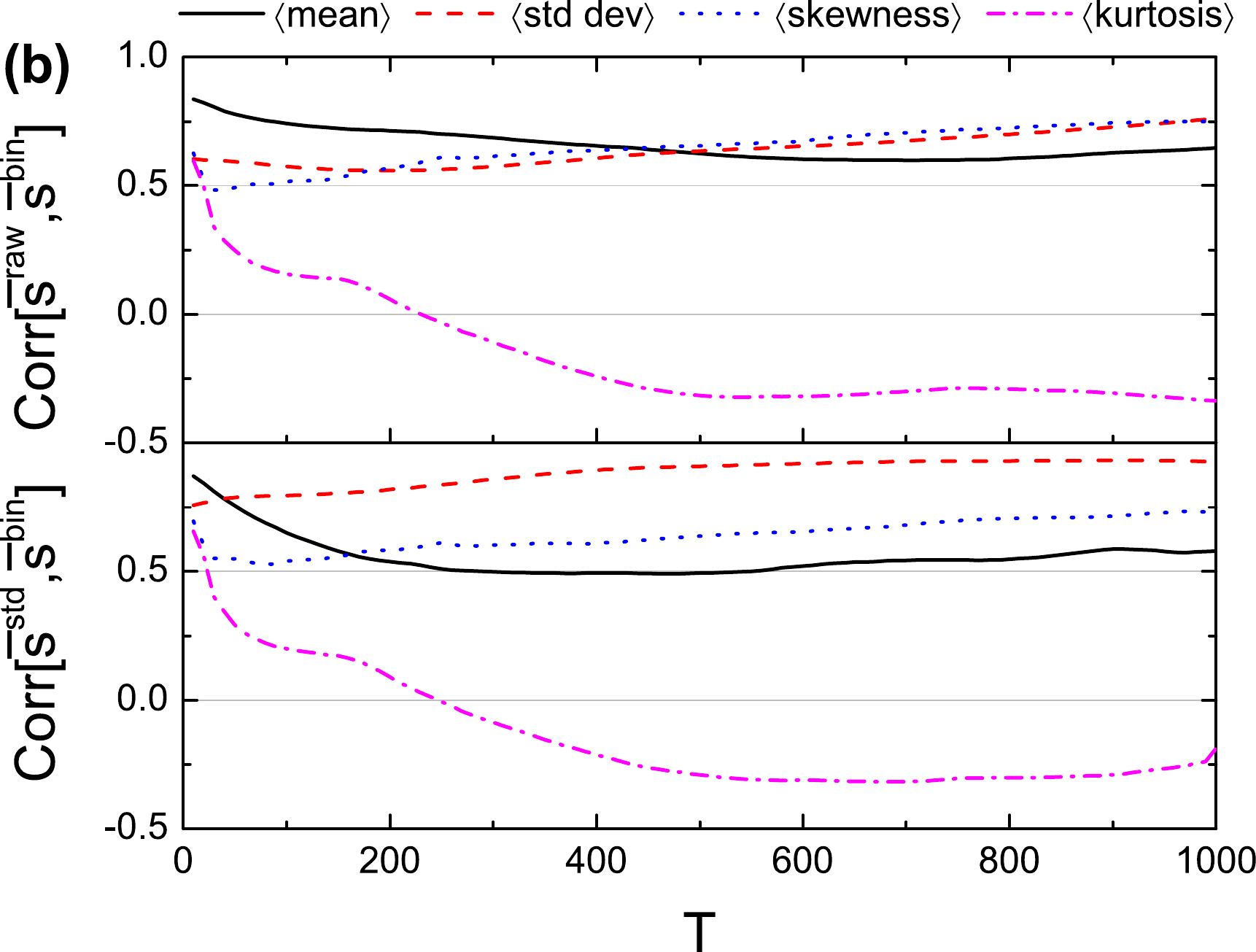}
}
 \caption{Historical dynamics of the first four temporal moments of the distribution of mean market return (a). Top-bottom: temporal mean, standard deviation, skewness and kurtosis calculated using SMA window of 250 days (approximately one trading year) for the raw ($\overline{s}^\mathrm{raw}$, blue), standardized ($\overline{s}^\mathrm{std}$, green) and binary ($\overline{s}^\mathrm{bin}$, red) returns of $71$ U.S. stocks (Table~\ref{tab:companies}). 95\% confidence intervals for the moments of the distribution of $\overline{s}^\mathrm{bin}$ are calculated using bootstrapping algorithm and denoted with the gray dashed lines. The number in each panel corresponds to overall historical correlation between the time series for $\overline{s}^\mathrm{std}$ and $\overline{s}^\mathrm{bin}$. Dependence of these overall correlations on the moving window size is shown in (b). Binary returns behave similar to raw and standardized returns, however information about kurtosis is completely lost.}
\label{fig:bin_vs_raw_vs_std_returns}
\end{figure*}

\begin{figure*}
\centering
\resizebox{0.4\textwidth}{!}{%
	\includegraphics{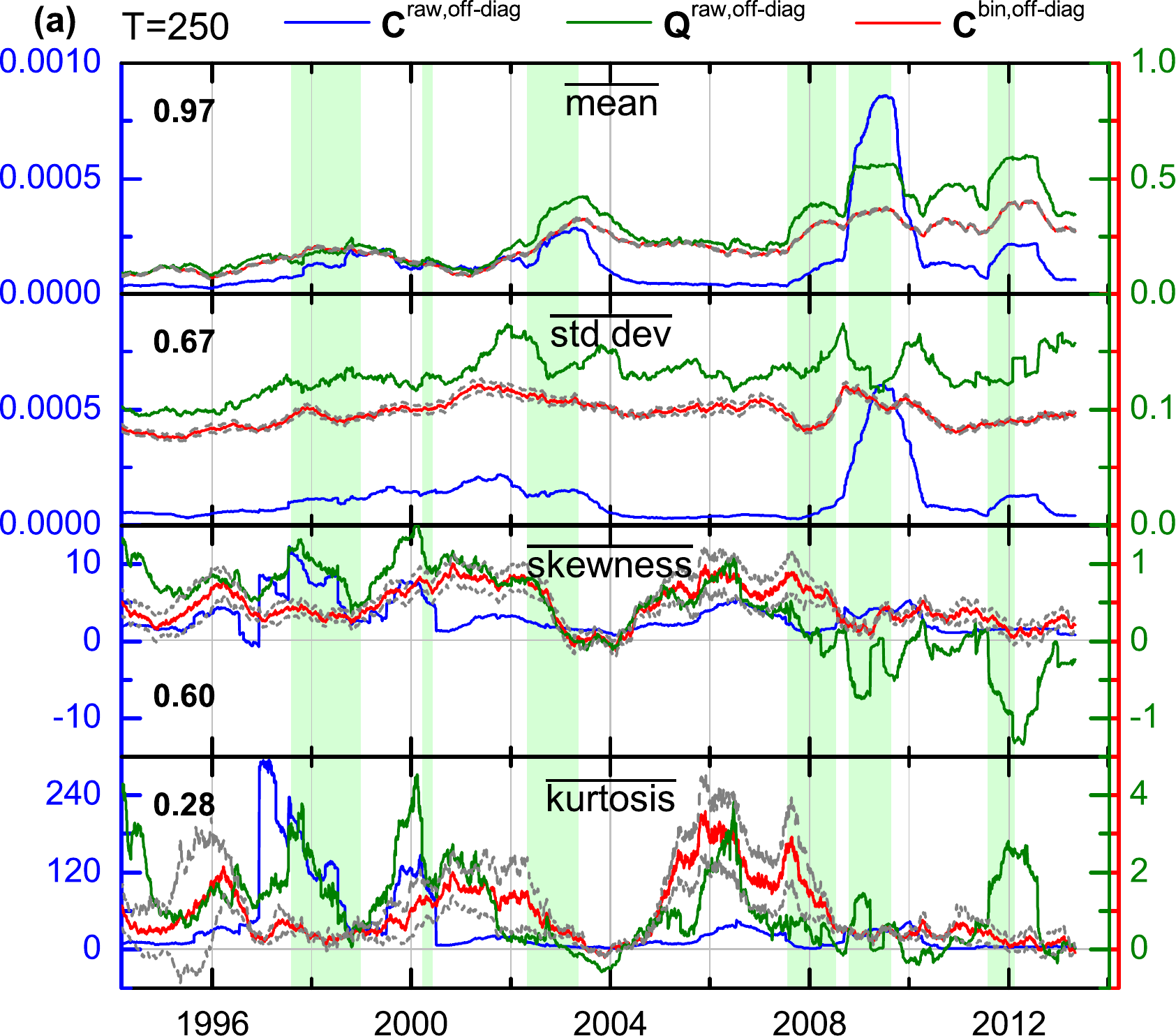}
}
\resizebox{0.4\textwidth}{!}{%
	\vspace{3mm}
 	\includegraphics{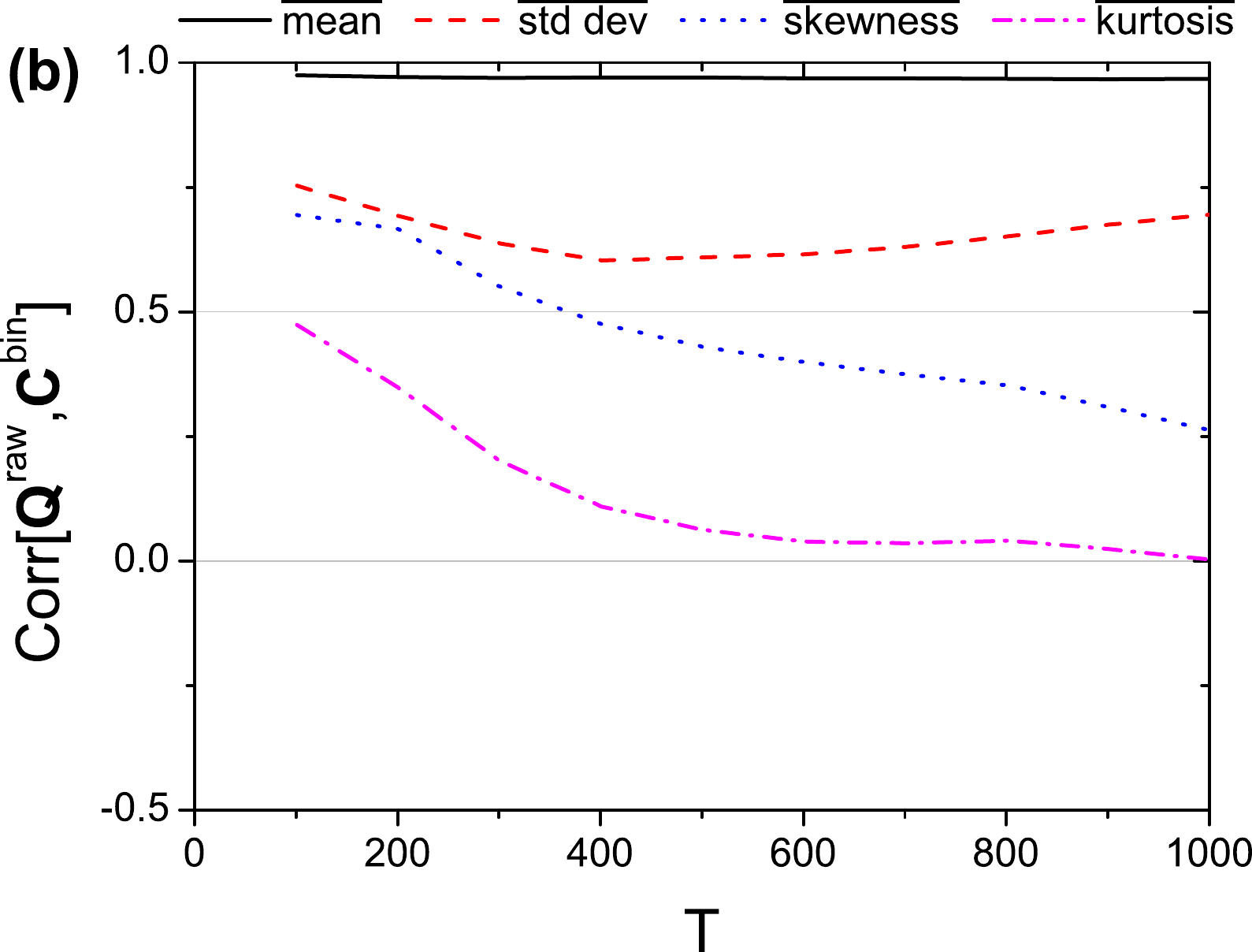}
}
 \caption{Historical dynamics of the first four moments of the distribution of off-diagonal elements of covariance ($\mathbf{C}^\mathrm{raw}$, blue) and correlation ($\mathbf{Q}^\mathrm{raw}$, green) matrices of raw returns, and covariance matrix of binary returns ($\mathbf{C}^\mathrm{bin}$, red) calculated using SMA window of 250 days (a). Top-bottom: mean, standard deviation, skewness and kurtosis of the off-diagonal elements of the matrices. 95\% confidence intervals for the moments of distribution of $C_{ij}^\mathrm{bin}$ are calculated using bootstrapping algorithm and denoted with the gray dashed lines. The number in each panel corresponds to overall historical correlation between the time series for $\mathbf{Q}^\mathrm{raw}$ and $\mathbf{C}^\mathrm{bin}$. Dependence of these overall correlations on the moving window size is shown in (b). Binarization makes covariance matrix similar to the correlation matrix of raw returns.}
\label{fig:bin_cov_vs_raw_corr}
\end{figure*}

%-------------------------------------------------------------------------------
The top panel in Fig.~\ref{fig:bin_vs_raw_returns_FT}(a) shows historical data for the average stock return. In order to extract long-term trends from the time series, we employ a simple moving window (or simple moving average, SMA) approach. It acts like a low-pass filter, averaging out high frequency components, which are usually related to noise. 
Within the SMA approach, data is divided into chunks (or windows) of size $T$, assuming the time series to be stationary on this scale. In this case, $t$th chunk corresponds to $T$ equally weighted previous values of daily log-returns (including the current one). For each set of chunks, one can calculate different statistical characteristics, such as \emph{average}
\begin{equation}
	\langle s_{i} \rangle = \frac{1}{T}\sum\limits_{t'=0}^{T-1}s_{i}(t'),
	\label{eq:average}
\end{equation}
and {\it covariance matrix}, $\mathbf{C}$ ($N\times N$), with the elements
\begin{equation}
	C_{ij} = \langle s_{i}s_{j} \rangle - \langle s_{i} \rangle \langle s_{j} \rangle,
	\label{eq:cov}
\end{equation}
where $T\geq N$ is assumed for the covariance matrix to be positive definite. Hereinafter, angular brackets $\langle\quad\rangle$ denote averaging over time (historical values), while bar $\bar{\quad}$ denotes averaging over index (vector or matrix elements). It is also possible to investigate nonlinear dependence between time series using more sophisticated statistical concepts \cite{lee2000statistics} or nonlinear data transformations, however only the simplest linear case is considered in the current paper. Series \emph{variance} is autocovariance, $\sigma^2_{i} \equiv C_{ii}$, where $\sigma$ denotes \emph{standard deviation} or \emph{volatility} in finance. Usually, SMA volatility serves as the simplest risk measure quantifying stability of returns. In order to quantify deviation from the normal distribution, it is also useful to define higher-order moments, such as {\it skewness}
\begin{equation}
	\left\langle\mathrm{Skew}\left(s_{i}\right)\right\rangle = \left\langle \left(\frac{s_i - \langle s_{i} \rangle}{\sigma_i}\right)^3 \right\rangle
	\label{eq:skewness}
\end{equation}
and {\it kurtosis} (also known as excess kurtosis)
\begin{equation}
	\left\langle\mathrm{Kurt}\left(s_{i}\right)\right\rangle = \frac{\left\langle \left(s_i - \langle s_{i} \rangle\right)^4 \right\rangle}{\sigma_i^4} - 3,
	\label{eq:kurtosis}
\end{equation}
which both equal zero in the Gaussian case. Indeed, SMA  filter allows one to extract long-term trends in the market and, for a large value of $N$, various moments of the distribution of returns can be used to identify market crashes (Figs.~\ref{fig:bin_vs_raw_vs_std_returns} and \ref{fig:bin_cov_vs_raw_corr}). Henceforth, we will call the considered portfolio ``market" as it represents a big number of the top companies from S\&P500 index. Also, we provide corresponding confidence intervals for these moments using basic bootstrapping algorithm (sampling with replacement) which are depicted in Figs.~\ref{fig:bin_vs_raw_vs_std_returns} and \ref{fig:bin_cov_vs_raw_corr}. However, a more comprehensive discussion on the moments estimates for the non-stationary time series, besides moving window and correlated variable effects  \cite{pindyck1998econometric}, would also require considering nontrivial long-term memory effects and intraday correlations \cite{PhysRevE.70.026110,0295-5075-90-6-68001} inherent to financial time series, what goes beyond the scope of the current investigation.

As mentioned in the Introduction, to reduce the amount of data required for our inference, we focus on a binarized version of the returns and not on the raw data. We thus define the binarized version of the returns as
\begin{equation}
	s_{i}^\mathrm{bin} = \mathrm{sign}\left(s_{i}^\mathrm{raw}\right).
	\label{eq:bin}
\end{equation}
This procedure also mitigates the scaling problem of multiple series since it assigns the same value of stock return independently of its absolute value. Another common technique to deal with this problem is standardization
\begin{equation}
	s^\mathrm{std}_i = \frac{s_i^\mathrm{raw} - \langle s_{i}^\mathrm{raw} \rangle}{\sigma_i}.
	\label{eq:norm}
\end{equation} 
In this case, {\it correlation matrix}, $\mathbf{Q}$ ($N\times N$), is a normalized covariance matrix with the elements 
\begin{equation}
Q_{ij} = \frac{C_{ij}}{\sigma_{i}\sigma_{j}}.
	\label{eq:norm_cov}
\end{equation}
Standardization procedure is known to preserve market behavior and widely used for the statistical analysis of the financial time series \cite{PhysRevE.84.026109}. Effects of these transformations defined by Eqs.~(\ref{eq:bin})--(\ref{eq:norm_cov}) are shown in Figs.~\ref{fig:bin_vs_raw_returns_FT}--\ref{fig:bin_cov_vs_raw_corr} and will be discussed further in Section \ref{sec:Results} where some simple statistical properties of the binarized return versus the raw and standardized returns are compared. Before this, however, we briefly describe the inference procedure.

%-------------------------------------------------------------------------------
\subsection{Equilibrium Boltzmann learning methods}
\label{sec:ising_inference}
%-------------------------------------------------------------------------------
We harness inference methods based on maximizing of the model's likelihood $\mathcal{L}(\mathbf{h}, \mathbf{J} \mid \mathbf{s}^\mathrm{data})$. In the equilibrium case, \emph{exact learning} of the Hamiltonian [Eq.~(\ref{eq:distrib_eq})] parameters implies solving Eq.~(\ref{eq:corr_eq}) in a self-consistent way, where corrections $\delta h_{i}$ and $\delta J_{ij}$ on each learning step can be calculated as
\begin{equation}
\begin{array}{lcl}
	\delta h_{i} &=& \eta_h\left(\langle s_{i} \rangle_\mathrm{data} - \langle s_{i} \rangle_\mathrm{model}\right),\\
	\delta J_{ij} &=& \eta_J\left(\langle s_{i}s_{j} \rangle_\mathrm{data} - \langle s_{i}s_{j} \rangle_\mathrm{model}\right).
\end{array}
	\label{eq:Eq-Exact}
\end{equation}
Here, $\eta_h$ and $\eta_J$ are learning rates, $\langle\cdot\rangle_\mathrm{data}$ are empirical (observed) moments and $\langle\cdot\rangle_\mathrm{model}$ are the moments sampled from the model using Monte Carlo (MC) methods. The exact learning algorithm always yields optimal values for $\mathbf{h}$ and $\mathbf{J}$ if there are no hidden nodes in the system \cite{125867}.

Being in general slow, the exact learning algorithm might be substituted by the approximate inference methods \cite{PhysRevE.58.2302} which are based on expansion of the free energy of a system for small fluctuations around its mean value. The first-order (\emph{na\"ive}) approximation within the mean field theory (nMF) gives
\begin{equation}
\begin{array}{lcl}
	\mathbf{J}^\mathrm{nMF} = \mathbf{A}^{-1} - \mathbf{C}^{-1}, \\
	h^\mathrm{nMF}_{i} = \tanh^{-1} \langle s_{i} \rangle - \sum\limits_{j=1}^{N} J_{ij}^{\mathrm{nMF}} \langle s_{i} \rangle ,
\end{array}
\label{eq:Eq-nMF}
\end{equation}
where $A_{ij} = (1 - \langle s_{i} \rangle^2)\delta_{ij}$ and $\delta_{ij}$ is the Kronecker delta. Here, taking into account the diagonal element $J_{ii}$ (which is usually discarded) for the calculation of corresponding $h_i$ improves accuracy of the approximation, being known as the diagonal-weight trick \cite{PhysRevE.58.2302}. The second-order correction to the nMF approximation requires solving \emph{Thouless-Anderson-Palmer} (TAP) equations 
\begin{equation}
\begin{array}{lcl}
	\left(\mathbf{C}^{-1}\right)_{ij} = - J^{\mathrm{TAP}}_{ij} - 2 \left( J^{\mathrm{TAP}}_{ij} \right)^2\langle s_{i} \rangle \langle s_{j} \rangle ,\\
	h^\mathrm{TAP}_{i} = h^\mathrm{nMF}_{i} - \langle s_{i} \rangle \sum\limits_{j=1}^{N} \left(J_{ij}^{\mathrm{TAP}}\right)^2 \left(1 - \langle s_{i} \rangle^2\right) ,
\end{array}
	\label{eq:Eq-TAP}
\end{equation}
where Eq.~(\ref{eq:Eq-nMF}) should be used instead for the calculation of the external fields if the diagonal-weight trick is used.

Another class of approximations can be derived using expansion of the free energy in pairwise correlations. 
The simplest \emph{independent-pair} (IP) approximation assumes independence of every stock pair from the rest of the system. In this case, couplings and external fields can be found as \cite{PhysRevE.79.051915}
\begin{equation}
\begin{array}{lcl}
	J^\mathrm{pair}_{ij} = \frac{1}{4}\ln\left[\frac{\left(1 + m_i + m_j + C^\ast_{ij}\right)\left(1 - m_i - m_j + C^\ast_{ij}\right)}{\left(1 - m_i + m_j - C^\ast_{ij}\right)\left(1 + m_i - m_j - C^\ast_{ij}\right)}\right],\\
	h^\mathrm{pair}_{i} = \frac{1}{2}\ln\left(\frac{1+m_i}{1-m_i}\right)-\sum\limits_j^NJ^\mathrm{pair}_{ij}m_j + O\left(\beta^2\right)
\end{array}
	\label{eq:Eq-pair}
\end{equation}
where $C^\ast_{ij}=C_{ij}+m_im_j$. \emph{Sessak and Monasson} (SM) derived higher-order corrections to the IP approximation in a closed form using other terms in the perturbative correlation expansion \cite{1751-8121-42-5-055001} 
\begin{equation}
\begin{array}{lcl}
	J^\mathrm{SM}_{ij} = J^\mathrm{nMF}_{ij} + J^\mathrm{pair}_{ij} - \frac{C_{ij}}{\left(1 - m_i^2\right)\left(1 - m_j^2\right) - \left(C_{ij}\right)^2},\\
	h^\mathrm{SM}_{i} = h^\mathrm{pair}_{i}
\end{array}
	\label{eq:Eq-SM}
\end{equation}

It is also worth noting that there have been developed a few other approximate inference schemes tailored for different system regimes, such as a pseudo-maximum likelihood inference using all data \cite{PhysRevLett.108.090201}, minimum probability flow \cite{PhysRevLett.107.220601} or mean field approximations for low temperatures \cite{PhysRevLett.109.050602}. For example, pseudo-maximum likelihood inference \cite{PhysRevLett.108.090201}, being more suitable for sparse connections, can be studied in a context of the correlation clustering structures described in Subsection \ref{sec:clustering_structure}.

%-------------------------------------------------------------------------------
\section{Results and discussion}
\label{sec:Results}
%-------------------------------------------------------------------------------
%-------------------------------------------------------------------------------
\subsection{Effect of binarization}
\label{sec:effect_of_bin}
%-------------------------------------------------------------------------------
The mapping defined by Eq.~(\ref{eq:bin}) obviously affects information contained in the time series. In order to estimate this effect, we compare time series for the average raw and binary returns first. As Fig.~\ref{fig:bin_vs_raw_returns_FT} shows, correlation coefficient between historical values of  $\overline{s}^\mathrm{raw}$ and $\overline{s}^\mathrm{bin}$ is very high ($0.91$). Moreover, correlation between amplitudes of their Fourier transforms is also high ($0.80$), suggesting that signatures of economic cycles and frequency of market crashes are preserved in the binarized time series. However, maximum magnitude of binary returns is obviously bounded by the definition. 

Comparison of the filtered time series using SMA also suggests the idea that the information about the market trends is preserved in the binarized time series. With this aim, we compare historical evolution of the first four moments of the distribution of average binarized versus raw and standardized returns. The results presented in Fig.~\ref{fig:bin_vs_raw_vs_std_returns} indicate that the binary returns behave similarly to the raw and standardized returns, preserving dynamics of the first two moments and less so about the third moment, while information about kurtosis is lost for all $T$ [Fig.~\ref{fig:bin_vs_raw_vs_std_returns}(b)]. Regarding the pairwise correlations, Figure~\ref{fig:bin_cov_vs_raw_corr} shows that the covariance matrix of the binarized returns becomes similar to the correlation matrix of the raw returns. Indeed, their off-diagonal elements follow similar distributions with very high correlation between means. For higher-order moments, correlation decreases with $T$ [Fig.~\ref{fig:bin_cov_vs_raw_corr}(b)]. 

\begin{figure*}
\centering
\resizebox{0.4\textwidth}{!}{%
	\includegraphics{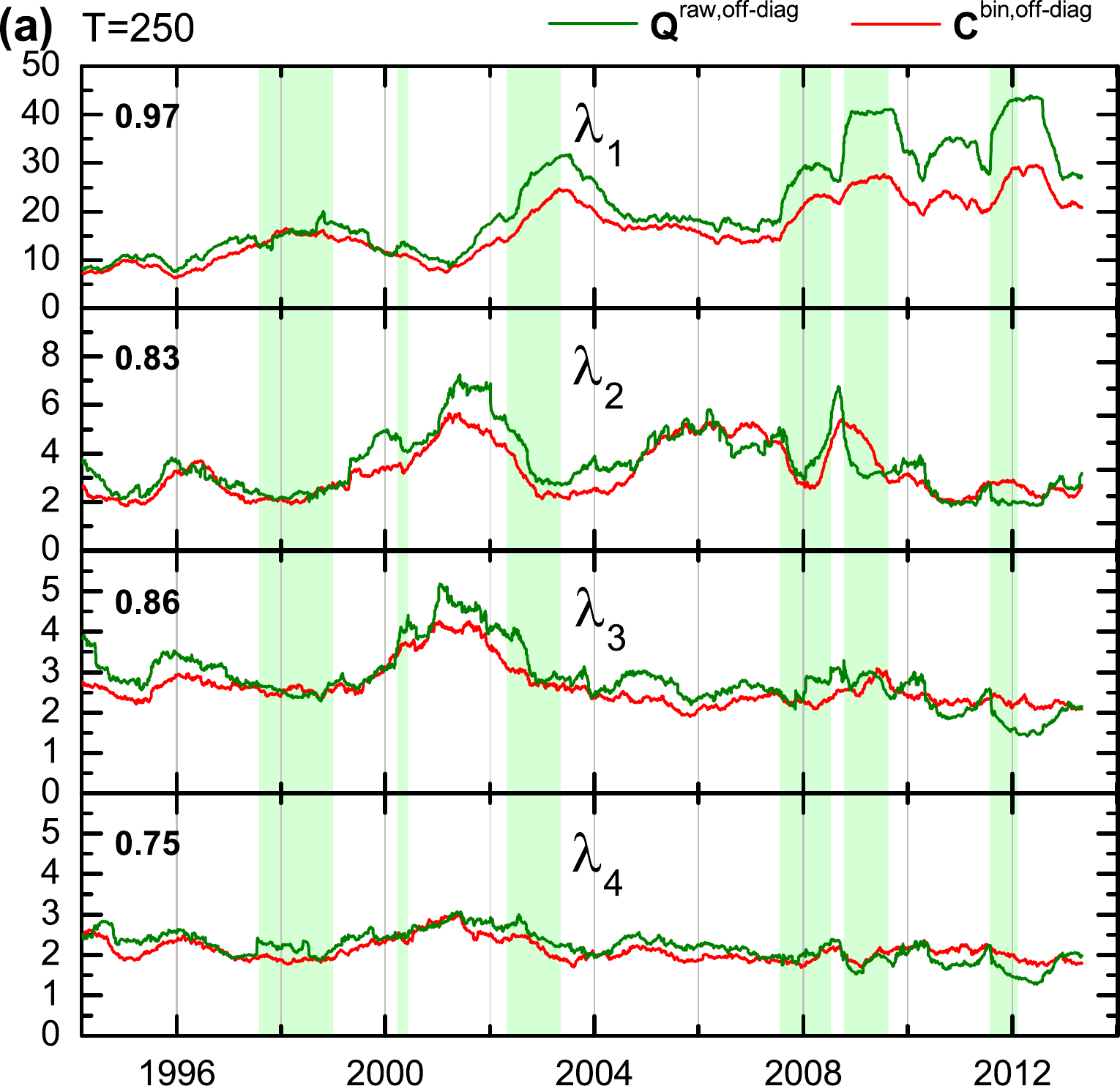}
}
\resizebox{0.4\textwidth}{!}{%
	\vspace{3mm}
 	\includegraphics{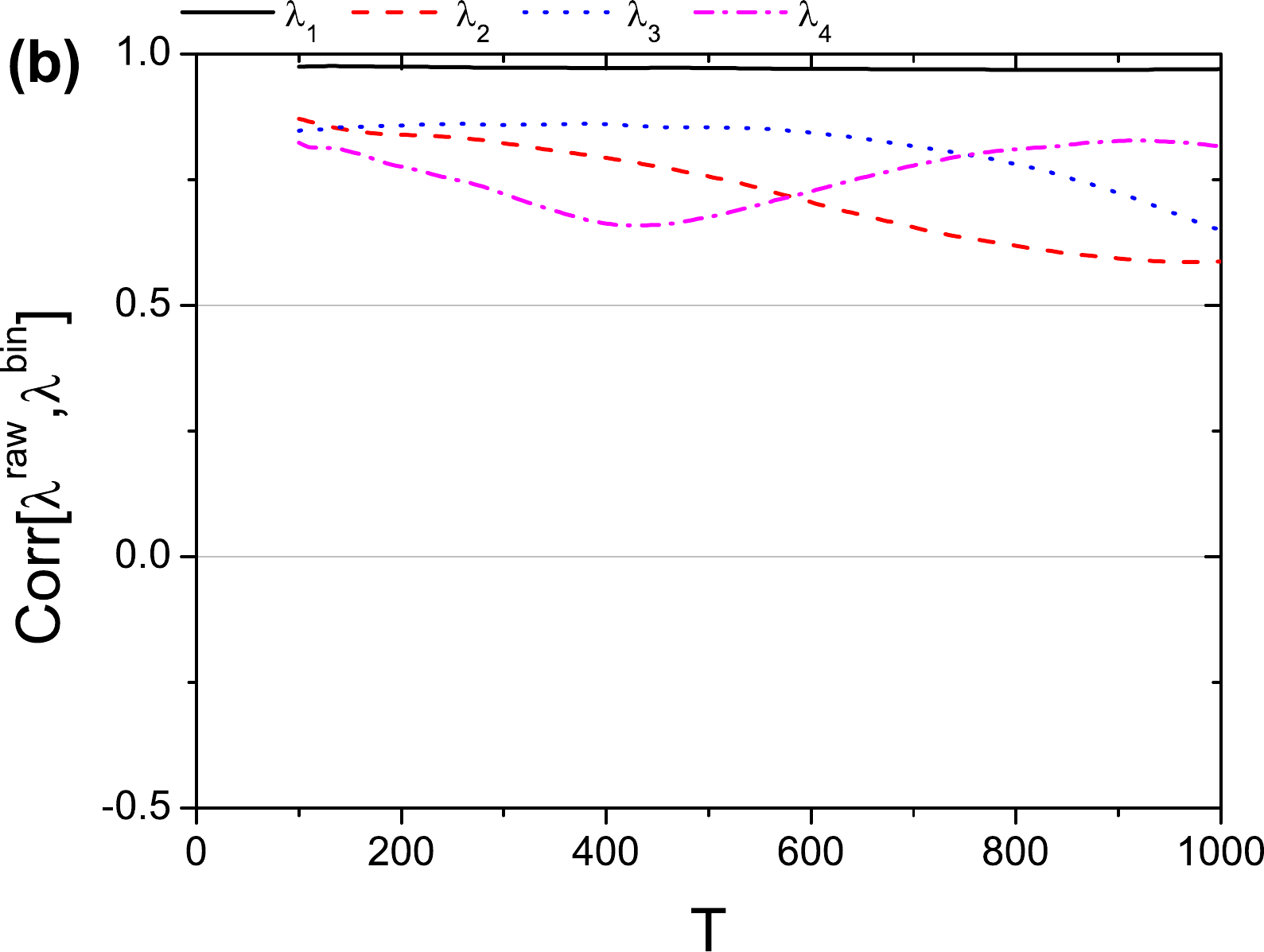}
}
 \caption{Historical dynamics of the four biggest eigenvalues of the correlation matrix of raw returns (green) and covariance matrix of binary returns (red) calculated using SMA window of 250 days (a). The number in each panel corresponds to overall historical correlation between the time series. Dependence of these overall correlations on the moving window size is shown in (b). Binarization preserves biggest eigenvalues of the correlation matrix.}
\label{fig:bin_cov_vs_raw_corr_eigenvals}
\end{figure*}

Another way to explore effects of the binarization procedure is to study eigenvalue distribution of correlation matrices for the original and transformed time series, since it is known to be different from
the distribution of a random correlation matrix \cite{PhysRevLett.83.1467}. With this aim, we compare historical dynamic of the four biggest eigenvalues, which do not match the Wigner law. As Fig.~\ref{fig:bin_cov_vs_raw_corr_eigenvals} shows, all four values are well preserved and the biggest one corresponding to the ``market mode'' is in remarkable agreement after binarization independently of moving window size.

Having validated that binary returns indeed capture market historical behavior similarly to standardized returns, we proceed to the Boltzmann model inference and evolution of its parameters.

%-------------------------------------------------------------------------------
\subsection{Comparison of approximate and exact learning methods}
\label{sec:ising_inference_comparison}
%-------------------------------------------------------------------------------
Following the SMA approach with $T=250$ trading days (approximately one trading year), we study historical evolution of the external fields and couplings inferred for each chunk of the binarized time series. We employ four approximate (nMF, TAP, IP and SM) learning methods described in the previous section and compare them with the exact learning, where MC sampling is used for the latter. 

\begin{figure*}
\centering
\resizebox{0.85\textwidth}{!}{%
 \includegraphics{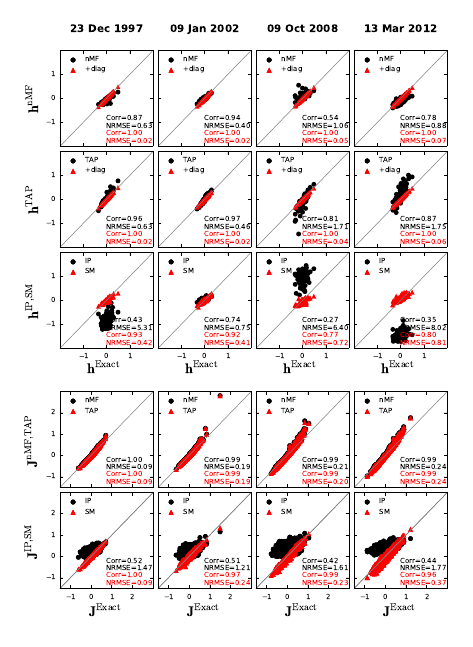}
}
 \caption{Comparison of external fields (three top rows) and couplings (two bottom rows) inferred using the exact and approximate (nMF, TAP, IP and SM) learning algorithms for four different historical dates and SMA window of $250$ trading days. The diagonal-weight trick is essential for correct inference of external fields in the mean field (nMF and TAP) cases. Both mean field approximations overestimate couplings with big absolute values. Accuracy of the IP algorithm is very low for both external fields (third row) and couplings (bottom row). SM corrections significantly improves the IP results and outperforms the TAP approximation for the positive outliers in the distribution of couplings.}
 \label{fig:Eq_Exact_vs_nMF_TAP}
\end{figure*}

Figure~\ref{fig:Eq_Exact_vs_nMF_TAP} shows comparison of the inferred parameters for four different historical dates. Without use of the diagonal-weight trick, inference of external fields in the nMF case works better when the dates far from market crashes are considered (top row in the two leftmost columns, corresponding to 22 Dec 1997 and 09 Jan 2002). TAP correction slightly improves the inference accuracy however producing overestimated values in comparison with the exact learning algorithm (second row). Both MF approximations perform worse for the data near market crashes (two top rows in the two rightmost columns, corresponding to 09 Oct 2008 and 13 March 2012), while the diagonal-weight trick allows to achieve almost perfect accuracy in both cases (red triangles in the first two rows), justifying the need of high-order corrections. The IP algorithm (third row) shows much worse performance for the all dates except 09 Jan 2002. Although higher-order SM corrections (third row) considerably improve accuracy of the IP external fields, it is still lower than in the MF cases. Couplings inferred using both MF approximations show the same trend, being more/less accurate far from/near crashes (fourth row). Although the bulk of the MF couplings is in an excellent agreement with the exact couplings even near crashes, positive outliers are overestimated. The couplings estimated using the IP algorithm are in a very poor agreement with the exact couplings for the all dates considered (bottom row), while the use of the SM correction yields considerable improvement (bottom row). Moreover, it correctly captures the biggest outliers, outperforming the MF algorithms. 

\begin{figure}
\centering
\resizebox{0.4\textwidth}{!}{%
 \includegraphics{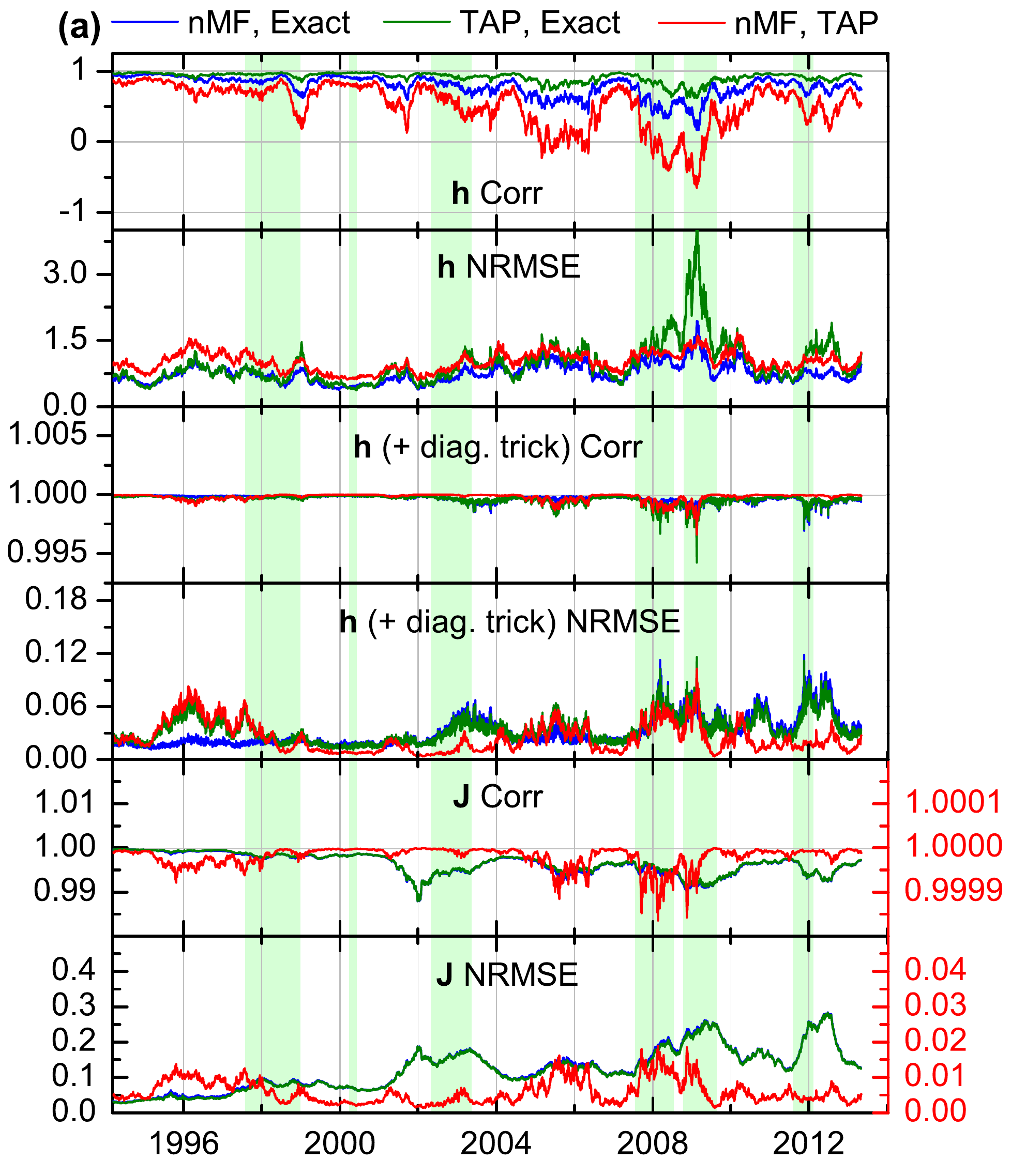}
}
\resizebox{0.36\textwidth}{!}{%
 \includegraphics{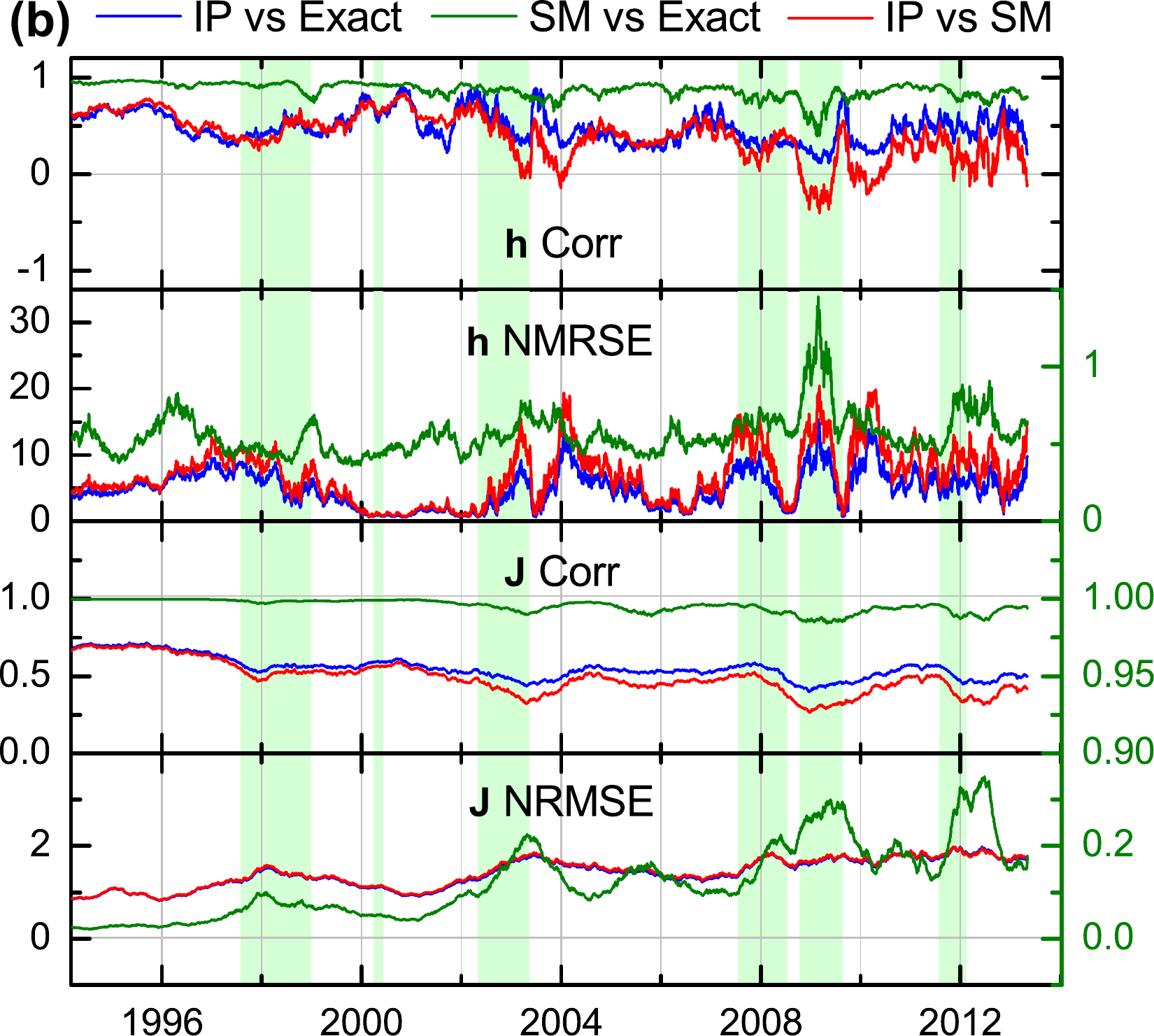}
}
 \caption{Comparison (normalized root mean square error and correlation coefficient) of the external fields and couplings inferred using different methods: (a) nMF versus exact (blue), TAP versus exact (green) and nMF versus TAP (red); (b) IP versus exact (blue), SM versus exact (green) and IP versus SM (red). Without use of the diagonal-weight trick, MF approximation quality of the external fields is very low and significantly drops during financial crashes [(a), two top panels], while making use of the trick improves it considerably [(a), two middle panels]. SM algorithm considerably improves quality of IP external fields, however being still lower than MF with the diagonal trick [(b), two top panels]. Couplings inferred using both MF methods are almost the same, with approximation quality being lower during the periods of financial crises [(a), two bottom panels]. The same behavior is observed for SM couplings [(b), two bottom panels].}
 \label{fig:Eq_hJ_corr_dist}
\end{figure}

Historical dynamics of the inference quality for approximate methods is depicted in Fig.~\ref{fig:Eq_hJ_corr_dist}, where normalized root mean square error
\begin{equation}
\mathrm{NRMSE}(\mathbf{x},\mathbf{y}) = \sqrt{\frac{\overline{\left(x_i - y_i\right)^2}}{\overline{\left(y_i - \overline{y}\right)^2}}} 
	\label{eq:nmrse}
\end{equation} 
and correlation between the model parameters inferred using different methods are presented. It confirms that the diagonal-weight trick considerably improves inference quality of the MF external fields, while couplings are systematically overestimated. Within the small correlation approximation, IP algorithm does not perform well for all historical dates, while quality of the SM couplings is comparable to the MF cases (however being still lower in general) and decreases for the periods where higher correlations are observed. Thus, the observed behavior indeed justifies use of TAP as a reliable approximation of true couplings and external fields (making use if the diagonal trick), which has been extensively used in the previous studies \cite{Bury20131375,bury1,1742-5468-2014-7-P07008,Maskawa2002563}. However, special care should be taken about overestimated positive outliers, where SM algorithm can be helpful.

\begin{figure}
\centering
\resizebox{0.23\textwidth}{!}{%
	\includegraphics{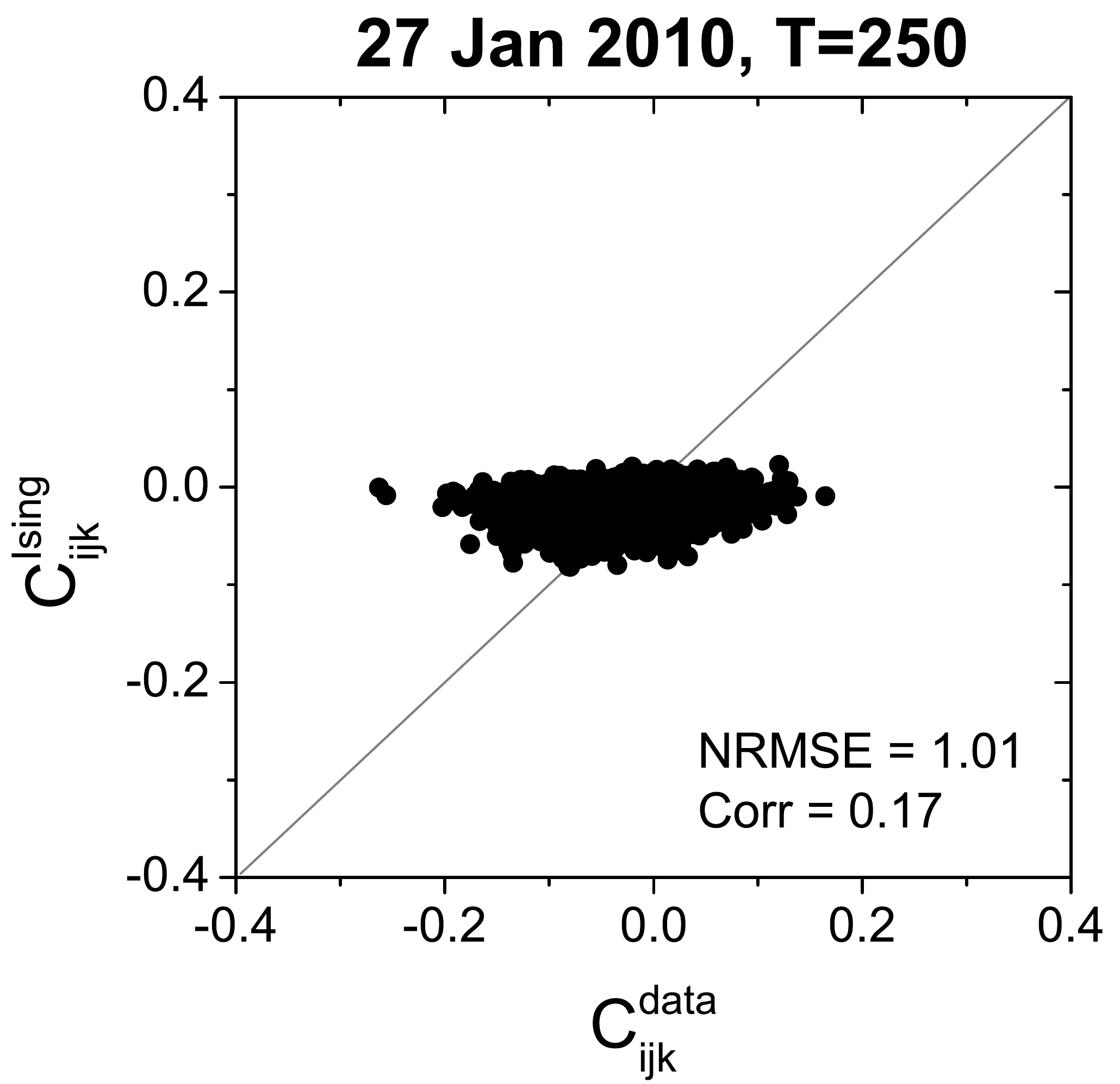}
}
\resizebox{0.23\textwidth}{!}{%
	\includegraphics{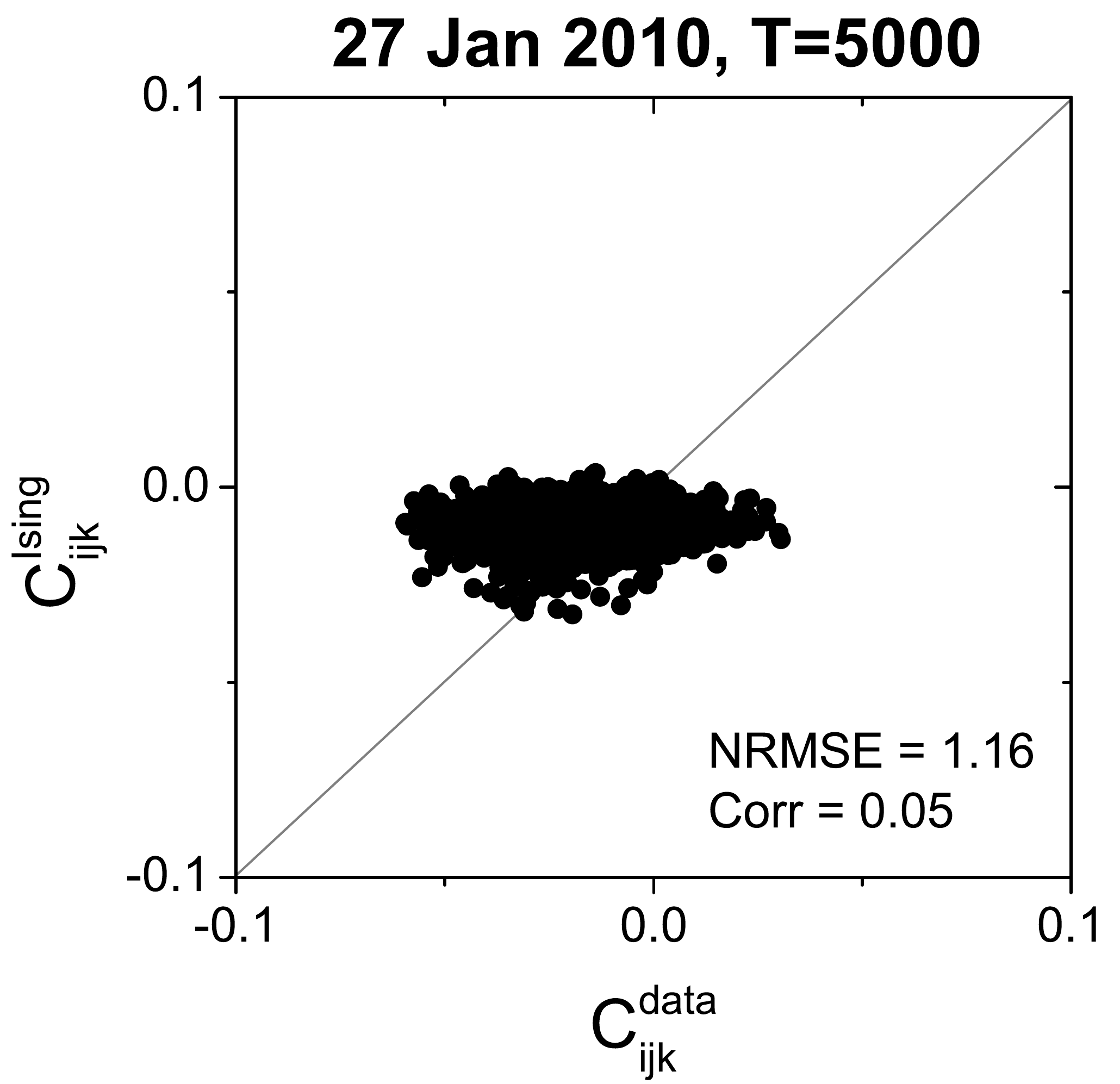}
}
 \caption{Comparison of the third-order covariances calculated using the real data and 50000 MC samples from the exact Ising model for two different moving window sizes. The Ising model is not capable of recovering observed individual third-order covariances independently of moving window size.}
 \label{fig:Eq_Exact_data_C3}
\end{figure}

\begin{figure}
\centering
\resizebox{0.4\textwidth}{!}{%
	\includegraphics{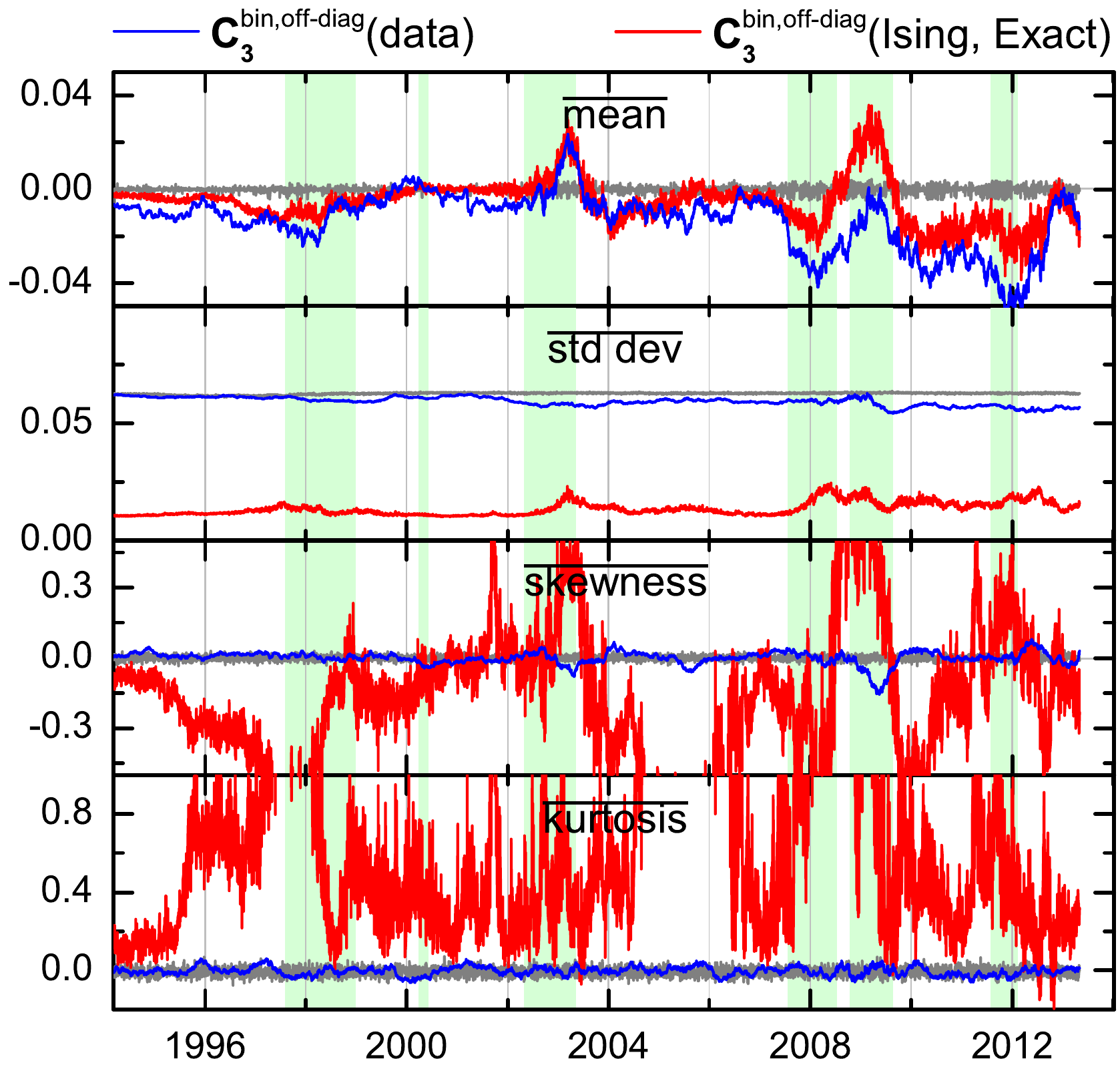}
}
 \caption{Historical evolution of the first four moments of distribution of the third-order covariances calculated using real (blue), randomly shuffled (gray) and 5000 MC samples (for each historical date) from the exact Ising model (red) data. The exact Ising model is only capable of capturing historical dynamics of the mean value of the empirical third-order covariances. However, all higher moments for the real data behave similarly to the ones calculated on randomly shuffled time series.}
 \label{fig:Eq_Exact_data_all}
\end{figure}

To validate results obtained by the exact learning algorithm, we sampled stock returns from the model using MC sampling. The single and pairwise empirical moments are nicely recovered from the exact model (figures are not shown), while the third-order covariances, $\left\langle \left(s_i - \left\langle s_i\right\rangle\right) \left(s_j - \left\langle s_j\right\rangle\right) \left(s_k - \left\langle s_k\right\rangle\right) \right\rangle$, are almost not captured by the model (Fig.~\ref{fig:Eq_Exact_data_C3}). Although the inability to capture third-order correlations could be due to the lack of data for estimating the real third-order covariances, this is unlikely to be the case as the degree of mismatch between the third-order covariances of the Ising model and the data is the same even when somehow bigger amounts of data are used for their estimation. Nevertheless, in spite of the fact that individual third-order covariances are not captured well, historical dynamics of their mean can be recovered from the model (Fig.~\ref{fig:Eq_Exact_data_all}).

%-------------------------------------------------------------------------------
\subsection{Distribution of inferred parameters}
\label{sec:parameters_distribution}
\begin{figure}
\centering
\resizebox{0.5\textwidth}{!}{%
 \includegraphics{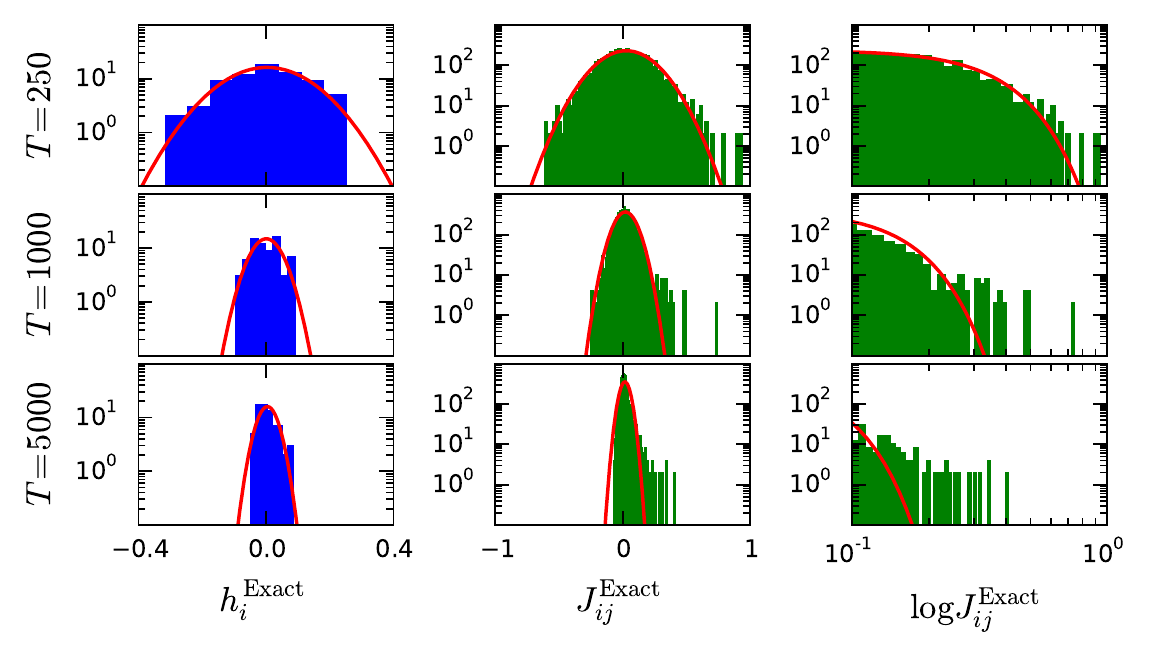}
}
 \caption{Histograms of the external fields and couplings inferred using the exact learning algorithm for 27 Jan 2010 for three different moving window sizes. The red curve denotes the Gaussian fit. For small moving window sizes, the bulk of couplings is distributed normally, while a heavy positive tail dominates for bigger sizes of moving window.}
 \label{fig:Eq_Exact_Hist}
\end{figure}

%-------------------------------------------------------------------------------
Figure~\ref{fig:Eq_Exact_Hist} shows histograms of the external fields and couplings inferred using the exact learning algorithm for three different window sizes: $T=250,1000$ and $5000$ trading days. The distribution of external fields is close to the Gaussian and does not possess outliers independently of $T$. The bulk of the couplings is also distributed normally, however a heavy positive tail is present. For bigger window sizes, the Gaussian bulk component of the inferred couplings becomes less prominent and the tail starts to dominate. Although this behavior has been previously reported in Ref.~\cite{bury1}, a role of the tail has remained unclear. We will address this question in the next subsection.

\begin{figure*}
\centering
\resizebox{0.24\textwidth}{!}{%
 \includegraphics{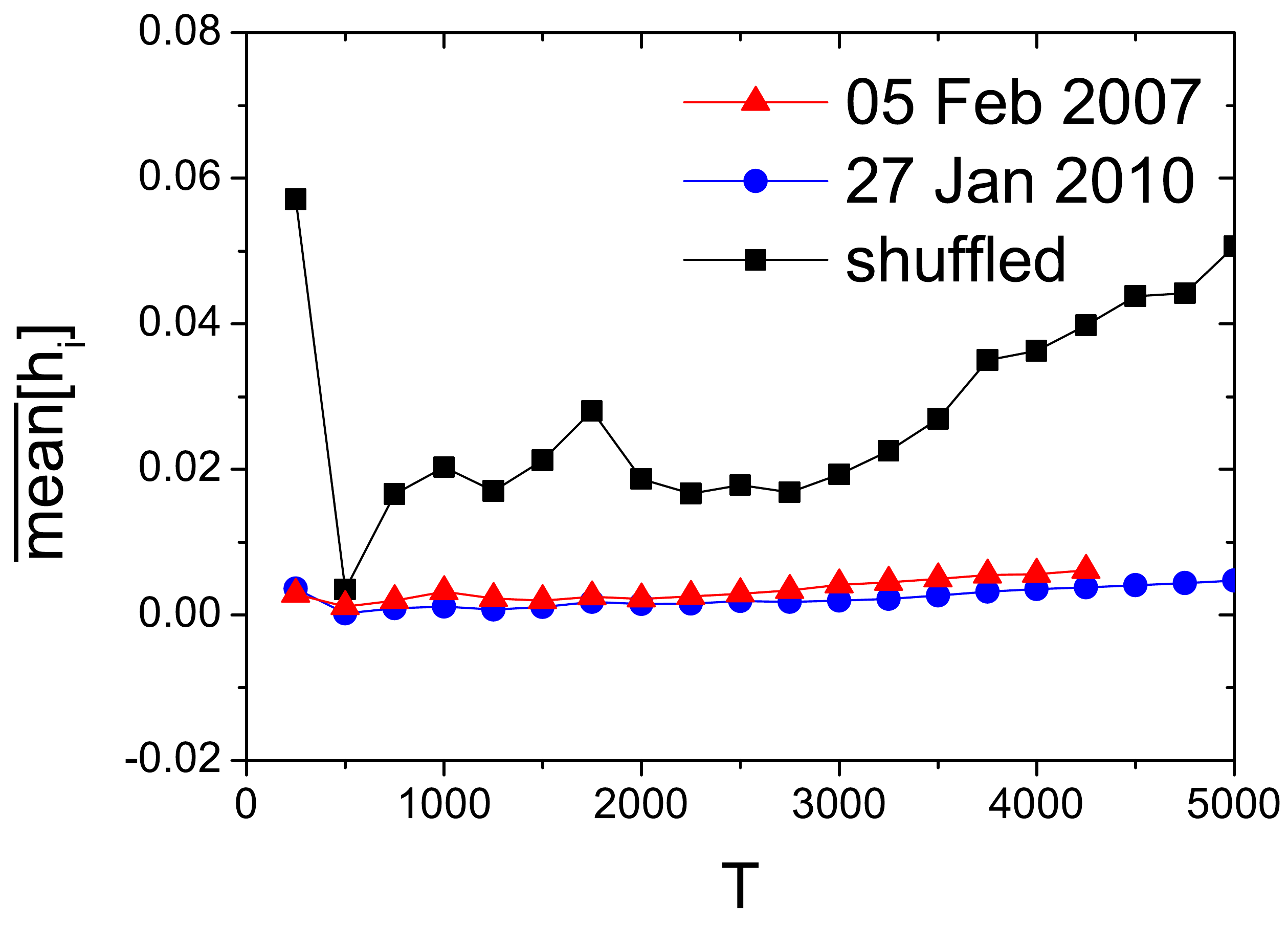}
}
\resizebox{0.24\textwidth}{!}{%
 \includegraphics{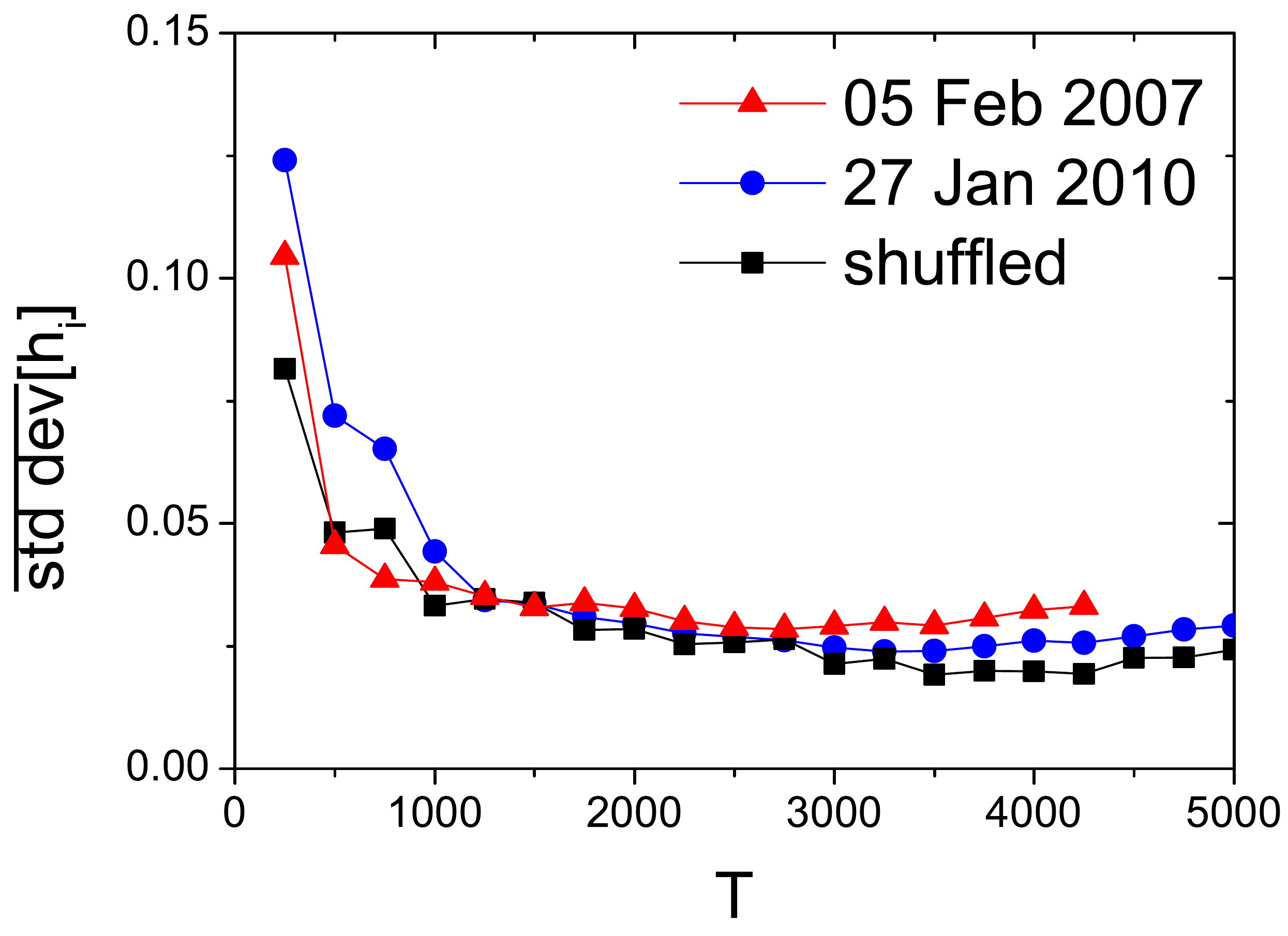}
}
\resizebox{0.23\textwidth}{!}{%
 \includegraphics{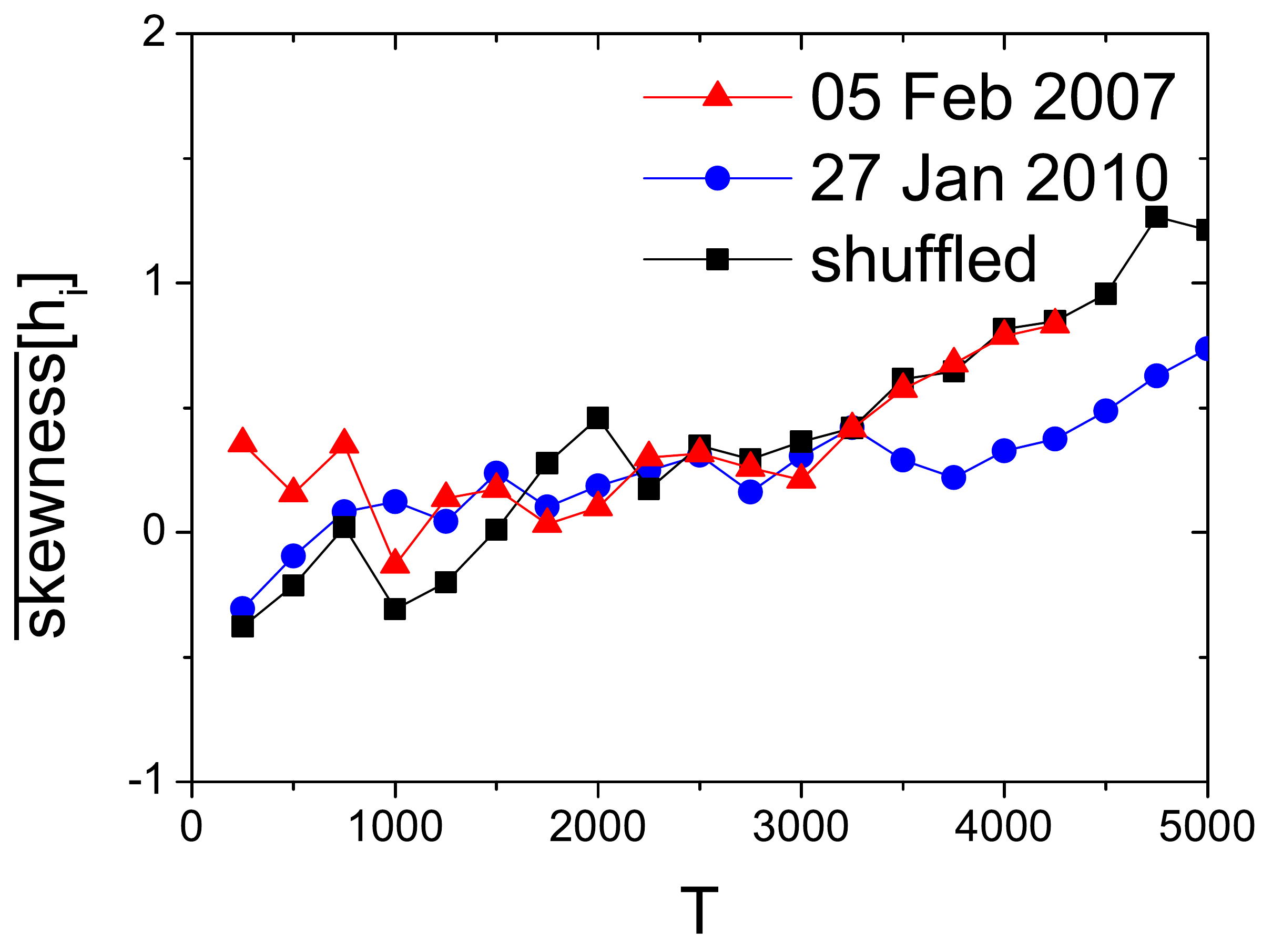}
}
\resizebox{0.23\textwidth}{!}{%
 \includegraphics{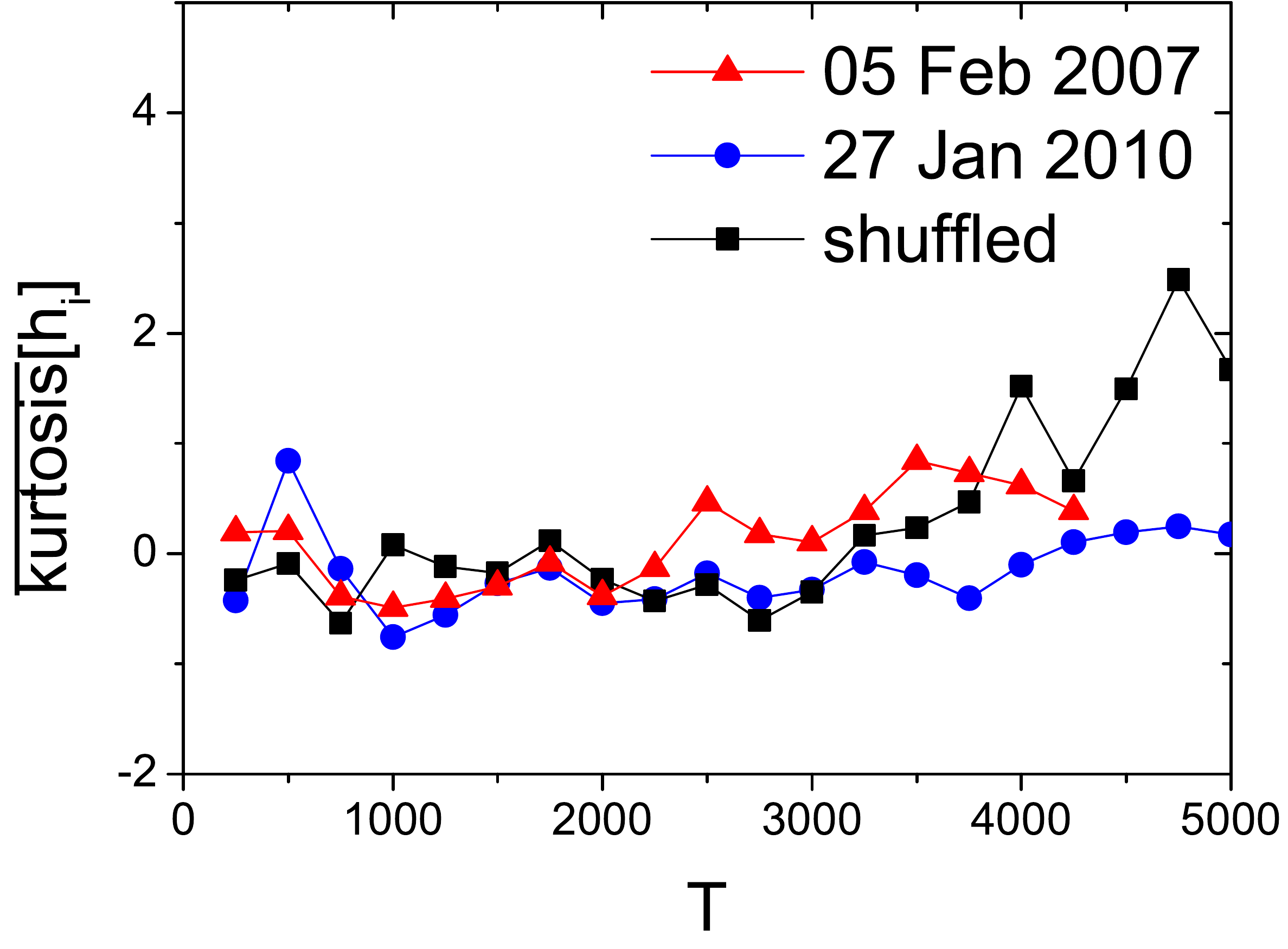}
}
\resizebox{0.24\textwidth}{!}{%
 \includegraphics{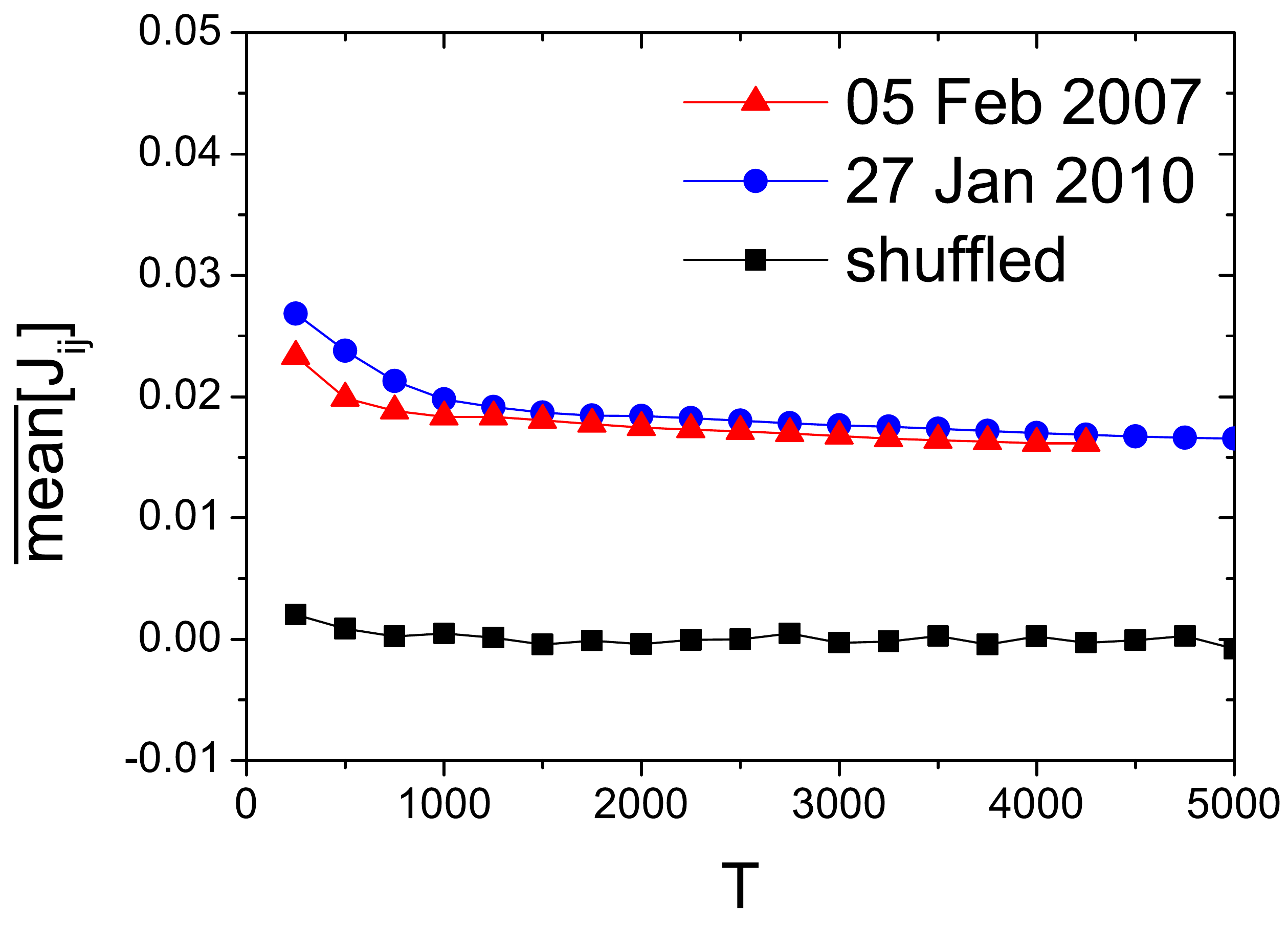}
}
\resizebox{0.24\textwidth}{!}{%
 \includegraphics{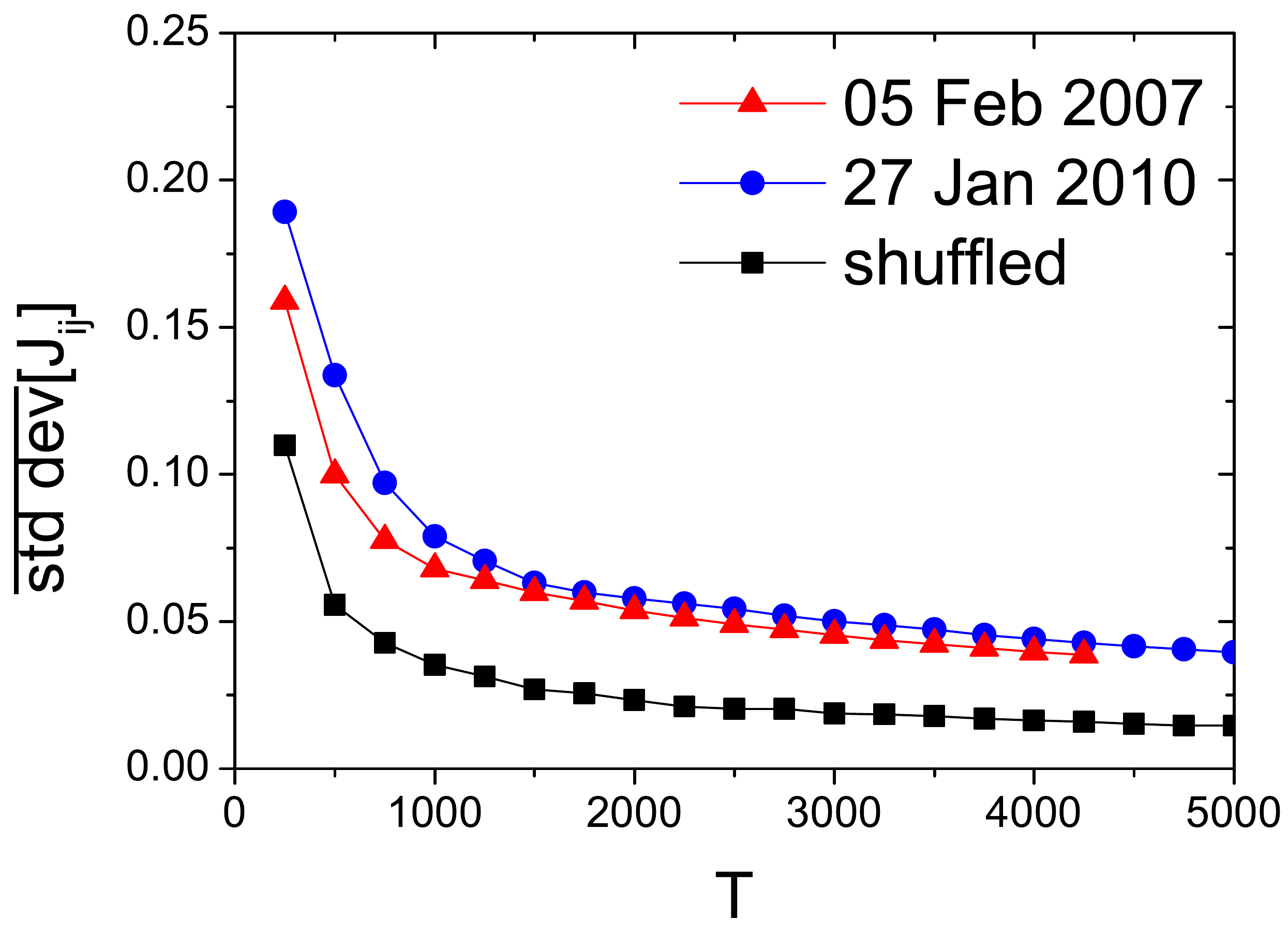}
}
\resizebox{0.23\textwidth}{!}{%
 \includegraphics{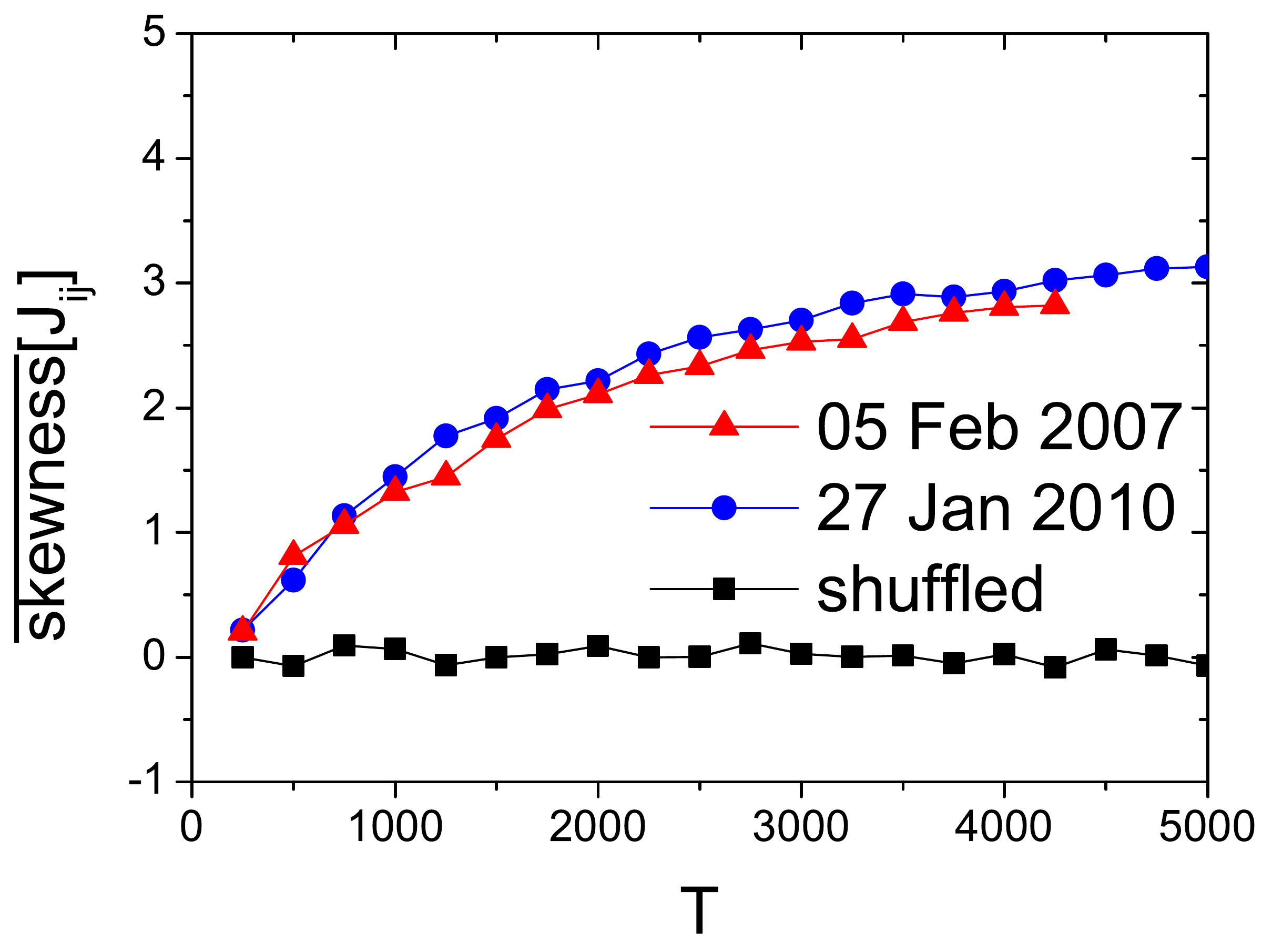}
}
\resizebox{0.23\textwidth}{!}{%
 \includegraphics{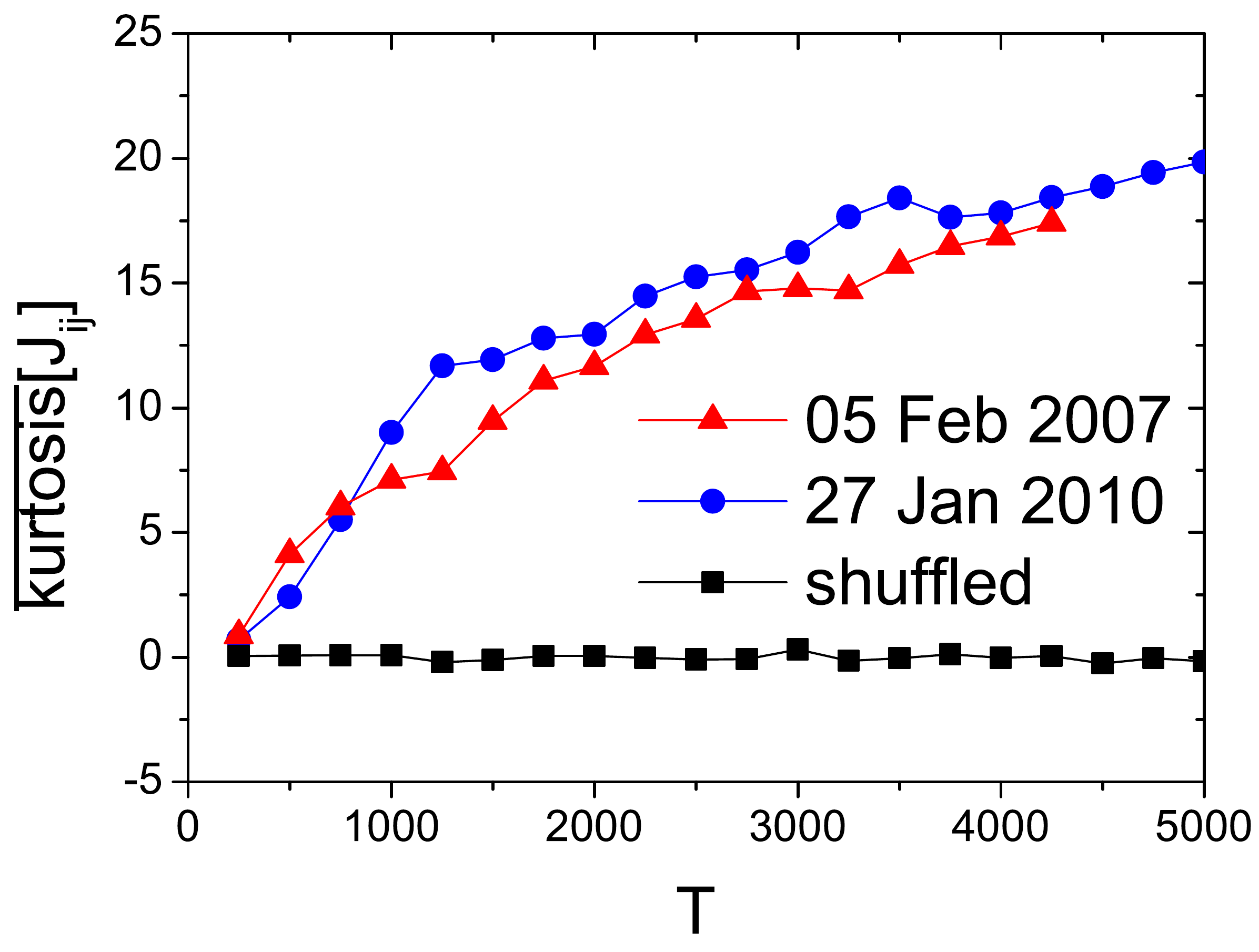}
}
 \caption{Scaling of the first four moments of the distribution of exact external fields (top row) and couplings (bottom row) with moving window size $T$ for two different historical dates and randomly shuffled time series for 27 Jan 2010. Higher value of the mean value of the external fields inferred on randomly shuffled time series compensates positive (ferromagnetic) contribution to the mean return of the mean value of couplings, which is zero for the shuffled time series.}
\label{fig:Eq_Exact_T}
\end{figure*}

In fact, as shown in Fig.~\ref{fig:Eq_Exact_T}, all moments of the distribution of $J_{ij}$ scale with $T$. On the contrary, there is no obvious dependence for $h_i$, which behavior is close to the external fields inferred on randomly shuffled time series (where each element is swapped with other randomly picked element using uniform distribution within the moving window chunk). Higher value of $\overline{h}$ inferred on randomly shuffled time series can be explained by the fact that it compensates positive (ferromagnetic) contribution to the mean return stemming from the positive mean of the inferred couplings: In the shuffled case, $\overline{J}$ becomes zero while all $\left\langle s_i^\mathrm{bin} \right\rangle$ remain unaffected, therefore this effect should be compensated by the external fields.

\begin{figure*}
\centering
\resizebox{0.32\textwidth}{!}{%
 \includegraphics{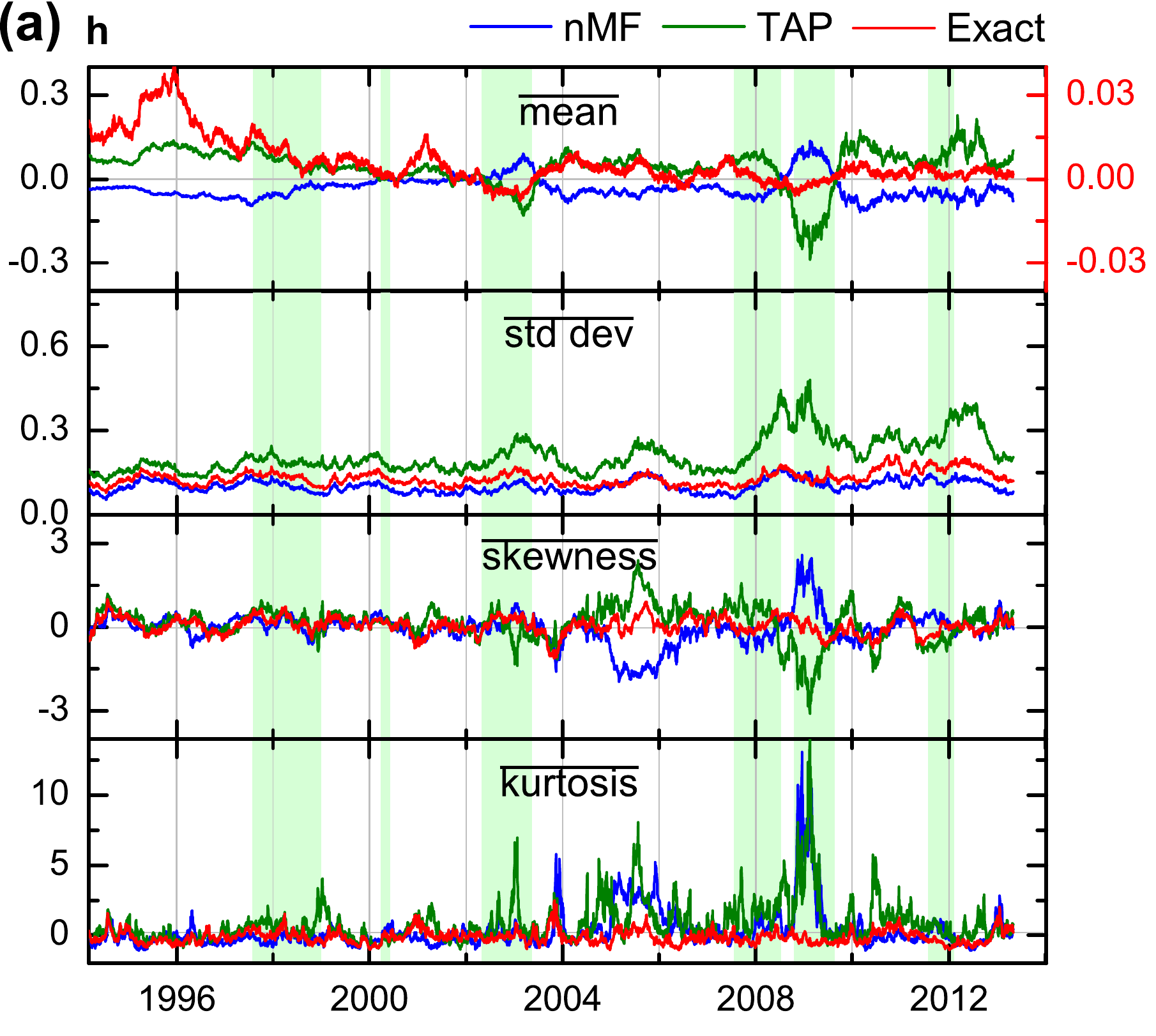}
}
\resizebox{0.295\textwidth}{!}{%
 \includegraphics{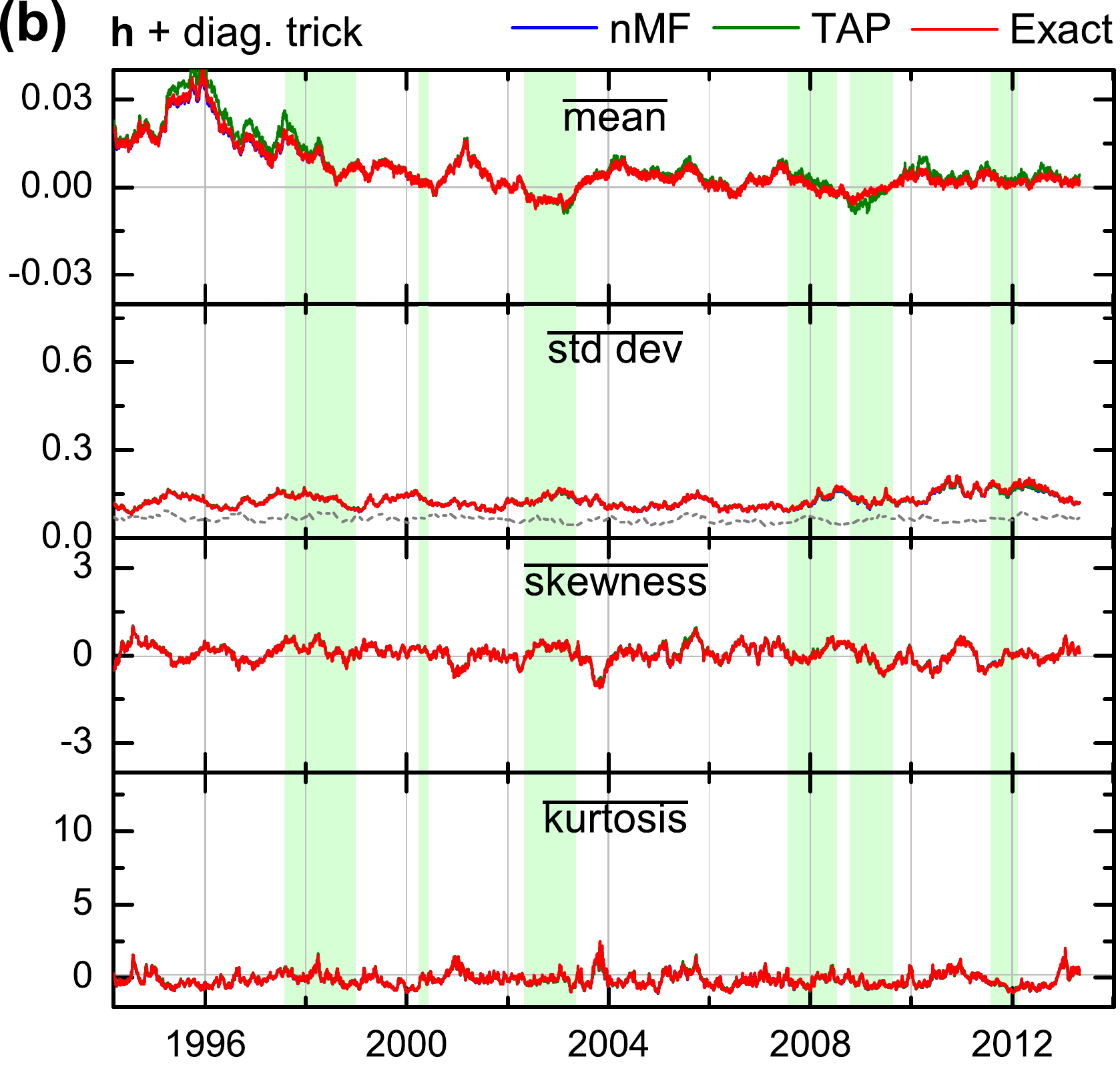}
}
\resizebox{0.3\textwidth}{!}{%
 \includegraphics{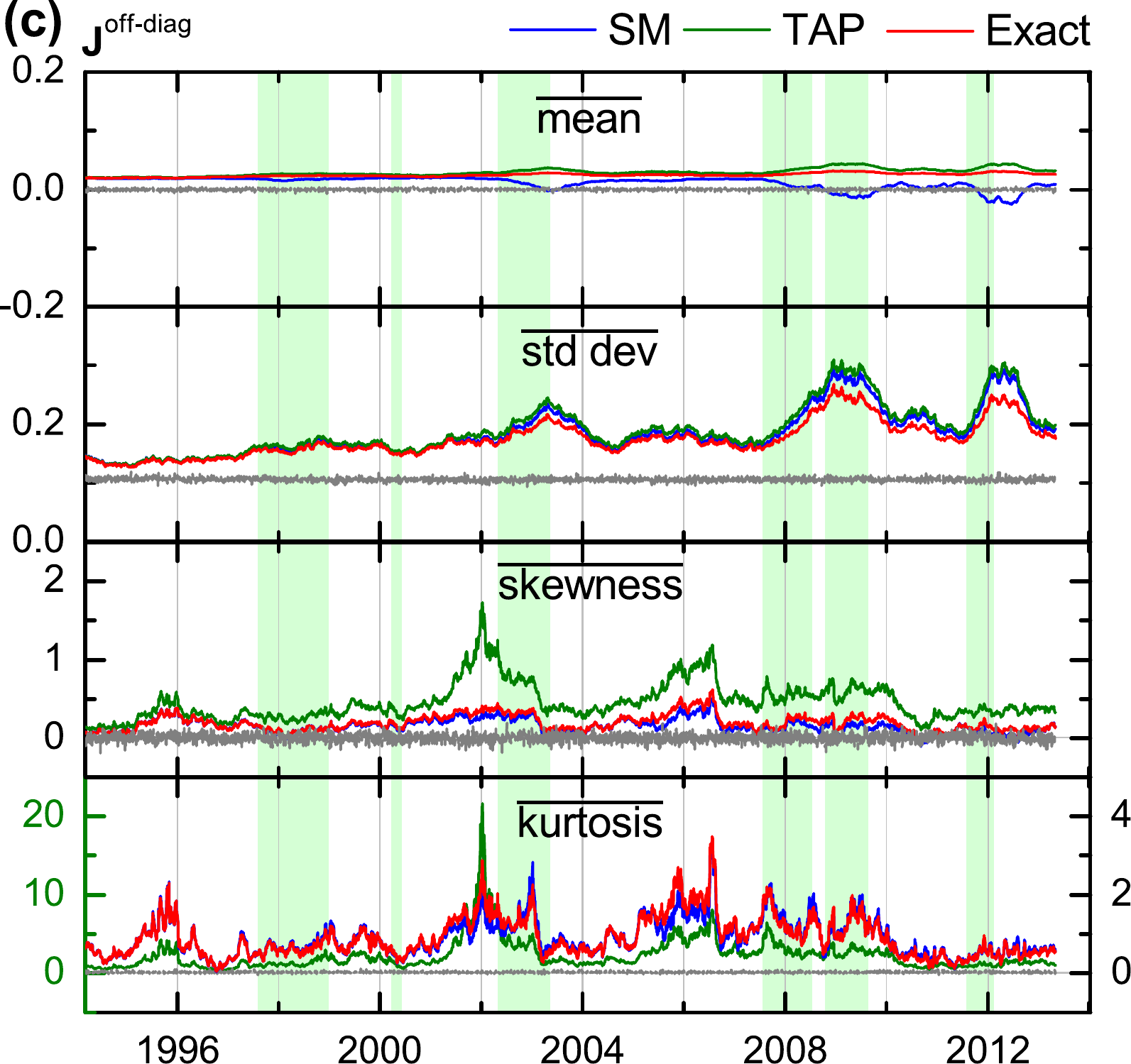}
}
 %\vspace{5cm}
 \caption{Historical dynamics of the first four moments of distribution of the external fields inferred without (a) and with (b) use of the diagonal-weight trick using the nMF (blue), TAP (green) and exact (red) inference algorithms; off-diagonal couplings (c) inferred using the SM (blue), TAP (green) and exact (red) inference algorithms. Moments of the distributions of the couplings and external fields inferred using the TAP algorithm on randomly shuffled returns are denoted with the gray dashed line. Both mean field approximations incorrectly estimate external fields without use of the diagonal-weight trick and tend to overestimate couplings. The SM approximation allows to correctly infer the biggest couplings, however their mean is incorrectly captured for the historical periods with high correlations. During periods of crisis, increase of couplings strength is observed.}
\label{fig:Eq_h_J_offdiag} 
\end{figure*}

Historical dynamics of the moments of the distributions shown in Fig.~\ref{fig:Eq_h_J_offdiag} indicates that the average external field is strongly correlated with the mean return ($0.90$), while higher moments, being almost stable over the historical period considered, seem not to convey any particular information about market behavior. 
Without use of the diagonal-weight trick, $\overline{h}^\mathrm{nMF}$ is completely inconsistent with $\overline{h}^\mathrm{Exact}$ [Fig.~\ref{fig:Eq_h_J_offdiag}(a)]. Although $h_i^\mathrm{TAP}$ behave more similar to $h_i^\mathrm{Exact}$, they are significantly overestimated. External fields inferred using the IP and SM algorithms show similar dynamics to $h_i^\mathrm{nMF}$ and $h_i^\mathrm{TAP}$ respectively, however being even more greatly overestimated (figures are not shown). At the same time, the diagonal-weight trick allows to achieve almost perfect accuracy in both MF cases [Fig.~\ref{fig:Eq_h_J_offdiag}(b)]. Distribution of couplings has stable small positive mean corresponding to ferromagnetic interaction, however with almost half of couplings being negative, i.e. the system is likely to exhibit frustrated configurations. The SM algorithm in general performs better than TAP for couplings estimation except for their mean value $\overline{J}^\mathrm{SM}$, which becomes inconsistent with $\overline{J}^\mathrm{Exact}$ for the historical periods with highly correlated market [Fig.~\ref{fig:Eq_h_J_offdiag}(c), top panel].

It is also interesting to note that the standard deviation of couplings far from the crashes almost linearly increases over the whole period considered from $0.14$ in 1996 to $0.2$ in 2013 [the second panel in Fig.~\ref{fig:Eq_h_J_offdiag}(c)]. This observation gains more meaning when we note that the standard deviation of the couplings in 1996 equals approximately to the standard deviation of the couplings inferred on randomly shuffled returns, while its value is almost twice bigger than one for the shuffled time series in 2013. Also, during the biggest market crashes, standard deviation of couplings  has jumps because interconnections on the market become tighter as a result of the herding behavior during a financial turmoil: Until system is not adapted to a new economic reality, prices tend to move collectively with overall market performance as a benchmark \cite{10.1371/journal.pone.0105874}. The heavy positive tail, which can be characterized by higher order moments, increases for some historical periods. A reason for it is unknown at the moment. 

%-------------------------------------------------------------------------------
\subsection{Industry-related clustering structure}
\label{sec:clustering_structure}
\begin{figure}
\centering
\resizebox{0.23\textwidth}{!}{%
 \includegraphics{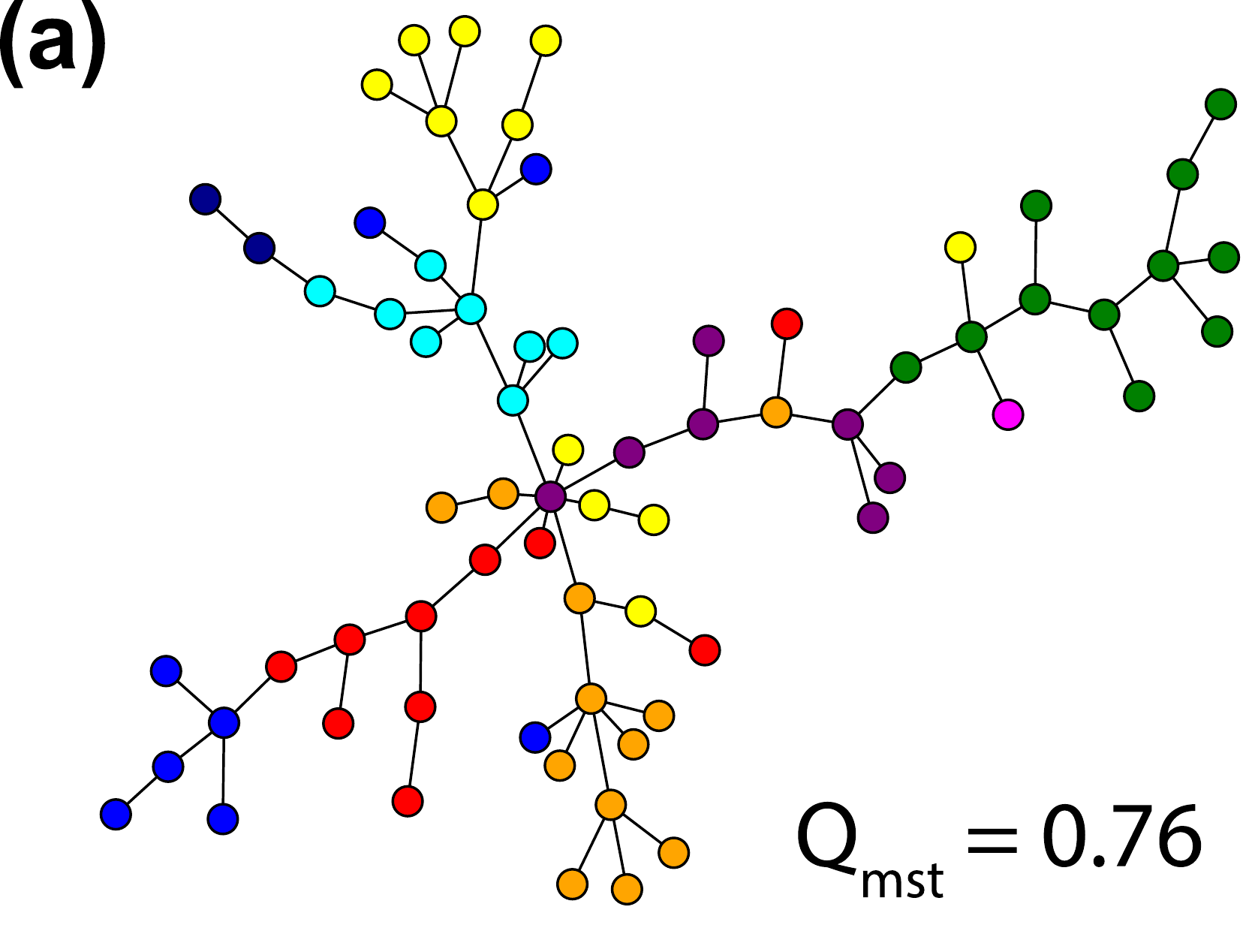}
}
\resizebox{0.23\textwidth}{!}{%
 \includegraphics{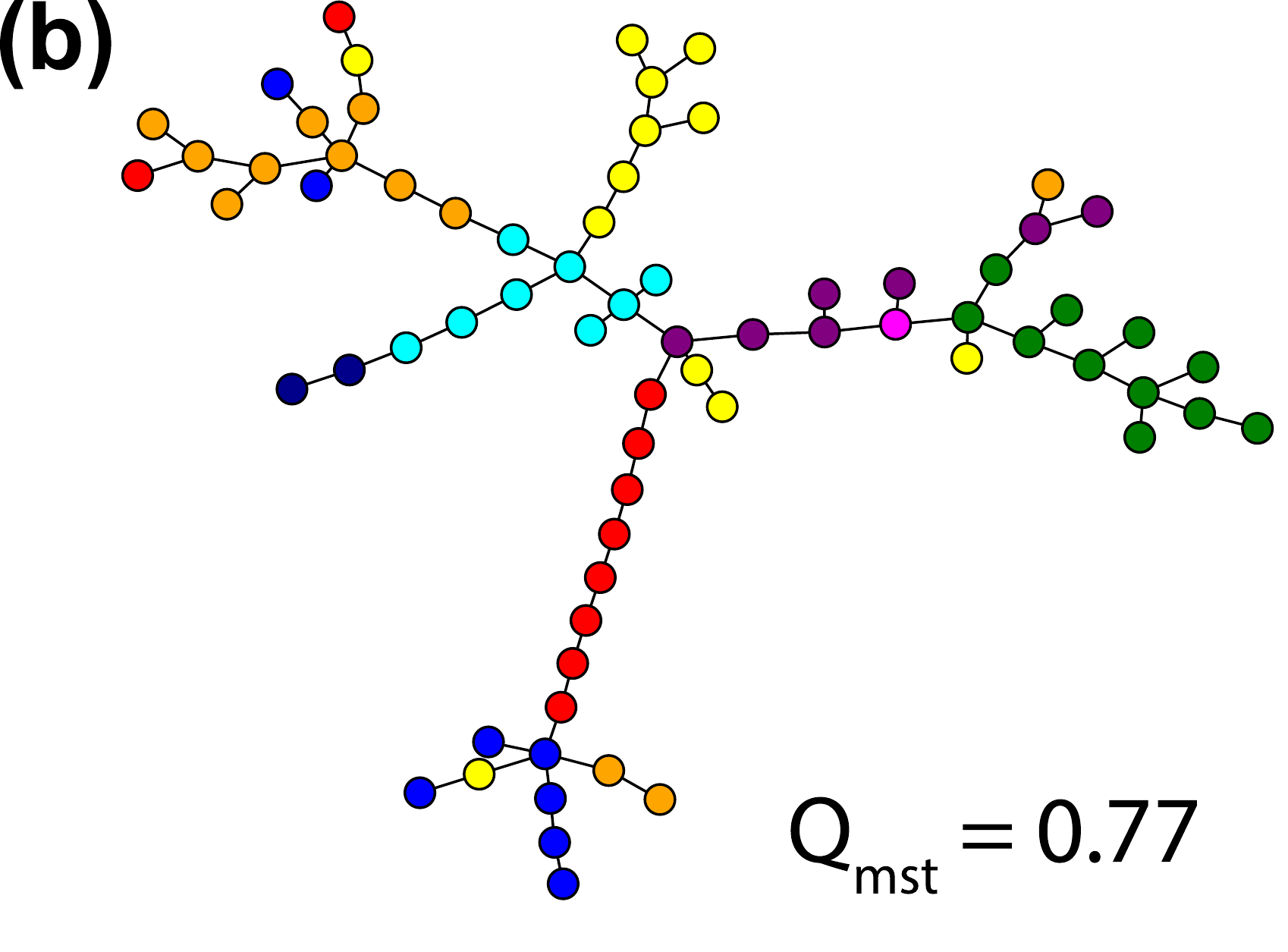}
}
 \caption{Minimum spanning tree for the covariance matrix (a) and corresponding exact couplings (b) for 27 Jan 2010 calculated using SMA window $T=4000$ trading days. Similar industry-related clustering structure is observed in both cases. The considered sectors are Healthcare (red), Consumer Goods (blue), Basic Materials (green), Financial (cyan), Industrial Goods (purple), Services (yellow), Technology (orange), Conglomerate (magenta) and Utilities (dark blue). The graphs are visualized using the NetworkX Python package \cite{hagberg-2008-exploring}.}
 \label{fig:mst}
\end{figure}

%-------------------------------------------------------------------------------
It is well known that a stock market possesses a hierarchical clustering structure which can be detected using correlations \cite{refId0,Coelho2007615} or couplings between stock prices \cite{Bury20131375} or trading volumes \cite{1742-5468-2014-7-P07008}. One of the most popular techniques employed to find related structures is based on the minimum spanning tree (MST) algorithm. For instance, Prim's algorithm---a basic MST construction algorithm---consists of the three following steps:
\begin{enumerate}
  \item Initialize a tree with a single stock, chosen arbitrarily from the all stocks.
  \item Find the stock not yet in the tree which has the strongest coupling (biggest value of $C_{ij}$ or $J_{ij}$) to a stock in the tree and include it to the tree.
  \item Repeat step 2 until all stocks are in the tree.
\end{enumerate}
Figure~\ref{fig:mst} shows examples of the MST constructed using covariance and coupling matrices of binary stock returns, which are similar to the previously reported in Refs.~\cite{Bury20131375} and \cite{1742-5468-2014-7-P07008}. The remarkable feature of these structures is a manifestation of industry sector clustering. Here, we define an industry sector cluster as a connected subset of the tree where all stocks belong to the same industry sector, i.e. each cluster member interacts to the other members only through the stocks from the same sector. Further, we denote cluster size as $N_{m,k}$, where $m=1,\dots,M$ is a sector index ($M=9$ is the total number of sectors listed in Table~\ref{tab:companies}) and $k=1,\dots,K_m$ is an index of the cluster ($K_m$ is a number of such clusters for the sector $m$).

\begin{figure}
\centering
\resizebox{0.4\textwidth}{!}{%
 \includegraphics{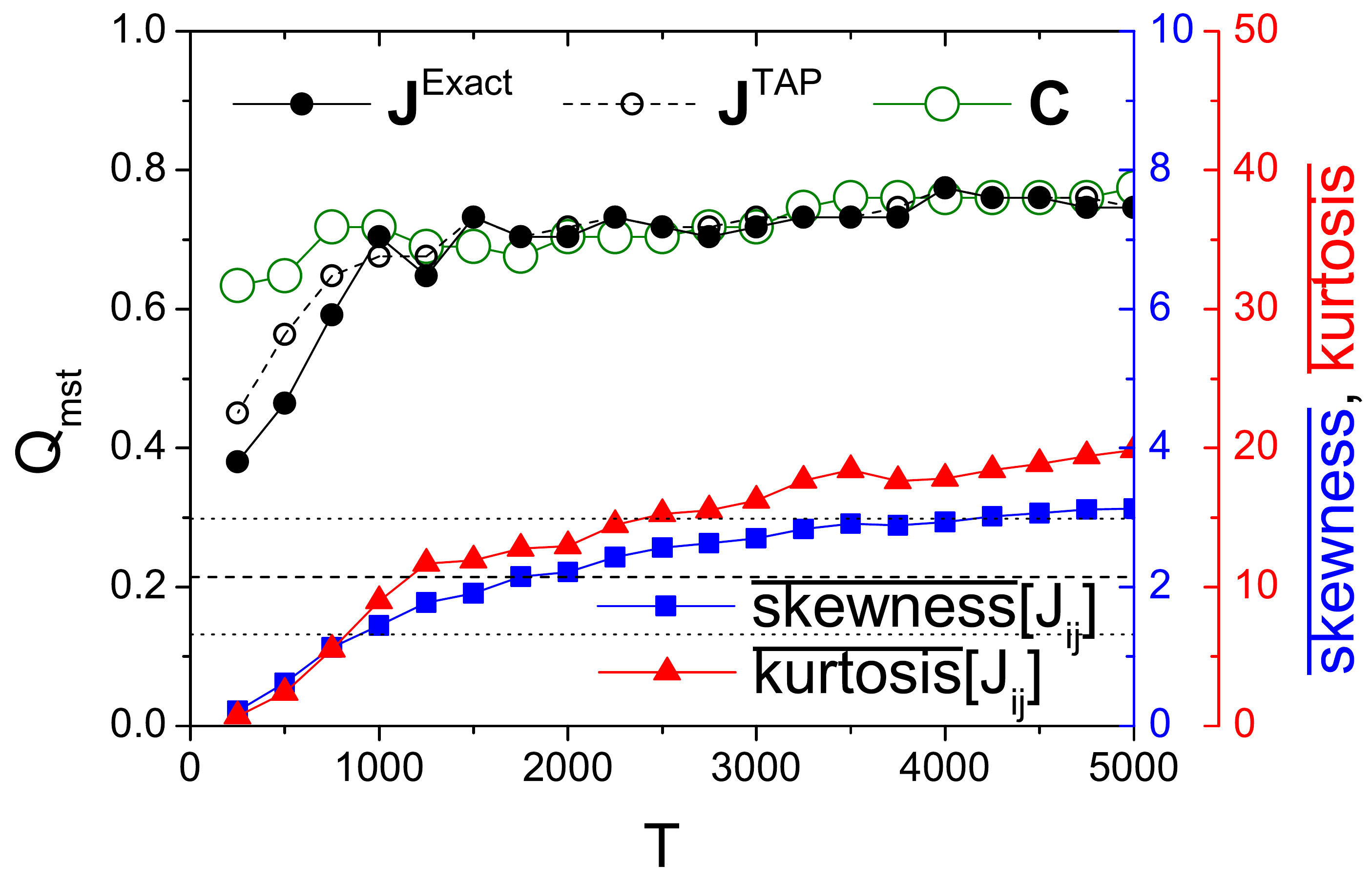}
}
\caption{Quality (industry sector clustering degree) of a minimum spanning tree, $Q_\mathrm{mst}$, depending on moving window size for exact and TAP couplings, and covariance matrix (27 Jan 2010). The gray dashed and dotted lines denote mean and 99.7\% confidence interval respectively for the quality of MST built on randomly shuffled time series. Quality of MST for couplings increases with deviation of their distribution from the Gaussian, characterized by skewness and kurtosis.}
\label{fig:mst_Q_T} 
\end{figure}

As mentioned above, there can be many clusters for each industry sector. In order to estimate overall industry clustering degree of the market, let us introduce a simple metric based on finding clusters of the maximum size
\begin{equation}
	Q_\mathrm{mst} = \frac{1}{N}\sum\limits_{m=1}^M \max_k N_{m,k}.
	\label{eq:Q_mst}
\end{equation}
Intuitively speaking, a small value of $Q_\mathrm{mst}$ corresponds to the case where stocks do not group with each other based on which industry sector they belong to. Its minimum value, $M/N$, is defined by the situation where the biggest cluster for each sector has only one stock, i.e. $K_m$ equals to the total number of stocks in the sector $m$. The maximum value of $Q_\mathrm{mst}$ is 1, which corresponds to the perfect industry clustering structure when there is only one cluster for each industry sector ($K_m=1$). This clustering measure shows interesting dependence on the size of moving window (Fig.~\ref{fig:mst_Q_T}), suggesting an increasing degree of sectoral connectedness of the stock market for bigger time windows as inferred by the Ising model. Also, the degree of connectedness increases with deviation of the distribution from the Gaussian measured by skewness and kurtosis.

\begin{figure}
\centering
\resizebox{0.45\textwidth}{!}{%
 \includegraphics{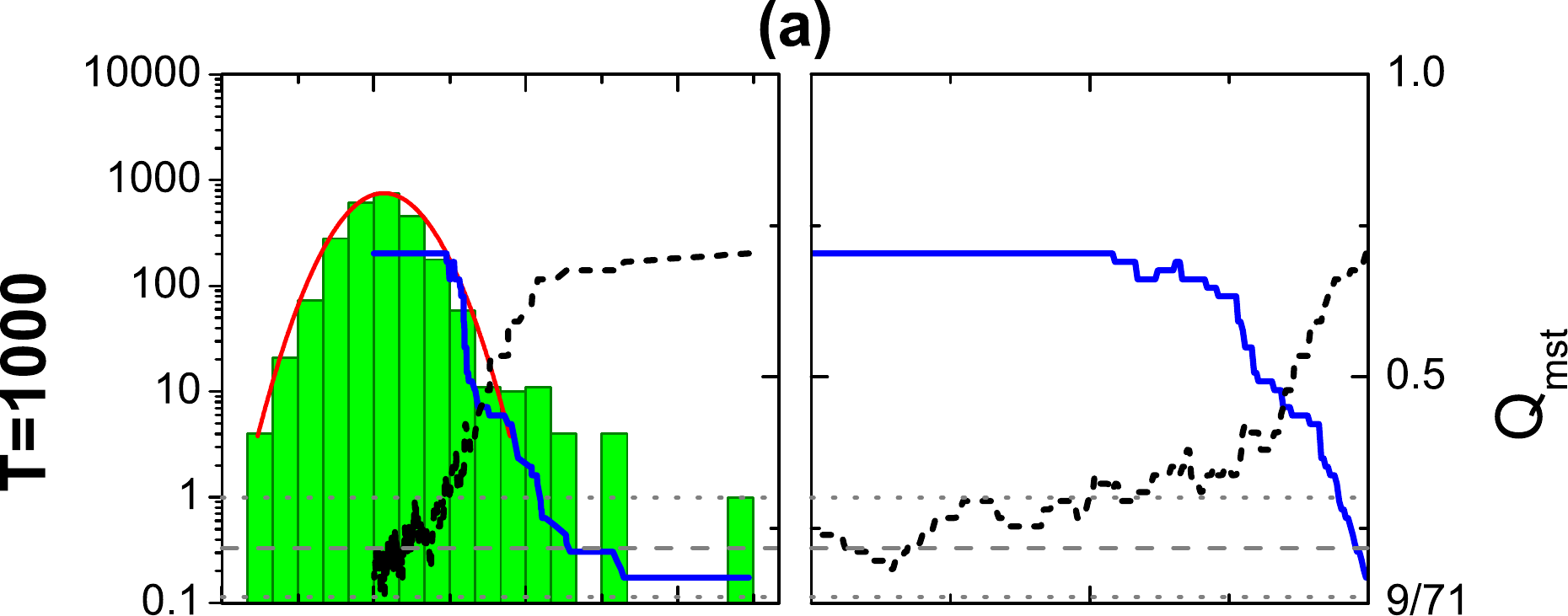}
}
\resizebox{0.45\textwidth}{!}{%
 \includegraphics{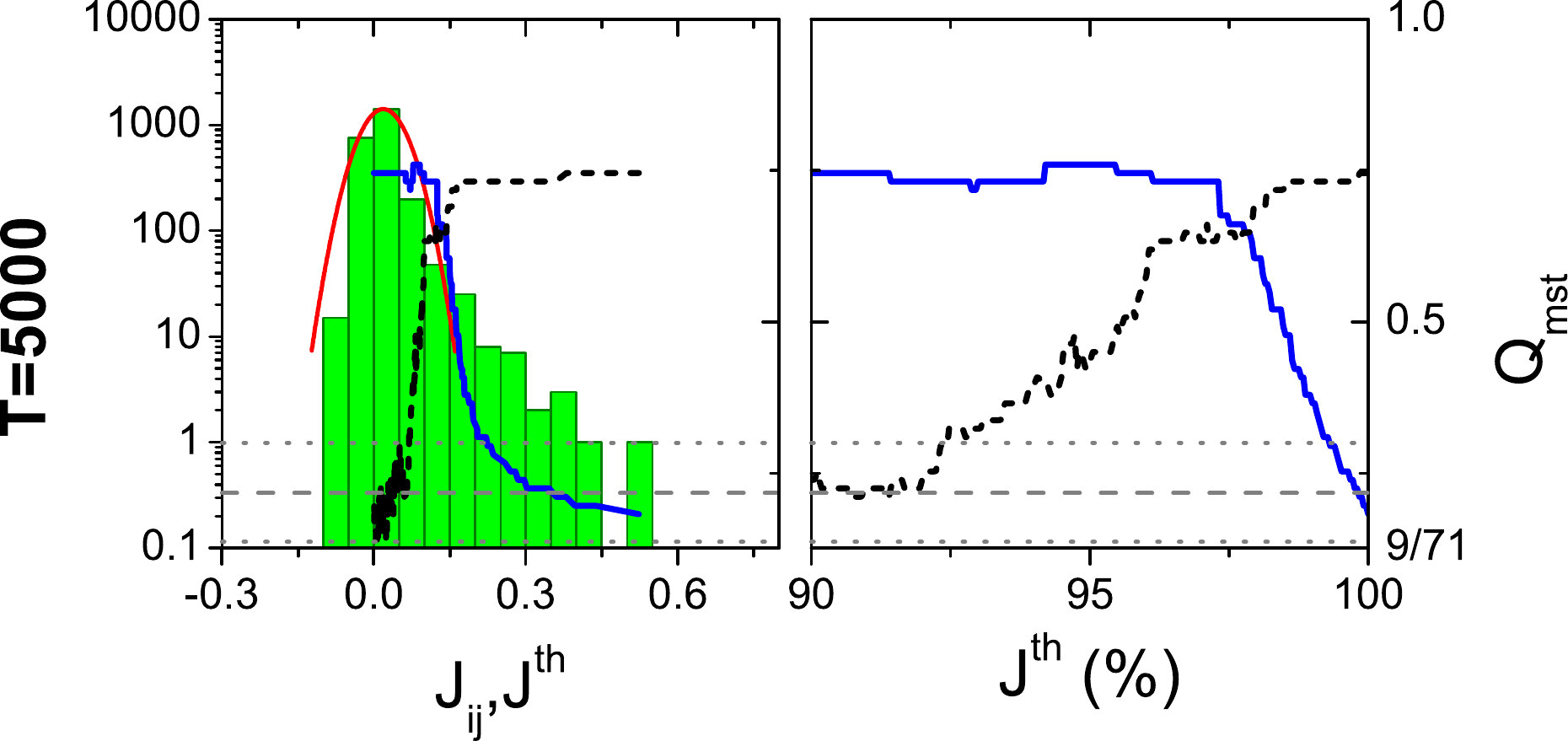}
}\\
\resizebox{0.45\textwidth}{!}{%
 \hspace{5mm}\includegraphics{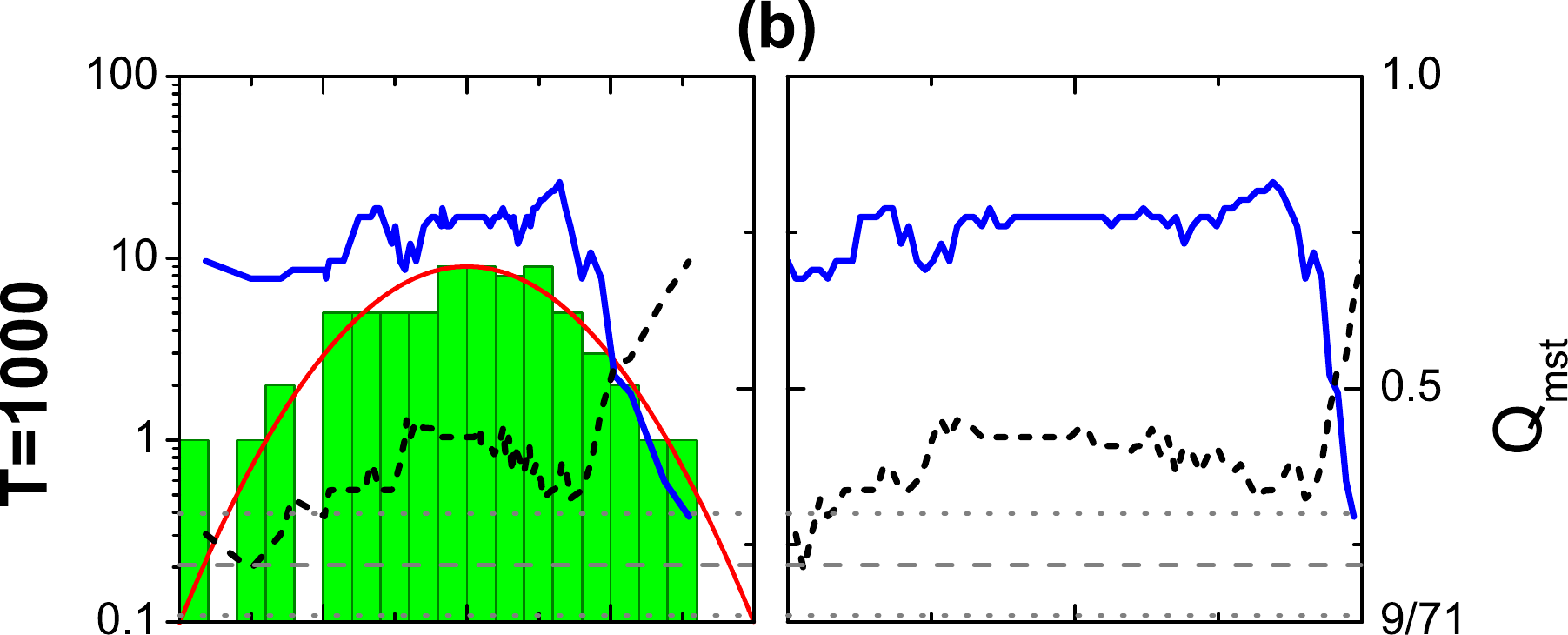}
}
\resizebox{0.45\textwidth}{!}{%
 \hspace{5mm}\includegraphics{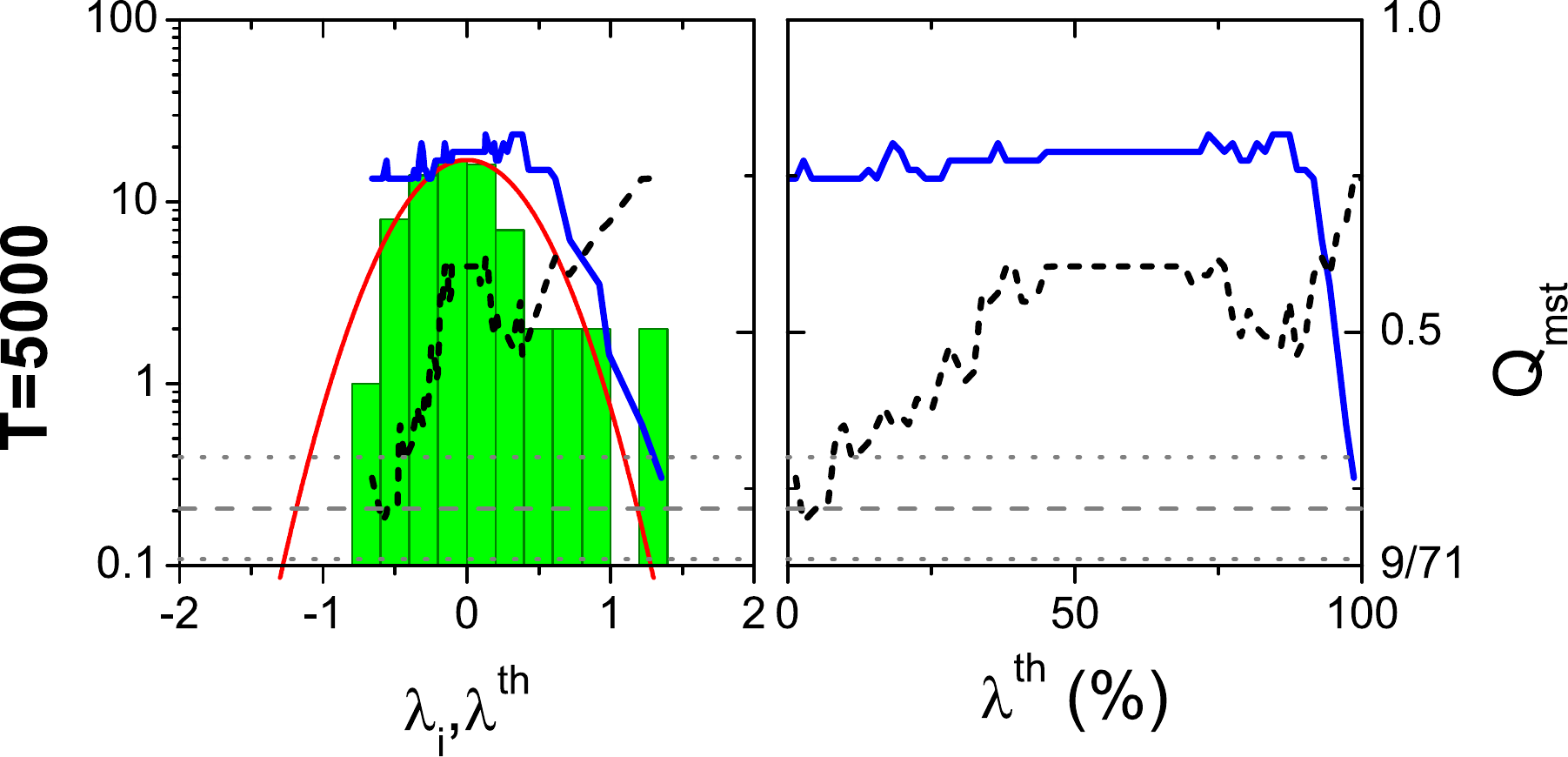}
}
 \caption{Cutoff analysis for the coupling matrix (a) and its eigenvalues (b) for two SMA windows $T=1000$ and $T=5000$ trading days (27 Jan 2010). Histograms of couplings and corresponding eigenvalues (left column), and quality (industry sector clustering degree) of a minimum spanning tree, $Q_\mathrm{mst}$, (left and right columns) depending on the positive (black dashed line) and negative (blue solid line) cutoffs, $J^\mathrm{th}$ and $\lambda^\mathrm{th}$, where all values above or below a cutoff, respectively, are discarded for MST construction. The red curve denotes the Gaussian fit. Discarding both biggest couplings and eigenvalues significantly affects quality of MST, while negative couplings and small eigenvalues do not contribute to the market clustering structure.}
\label{fig:mst_Q_j} 
\end{figure}

To further investigate the network structure and clustering degree of $\mathbf{J}$, we perform the clustering analysis based on two different filtering procedures, namely, considering only a subset of (i) couplings $J_{ij}$ and (ii) eigenmodes of $\mathbf{J}$ corresponding to different eigenvalues $\lambda_i$. With this aim, we choose thresholds $J^\mathrm{th}$ and $\lambda^\mathrm{th}$ and construct MST only using values $J_{ij}\lessgtr J^\mathrm{th}$ and $\lambda_i\lessgtr \lambda^\mathrm{th}$ respectively. Figure~\ref{fig:mst_Q_j}(a) shows that the biggest drop/increase of $Q_\mathrm{mst}$ occurs only when the a small percent of the strongest couplings is excluded/included for MST construction. In a similar way, it is also sensitive to discarding the biggest eigenvalues [Fig.~\ref{fig:mst_Q_j}(b)]. Thus, one might conclude that distribution of couplings indeed can be viewed as a mixture of the Gaussian bulk and heavy positive tail which contains all information about market clustering structure. From this perspective, use of sparse regularization methods for couplings inference, for example, $\ell_1$-regularization discussed in Ref.~\cite{ravikumar2010}, would be of practical interest.

Finally, it is also worth noting that intraday internal structure of couplings is neither stable (quenched) nor completely random (annealed), somehow preserving a clustering structure with the diameter however being smaller near market crashes (figures are not shown, see for example Ref.~\cite{Coelho2007615}). These non-trivial features of a coupling matrix will be studied in more detail in the future works. 

%-------------------------------------------------------------------------------
\subsection{Scaling of inferred parameters}
\label{sec:parameters_scaling}
%-------------------------------------------------------------------------------
In order to study extensive properties of the system, we investigate scaling properties of the parameter distributions with number of stocks, which are usually characterized by the scaling exponent $N^\alpha$. We estimate its average value for each distribution moment over scaling exponents for randomly selected subsets of stocks of different size.

\begin{figure*}
\centering
\resizebox{0.24\textwidth}{!}{%
 \includegraphics{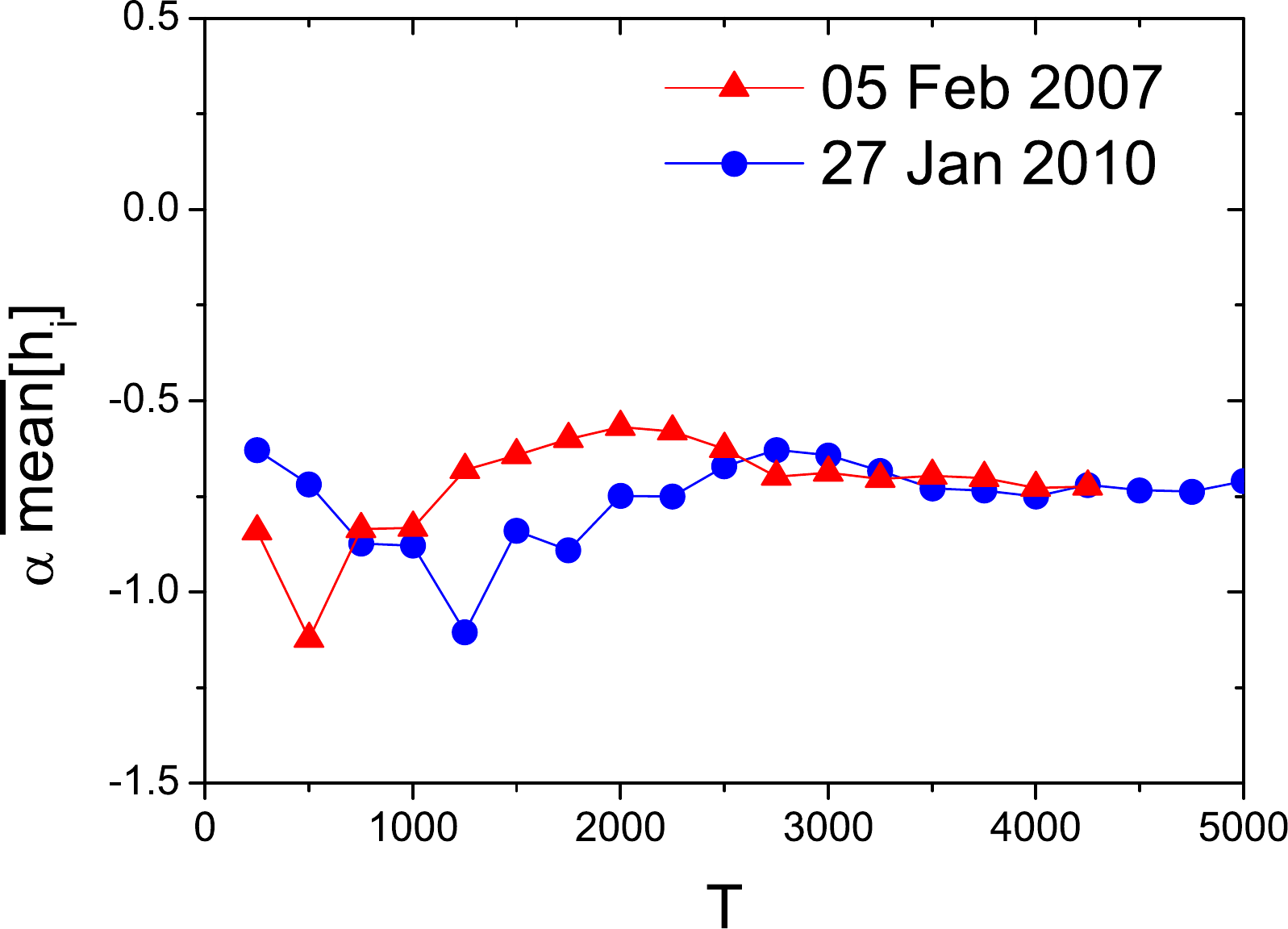}
}
\resizebox{0.24\textwidth}{!}{%
 \includegraphics{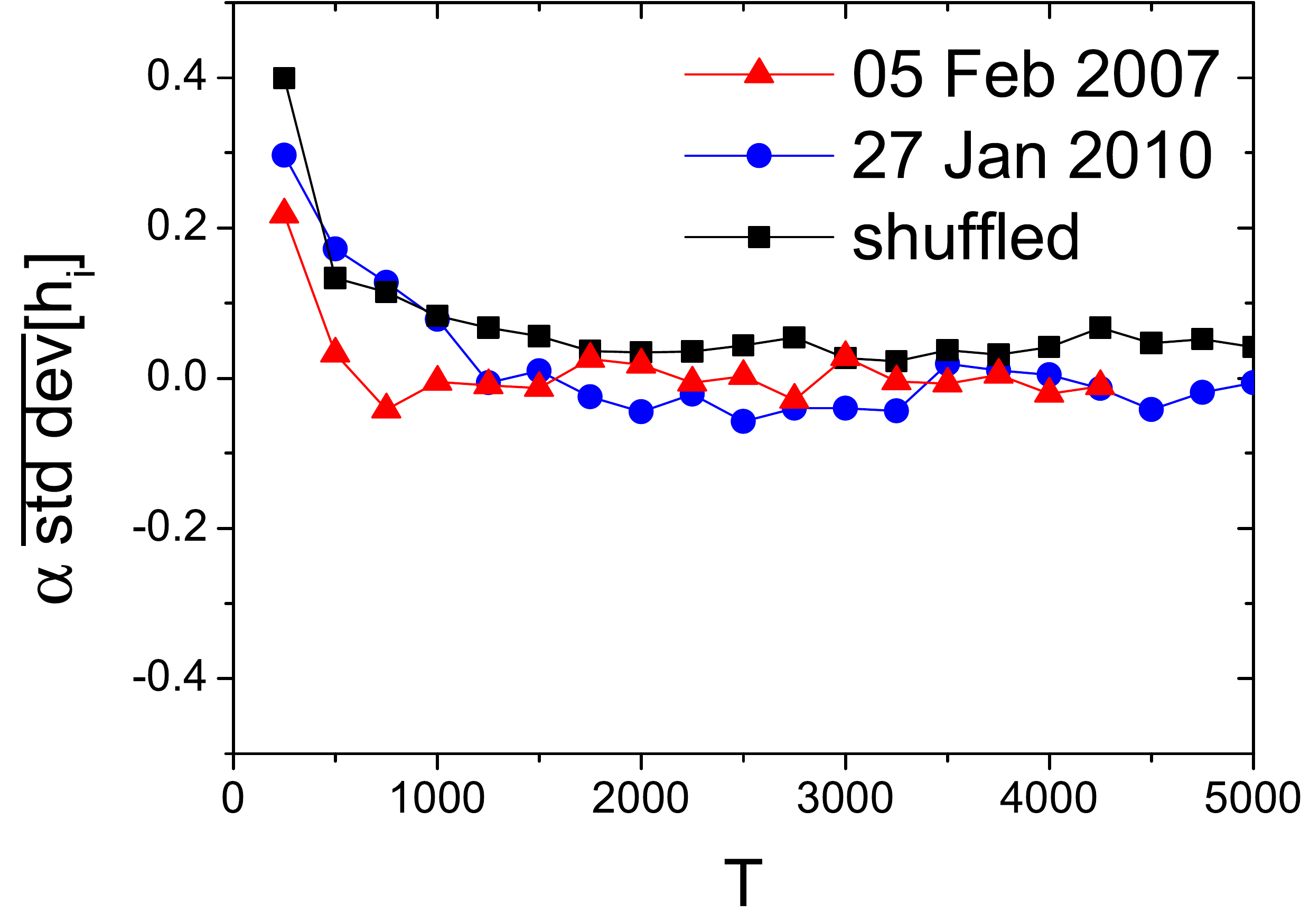}
}
\resizebox{0.24\textwidth}{!}{%
 \includegraphics{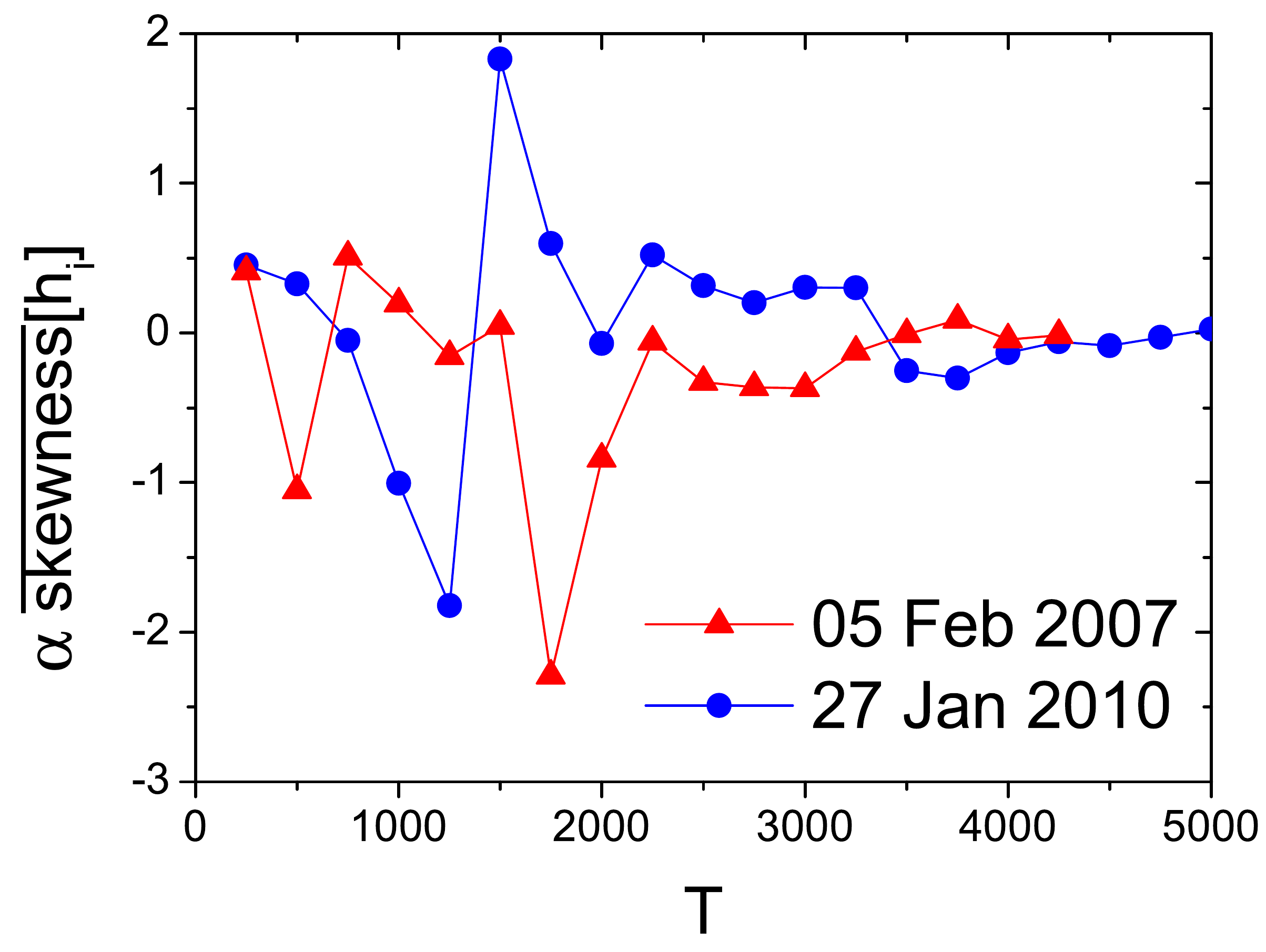}
}
\resizebox{0.24\textwidth}{!}{%
 \includegraphics{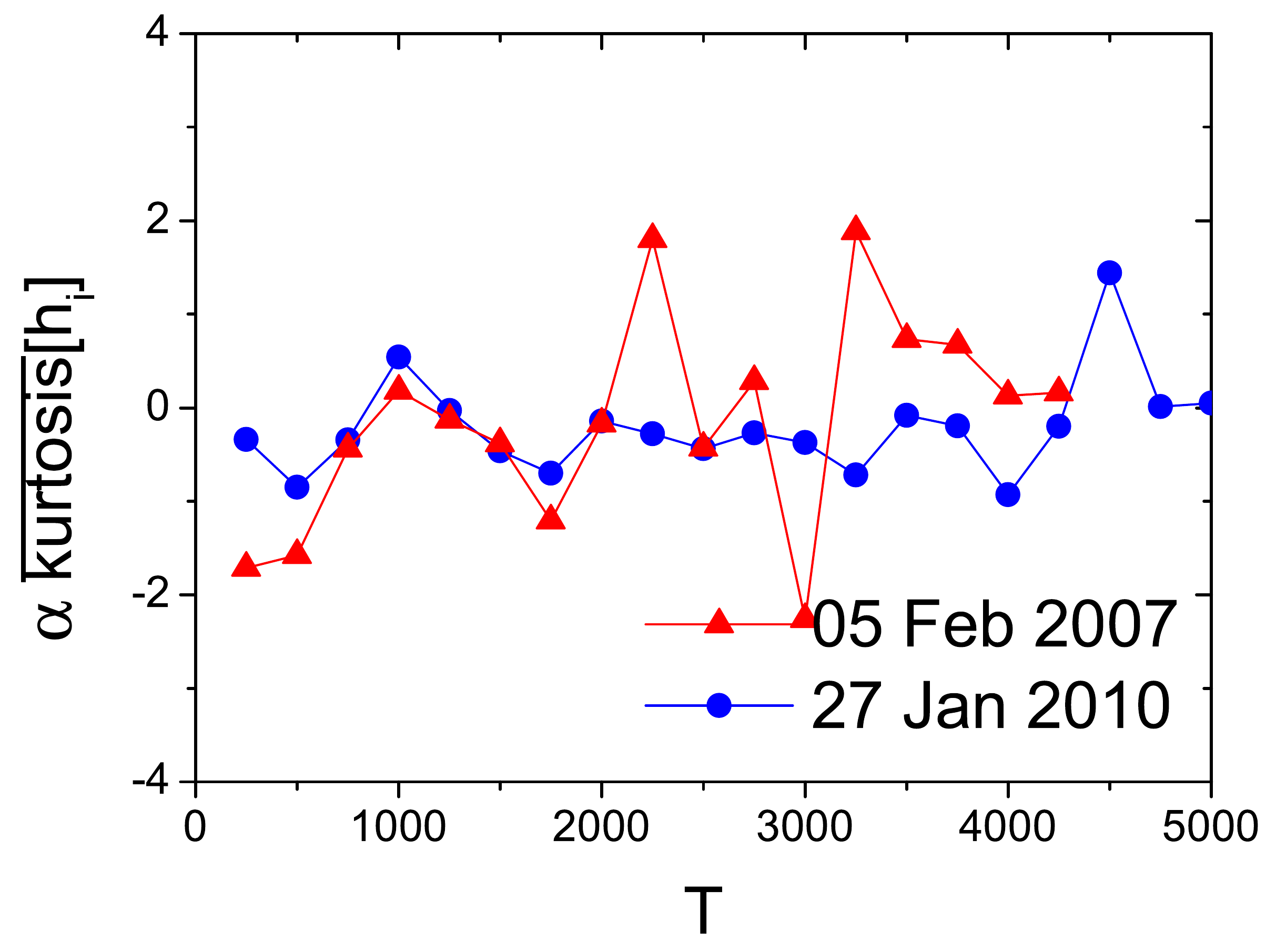}
}
\resizebox{0.24\textwidth}{!}{%
 \includegraphics{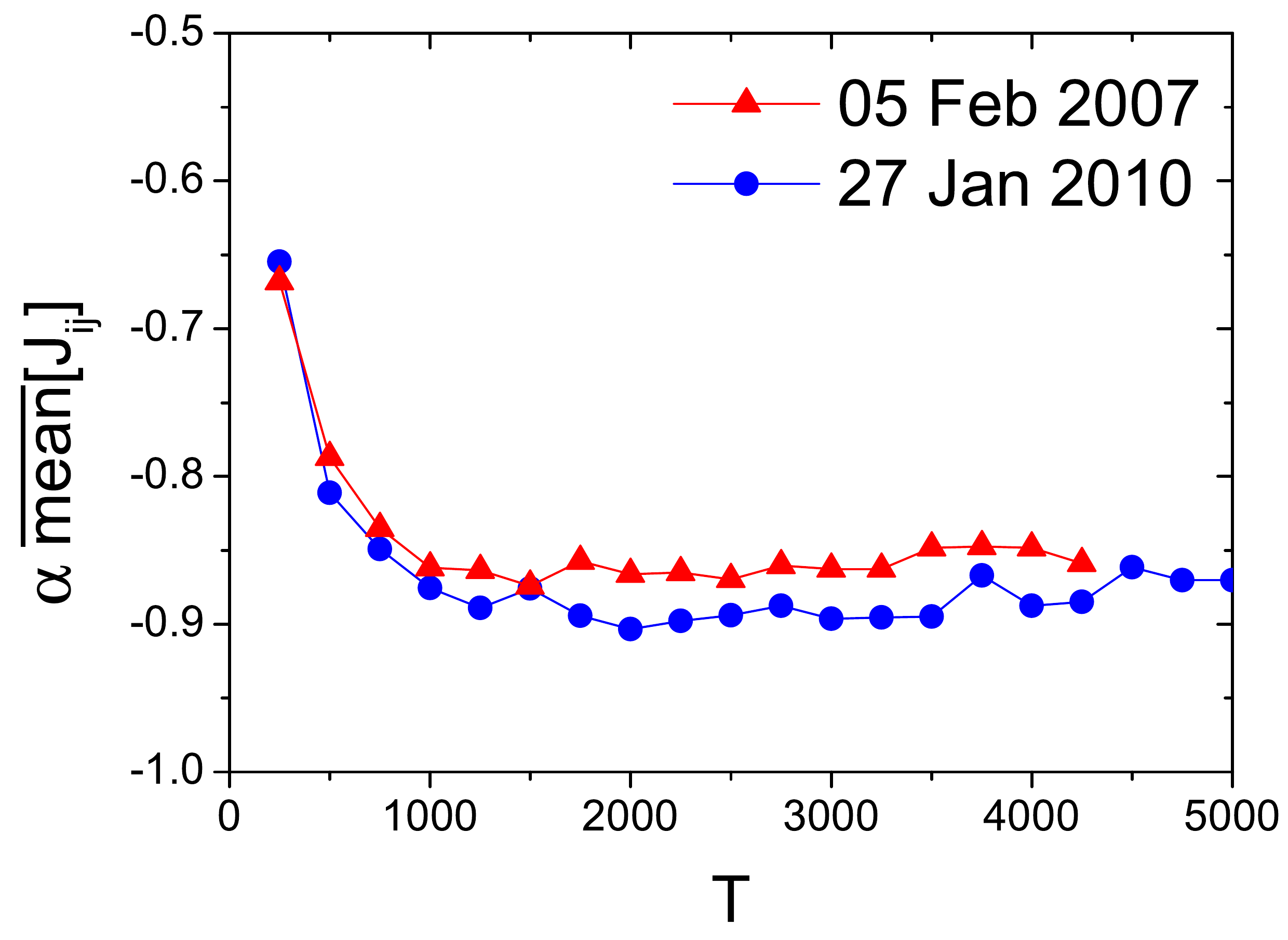}
}
\resizebox{0.24\textwidth}{!}{%
 \includegraphics{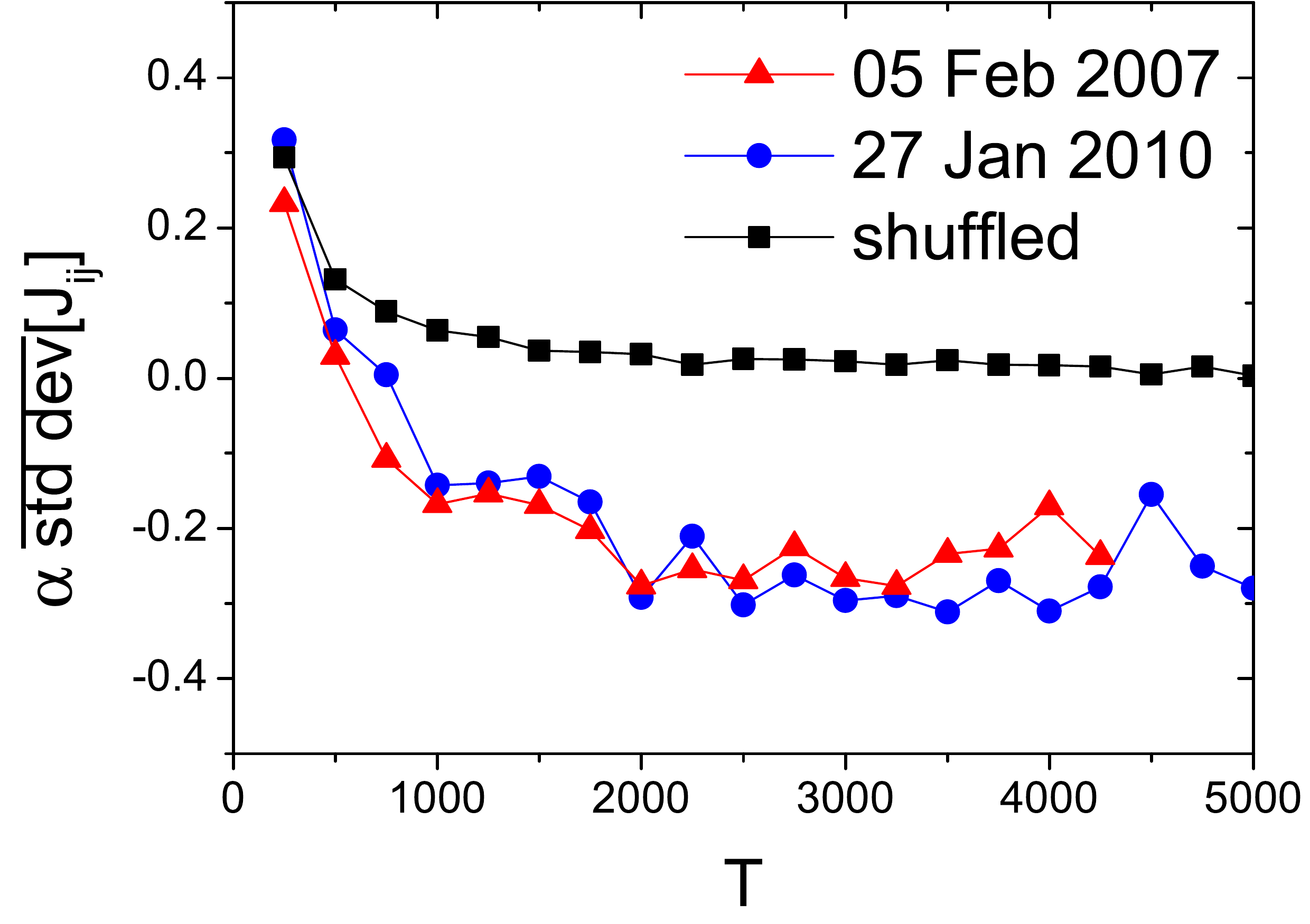}
}
\resizebox{0.24\textwidth}{!}{%
 \includegraphics{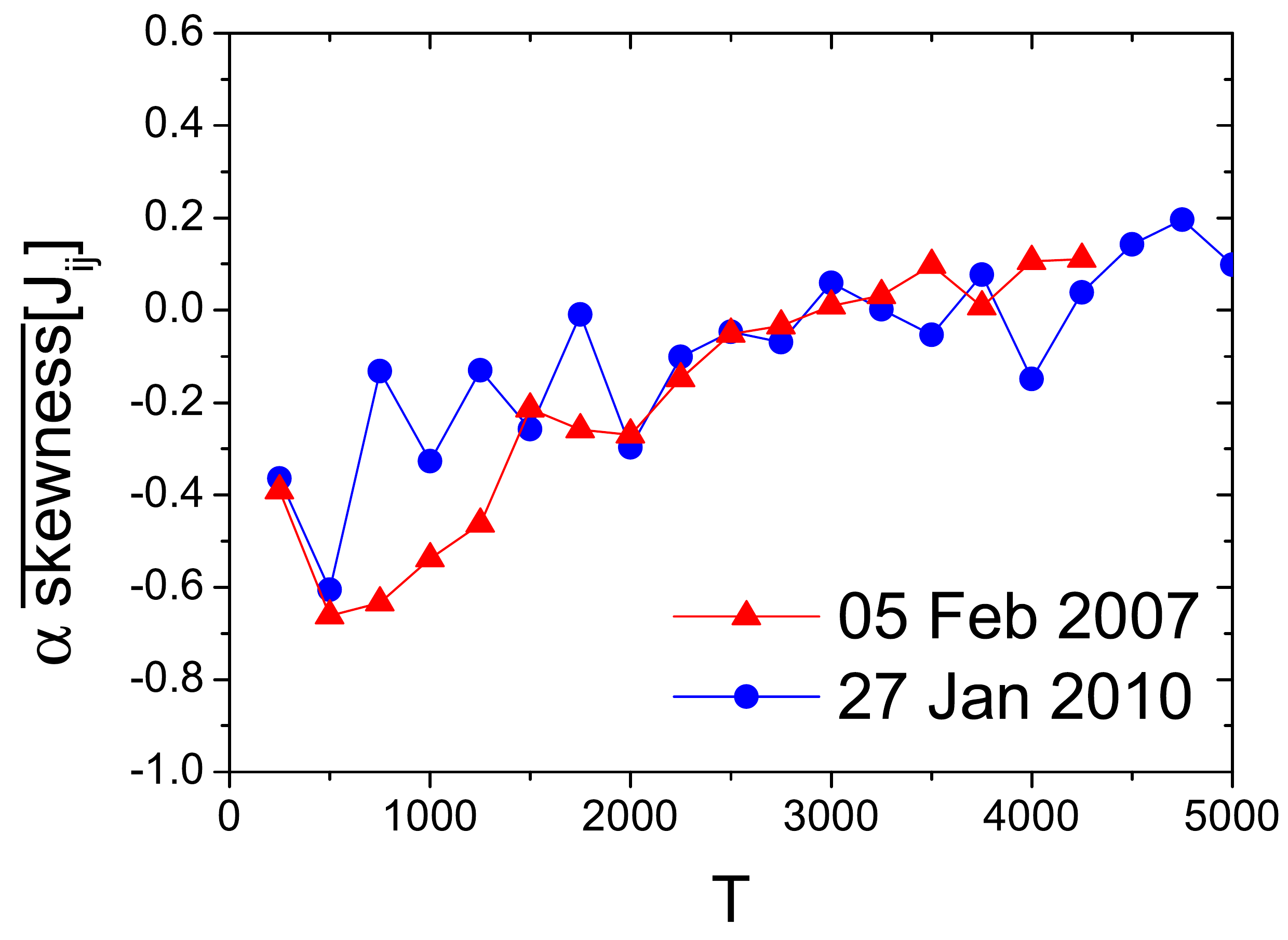}
}
\resizebox{0.24\textwidth}{!}{%
 \includegraphics{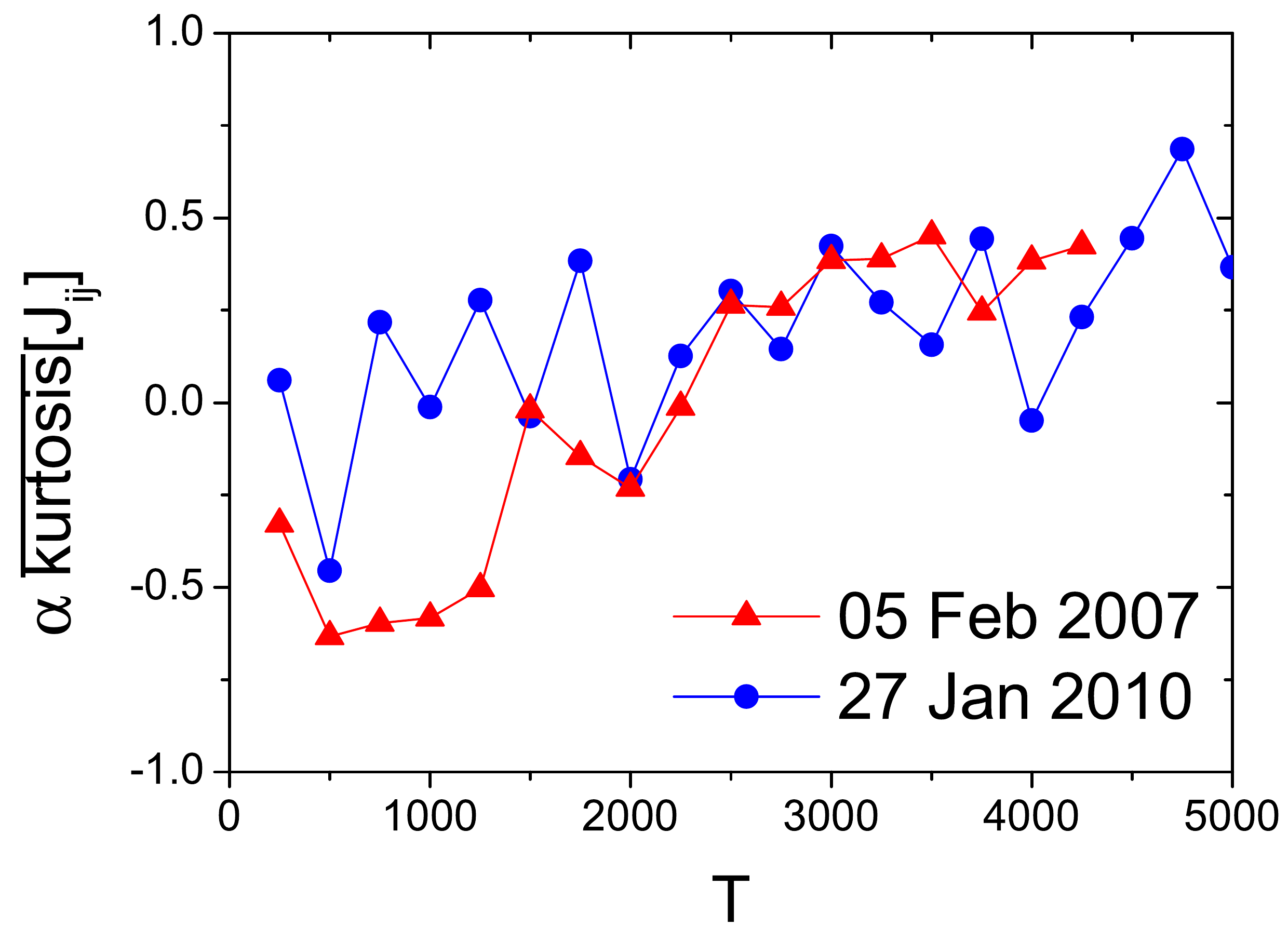}
}
 \caption{Average scaling exponent with number of stocks, $\alpha$, for the exact external fields (top row) and couplings (bottom row) distributions depending on moving window size, $T$, for two different historical dates and randomly shuffled time series for 27 Jan 2010. Scaling exponent of the mean of external fields is close to $-0.75$ for big values of $T$, while the higher moments behave similar to the external fields inferred on randomly shuffled time series. Empirical couplings scale similar to the couplings inferred on randomly shuffled time series for small window sizes, while their scaling properties become closer to the properties of the Gaussian distribution for bigger values of $T$.}
 \label{fig:Eq_scaling_T}
\end{figure*}

As Fig.~\ref{fig:Eq_scaling_T} (top row) shows, the distribution of external fields does not possess any particular scaling law, except the mean, which has $\alpha$ close to $-0.75$. The other moments scale similar to the corresponding moments of the external fields inferred on randomly shuffled time series. Scaling properties of distribution of couplings depends on the size of moving window (Fig.~\ref{fig:Eq_scaling_T}, bottom row). When $T$ is small, empirical couplings show similar scaling features to the couplings inferred on randomly shuffled returns. However, scaling of the mean and standard deviation becomes closer to the properties of the normal distribution with growth of $T$. This dependence might be related to the presence of finite-size effects, when use of a small number of historical values is not enough to estimate true correlations on the market. These results are similar to the scaling properties obtained in Ref.~\cite{bury1}, however their SMA window dependence has been shown for the first time.

\begin{figure*}
\centering
\resizebox{0.31\textwidth}{!}{%
 \includegraphics{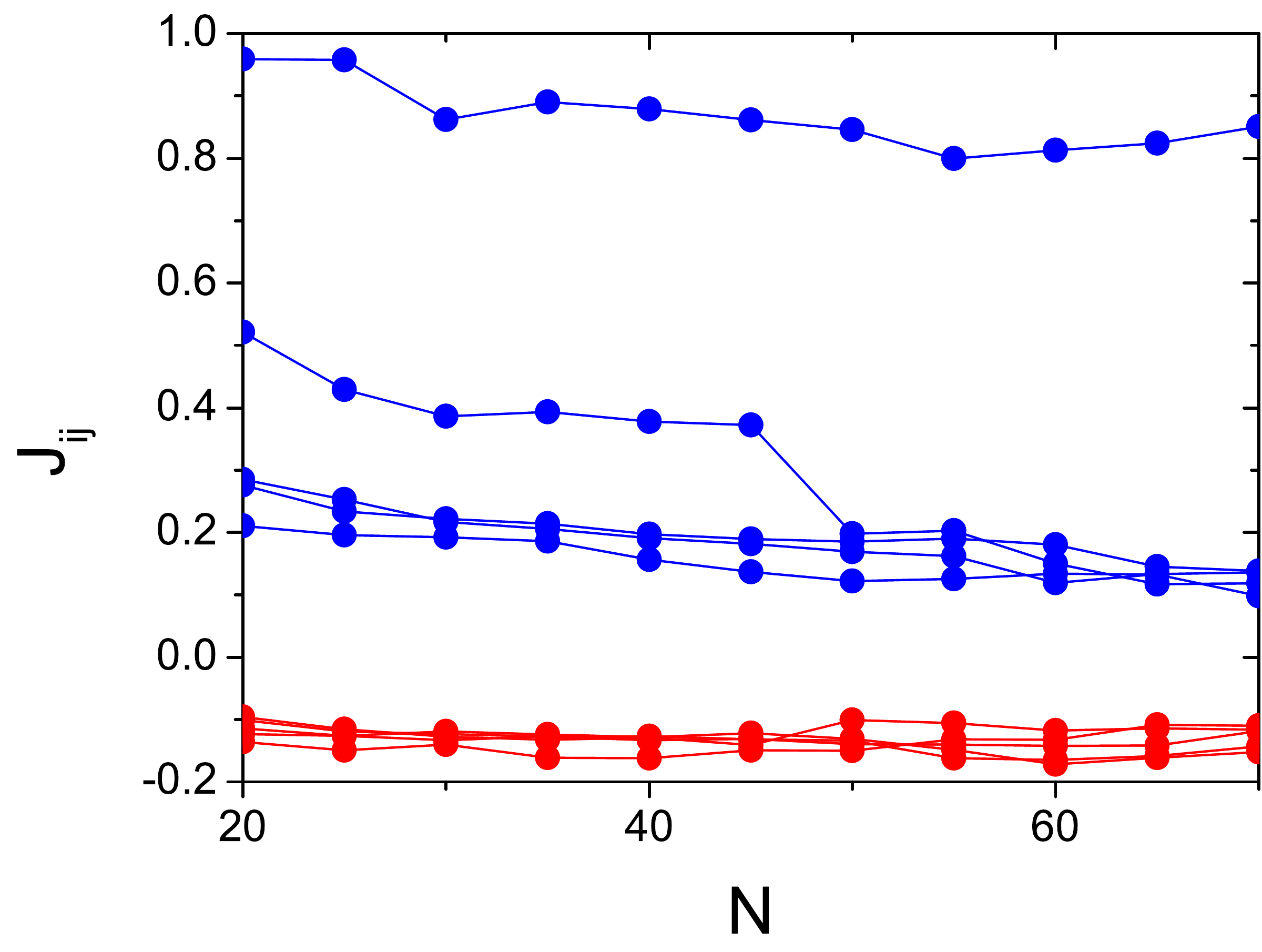}
}
\resizebox{0.29\textwidth}{!}{%
 \includegraphics{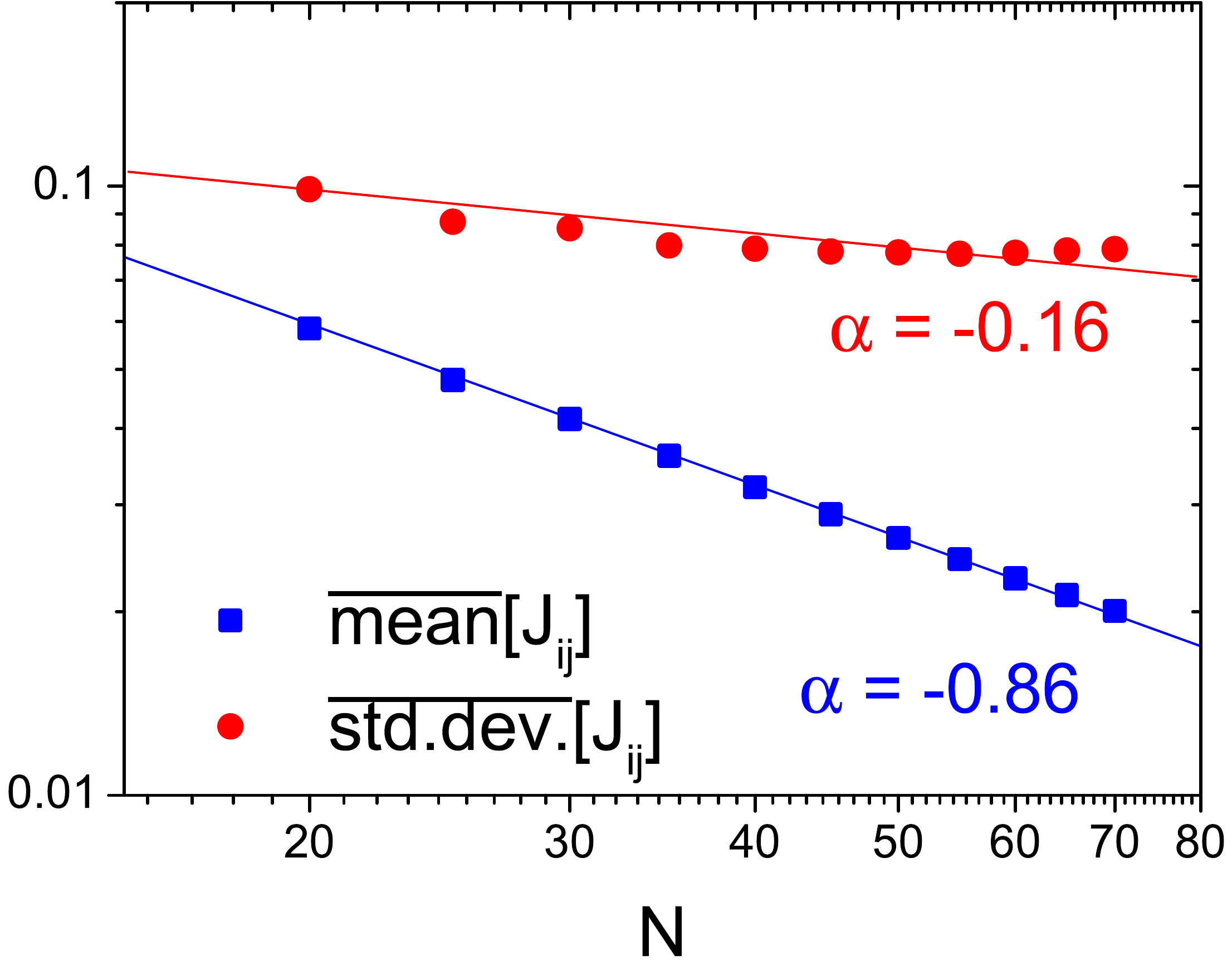}
}
\resizebox{0.31\textwidth}{!}{%
 \includegraphics{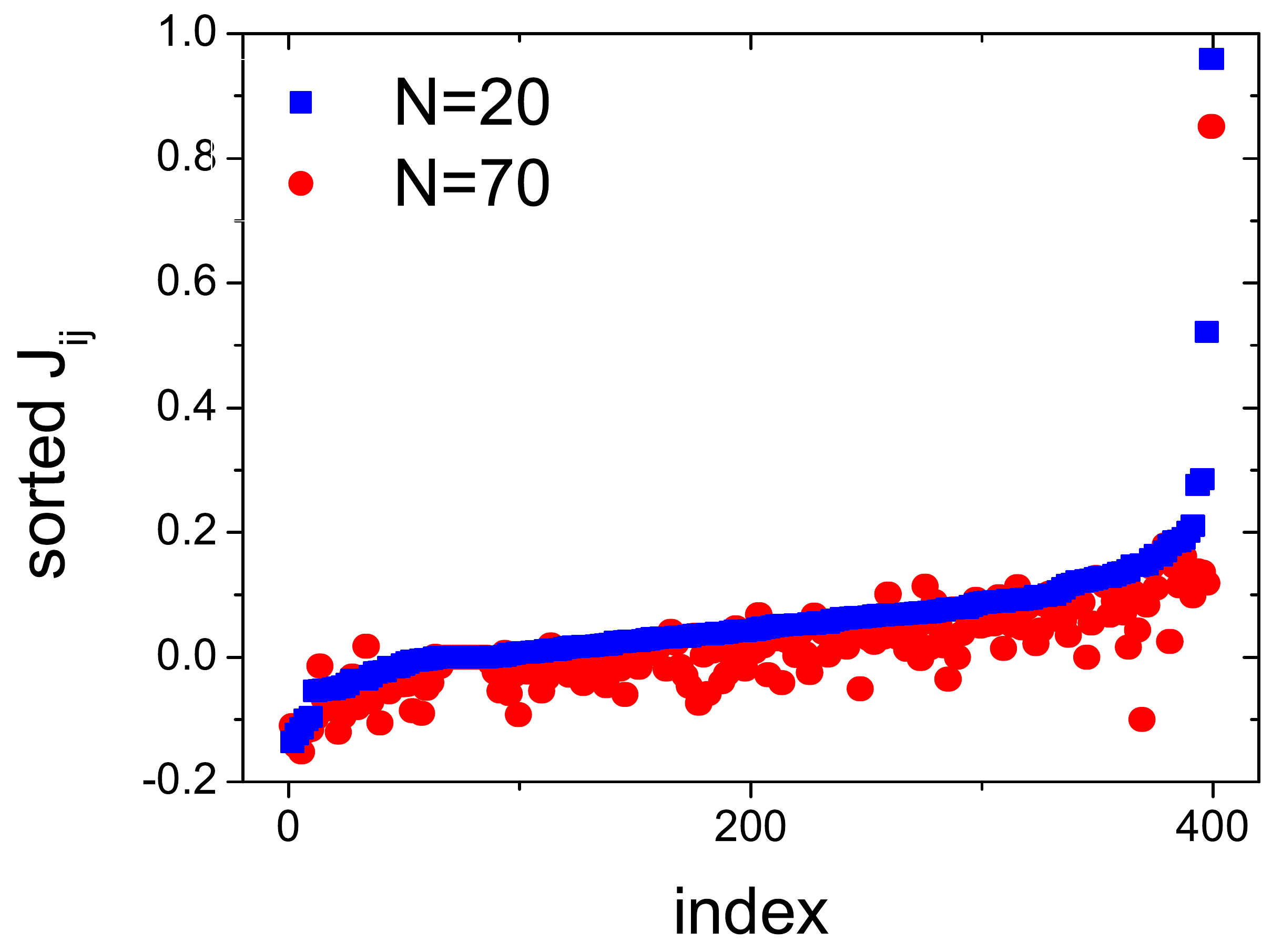}
}
\resizebox{0.31\textwidth}{!}{%
 \includegraphics{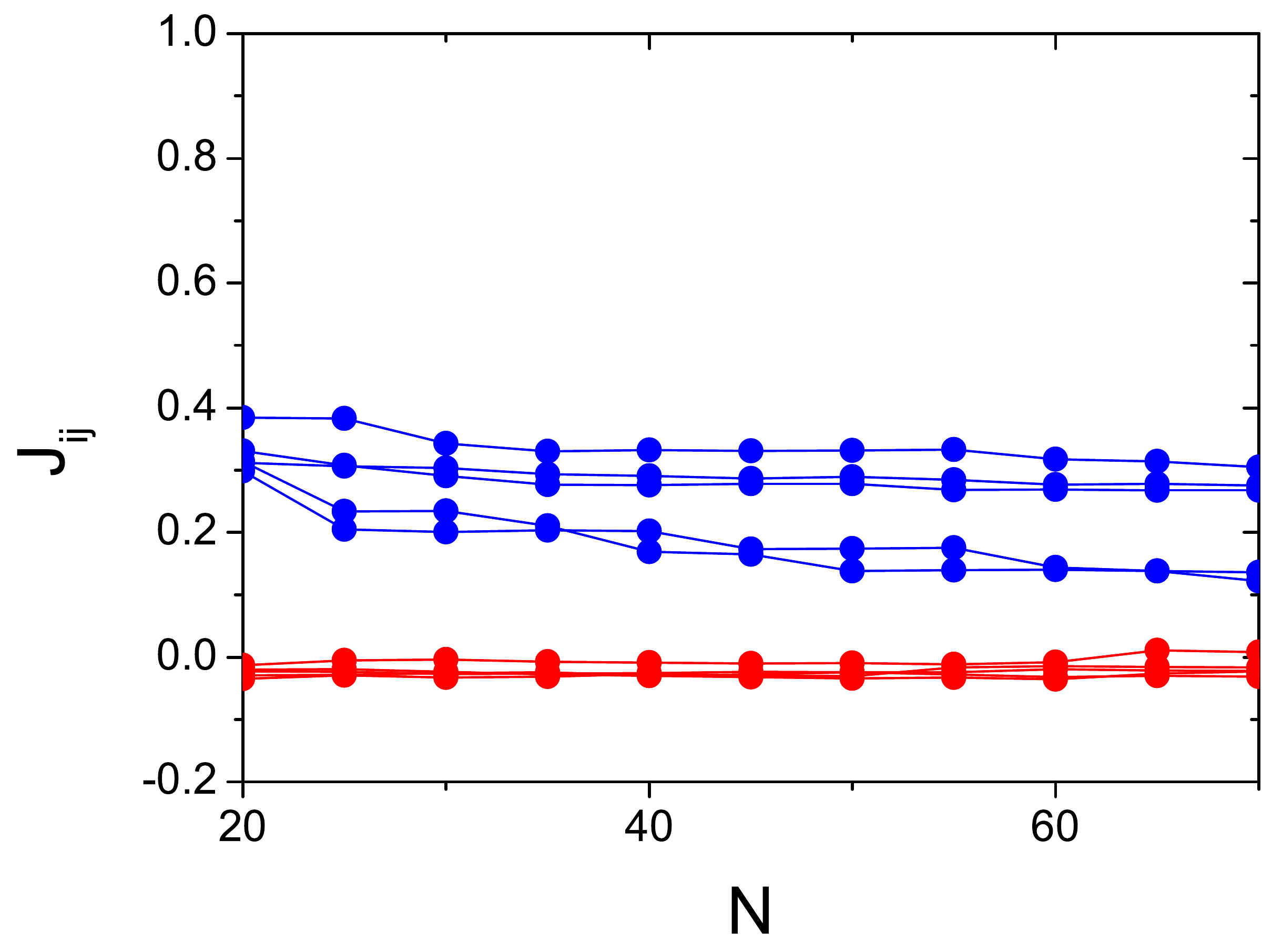}
}
\resizebox{0.29\textwidth}{!}{%
 \includegraphics{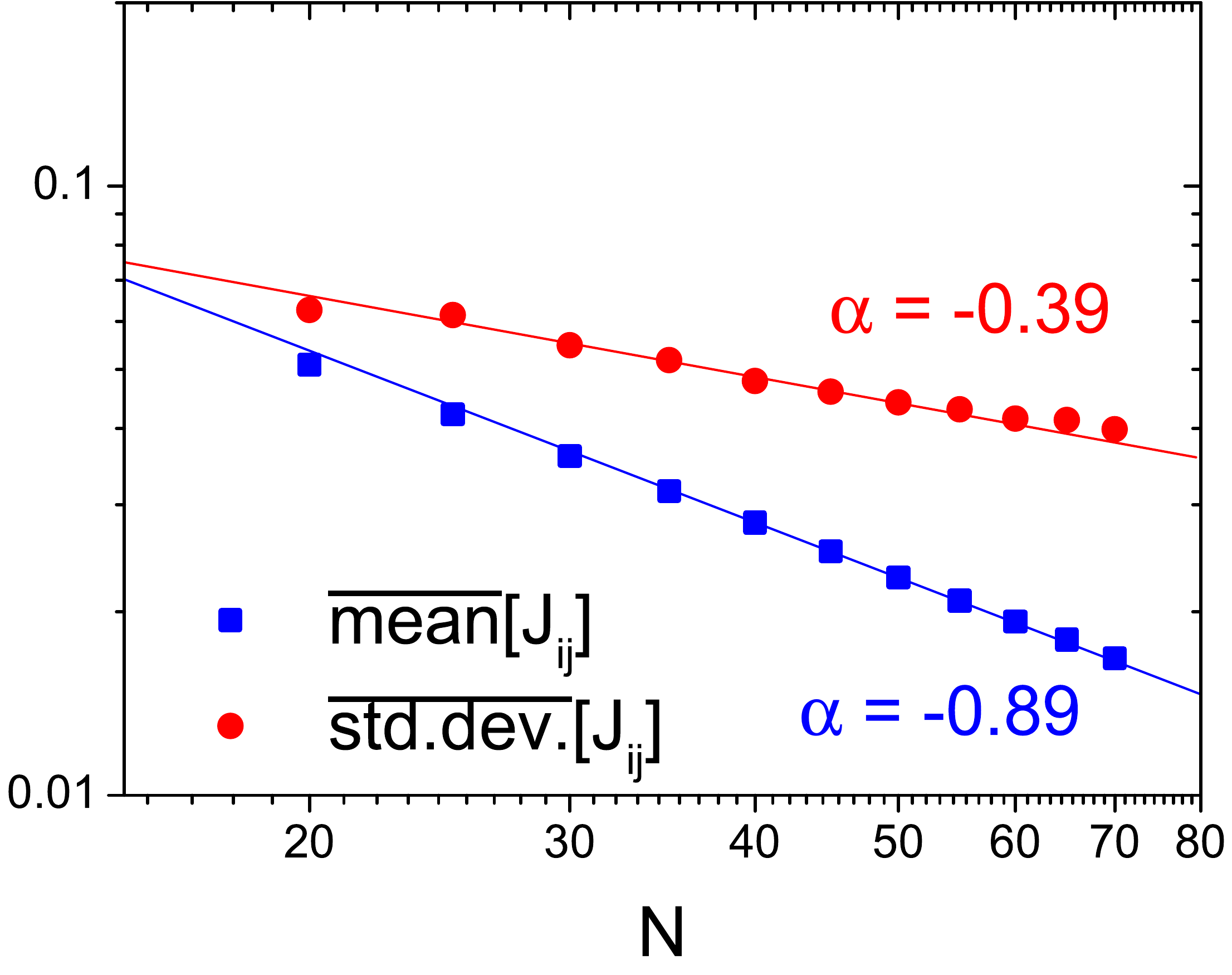}
}
\resizebox{0.31\textwidth}{!}{%
 \includegraphics{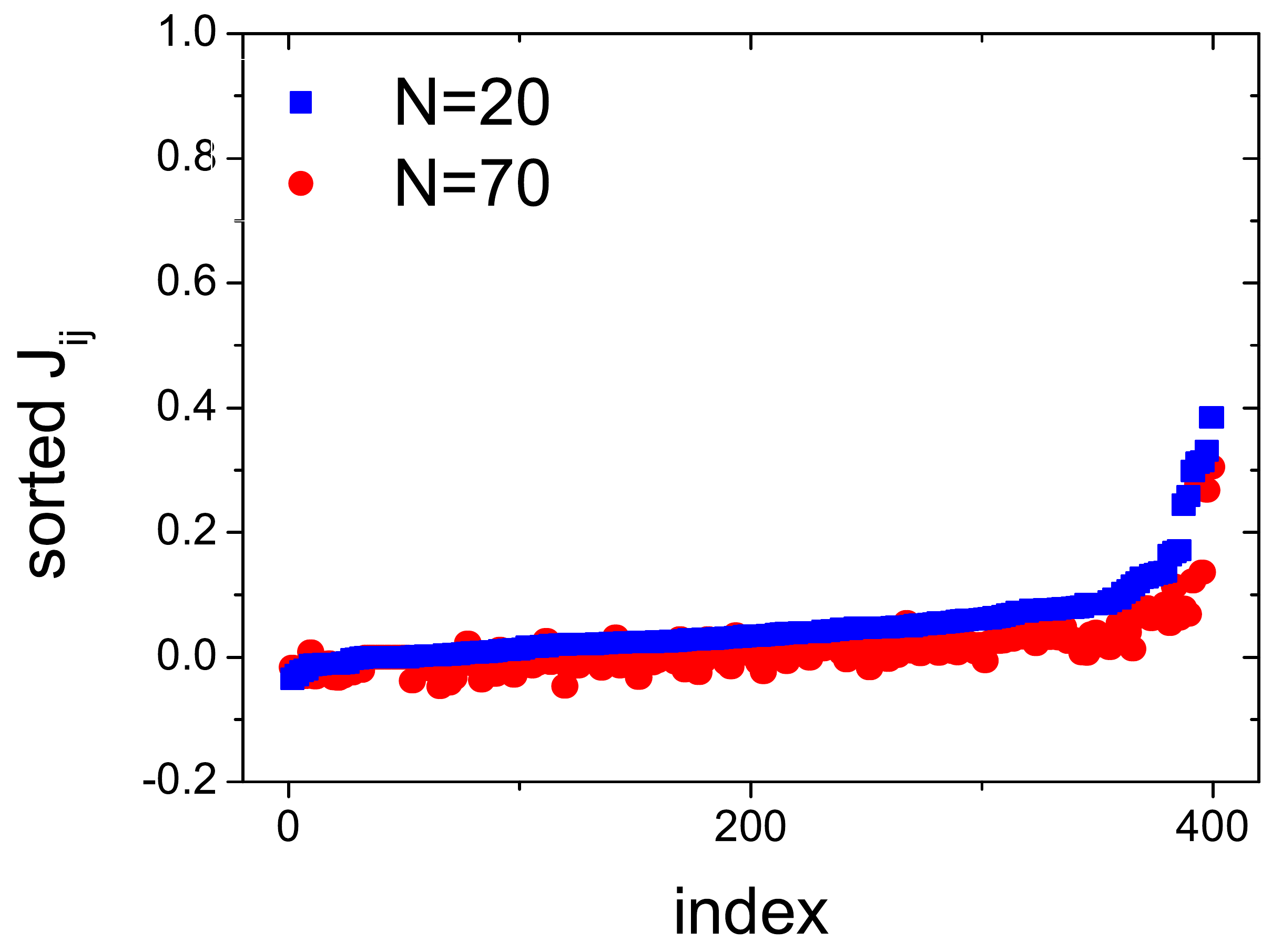}
}
 \caption{Scaling of the 380 couplings between a randomly selected subset of $20$ stocks depending on total number of stocks used for the inference (including original 20) for 27 Jan 2010 and $T=1000$ (top row) and $T=5000$ (bottom row): 10 biggest/smallest couplings (first column), mean and standard deviation of the couplings (second column) and their visualization (third column). Relative magnitude of the couplings decreases as the number or stocks grow, while the underlying network structure is preserved.}
 \label{fig:Eq_scaling_subset_h_J}
\end{figure*}

There are two extreme ways in which a change in the distribution of couplings can arise as one increases the size of the observed system. One is that the structure between the previous stocks entirely changes by adding new stocks. Alternatively, the couplings could only change their absolute magnitude, while they maintain their magnitude relative to one another. To better understand where in this spectrum our financial market exists, we performed analysis similar to Ref.~\cite{PhysRevE.79.051915}: We chose a random subset of $20$ stocks and analyzed couplings between them for different total number of stocks taken into account for the inference (including the original 20). Figure~\ref{fig:Eq_scaling_subset_h_J} shows that the biggest/smallest values of $J_{ij}$ remain the same with growth of $N$ and their scaling becomes closer to the normal distribution for bigger time windows. This behavior also suggests that important features of market connectivity are preserved with the number of stocks.

%-------------------------------------------------------------------------------
\subsection{External and internal influence}
\label{sec:external_vs_internal}
\begin{figure*}
\centering
\resizebox{0.75\textwidth}{!}{%
 \includegraphics{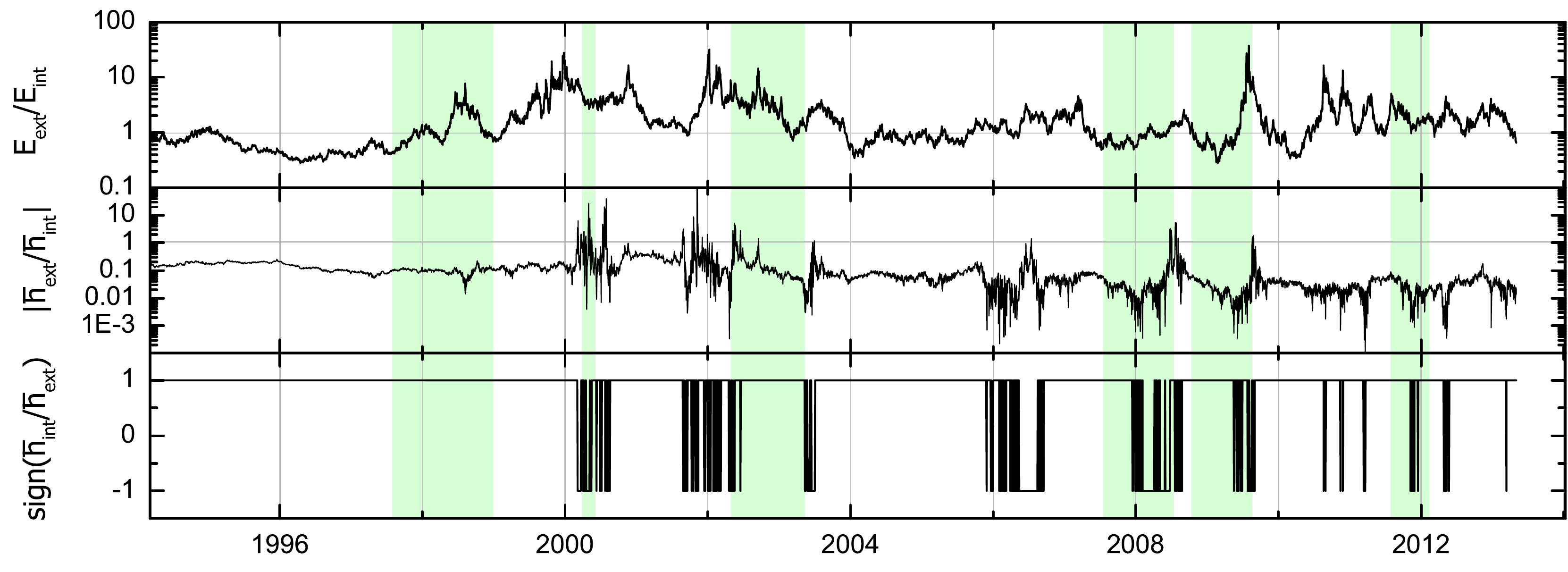}
}
 \caption{Comparison of external and internal biases for the exact Ising model. Top-bottom: ratio of external and internal energies contributing to the Hamiltonian, absolute value of the ratio of the mean external and internal biases and its sign. Major market crashes are preceded by growth of the external energy and discrepancy between the biases.}
 \label{fig:Eq_Exact_int_vs_ext}
\end{figure*}

%-------------------------------------------------------------------------------
Considering the two terms in the system's Hamiltonian [Eq.~(\ref{eq:distrib_eq})], it is also possible to define internal and external influences in the market. For this purpose, external fields can be interpreted as the influence of external factors, $\mathbf{h}^\mathrm{ext} \equiv \mathbf{h}$, while couplings define internal bias, $\mathbf{h}^\mathrm{int} = \langle\mathbf{s}^\intercal\rangle \mathbf{J}$ (in the MF sense) \cite{bury1}. In this case, external contribution corresponds to the individual stocks biases which come from outside the market, while internal one is solely defined in terms of internal market interactions. Similarly, one can also define two energy terms as $\mathcal{H} = E^\mathrm{ext} + E^\mathrm{int}$, where $E^\mathrm{ext,int} = - \left(\mathbf{h}^\mathrm{ext,int}\right)^\intercal \langle\mathbf{s}\rangle$. Figure~\ref{fig:Eq_Exact_int_vs_ext} shows that both energies have almost the same order of magnitude over the historical period considered, while near the major crashes $E^\mathrm{ext}$ is more than 10 times bigger than $E^\mathrm{int}$. The ratio between the mean biases also possesses interesting historical dynamics. Being in principle strongly correlated with the mean return ($0.9$ for $\overline{h}^\mathrm{ext}$ and $0.99$ for $\overline{h}^\mathrm{int}$) discrepancies between them might be used as a leading indicator of financial instabilities. Away from the periods of crisis, their ratio is almost stable, while divergent behavior is observed before the U.S. market crashes (two bottom panels in Fig.~\ref{fig:Eq_Exact_int_vs_ext}). Possible explanation of the observed behavior from a financial point of view is still an open question.

%-------------------------------------------------------------------------------
\section{Conclusion}
\label{sec:Discussion}
%-------------------------------------------------------------------------------
We have investigated various aspects of application of the pairwise interaction model to financial time series. The model, being parametrized by external fields and couplings, is used for the approximation of the joint equilibrium distribution of stock returns. Since the considered learning algorithms require use of binary variables, the logarithmic returns are binarized using the sign function. Effect of the binarization suggests that the distribution of binary returns captures the distributions of raw and standardized returns to a good degree, preserving information about market correlation structure, economic cycles and market crashes as well as industry-related market clustering structure.

The model parameters are inferred using approximate and exact learning algorithms. Mean field approximations nicely recover bulk of the couplings except outliers, which can be correctly inferred using small correlation expansions. External fields are in the almost perfect agreement with the exact algorithm if the diagonal-weight trick is used. The obtained results also suggest that the quality of approximate inference methods drops in the periods of financial crises due to increase of magnitude of the correlations observed. 
For the historical period considered, the distribution of external fields is close to the Gaussian and does not possess any outliers independently of moving window size. On the contrary, the distribution of couplings possesses a heavy positive tail, which starts to dominate over the Gaussian bulk for bigger moving window sizes. Mean of external fields decreases with the number of stocks while their standard deviation remains almost constant, corresponding to the standard deviation of the external fields inferred on randomly shuffled time series. Scaling properties of the distribution of couplings depends on the moving window size, becoming closer to the properties of the Gaussian distribution with its growth. 
Despite possible presence of finite-size effects, an industry-related clustering structure is observed for both exact and approximate couplings. The performed cutoff analysis suggests that the biggest positive couplings as well as eigenmodes with the biggest eigenvalues contain all information about this structure. Scaling properties of the couplings between a small random subset of stocks suggest that the underlying network structure is also preserved with the number of stocks. 
Finally, the pairwise interaction model also allows one for defining the external and internal biases which correspond to contribution of external fields and couplings respectively. Discrepancies between them might be used as a precursor of imminent financial instabilities and should be explained from a financial point of view. These non-trivial market properties as well as their historical evolution will be studied in the future works, where one of the natural extensions to incorporate dynamics is to consider the kinetic Ising model \cite{PhysRevLett.106.048702}.

%-------------------------------------------------------------------------------
\section{Acknowledgments}
%-------------------------------------------------------------------------------
This work is supported by Nordita, VR VCB 621-2012-2983, U.S. DOE, the Marie Curie Training Network NETADIS (FP7, grant 290038), the Kavli Foundation and the Norwegian Research Councils Centre of Excellent Scheme. We are also grateful to the anonymous referees for their valuable suggestions.

\bibliography{ising_stocks}

\begin{table*}
\caption{
\bf{List of the companies which stock prices are used for the calculations in the paper.}}
\small
\begin{tabular}{lll|lll}
\hline
\hline
\noalign{\smallskip}
Ticker & Name & Sector & Ticker & Name & Sector \\
\noalign{\smallskip}
\hline
\noalign{\smallskip}
ABT & Abbott Laboratories & Hea &
AIG & American International Group, Inc. & Fin \\
AMGN & Amgen Inc. & Hea &
APA & Apache Corp. & Bas \\
APC & Anadarko Petroleum Corp. & Bas &
AAPL & Apple Inc. & Con \\
AXP & American Express Company & Fin &
BA & The Boeing Company & Ind \\
BAC & Bank of America Corp. & Fin &
BAX & Baxter International Inc. & Hea \\

BMY & Bristol-Myers Squibb Company & Hea &
C & Citigroup, Inc. & Fin \\
CAT & Caterpillar Inc. & Ind &
CELG & Celgene Corporation & Hea \\
CL & Colgate-Palmolive Co. & Con &
CMCSA & Comcast Corporation & Ser \\
COP & ConocoPhillips & Bas &
COST & Costco Wholesale Corp. & Ser \\
CSCO & Cisco Systems, Inc. & Tec &
CVS & CVS Caremark Corp. & Ser \\

CVX & Chevron Corp. & Bas &
DD & E. I. du Pont de Nemours and Co. & Bas \\
DE & Deere \& Company & Ind &
DELL & Dell Inc. & Tec \\
DHR & Danaher Corp. & Ind &
DIS & The Walt Disney Company & Ser \\
DOW & The Dow Chemical Company & Bas &
EMC & EMC Corporation & Tec \\
EMR & Emerson Electric Co. & Tec &
EOG & EOG Resources, Inc. & Bas \\

EXC & Exelon Corp. & Uti &
F & Ford Motor Co. & Con \\
GE & General Electric Company & Ind &
HAL & Halliburton Company & Bas \\
HD & The Home Depot, Inc. & Ser &
HON & Honeywell International Inc. & Ind \\
HPQ & Hewlett-Packard Company & Tec &
IBM & International Business Machines Corp. & Tec \\
INTC & Intel Corp. & Tec &
JNJ & Johnson \& Johnson & Hea \\

JPM & JPMorgan Chase \& Co. & Fin &
KO & The Coca-Cola Company & Con \\
LLY & Eli Lilly and Company & Hea &
LOW & Lowe's Companies Inc. & Ser \\
MCD & McDonald's Corp. & Ser &
MDT & Medtronic, Inc. & Hea \\
MMM & 3M Company & Cng &
MO & Altria Group Inc. & Con \\
MRK & Merck \& Co. Inc. & Hea &
MSFT & Microsoft Corp. & Tec \\

NKE & Nike, Inc. & Con &
ORCL & Oracle Corporation & Tec \\
OXY & Occidental Petroleum Corp. & Bas &
PEP & Pepsico, Inc. & Con \\
PFE & Pfizer Inc. & Hea &
PG & The Procter \& Gamble Company & Con \\
PNC & The PNC Financial Services Group & Fin &
SLB & Schlumberger Limited & Bas \\
SO & Southern Company & Uti &
T & AT\&T, Inc. & Tec \\

TGT & Target Corp. & Ser &
TJX & The TJX Companies, Inc. & Ser \\
TXN & Texas Instruments Inc. & Tec &
UNH & UnitedHealth Group Incorporated & Hea \\
UNP & Union Pacific Corp. & Ser &
USB & U.S. Bancorp & Fin \\
UTX & United Technologies Corp. & Ind &
VZ & Verizon Communications Inc. & Tec \\
WFC & Wells Fargo \& Company & Fin &
WMT & Wal-Mart Stores Inc. & Ser \\

XOM & Exxon Mobil Corp. & Bas &
& & \\
\noalign{\smallskip}
\hline
\hline
\end{tabular}
\begin{flushleft}
Industry sectors are defined as Basic Materials (Bas), Conglomerate (Cng), Consumer Goods (Con), Financial (Fin), Healthcare (Hea), Industrial Goods (Ind), Services (Ser), Technology (Tec) and Utilities (Uti).
\end{flushleft}
\label{tab:companies}
\end{table*}

\end{document}